\documentclass[twocolumn,showpacs,prb,citeautoscript,amsmath,amssymb,floatfix,eqsecnum]{revtex4-1}
\usepackage{amsmath,amssymb,color}
\usepackage{bm}
\usepackage{slashed}
\usepackage{graphicx}
\usepackage{psfrag}
\usepackage{bbold}
\usepackage{bbm}
\newcommand{\bd}{\bm}
\usepackage{ulem}

\begin{document}

\title{Spin functional renormalization group for quantum Heisenberg ferromagnets:  \\
Magnetization and magnon damping in two dimensions
}

\author{Raphael Goll, Dmytro Tarasevych, Jan Krieg, and Peter Kopietz}
  
\affiliation{Institut f\"{u}r Theoretische Physik, Universit\"{a}t
  Frankfurt,  Max-von-Laue Strasse 1, 60438 Frankfurt, Germany}

 \date{November 21, 2019}

 \begin{abstract}
We use the spin functional renormalization group
recently developed by two of us [J. Krieg and P. Kopietz, Phys. Rev. B {\bf{99}}, 060403(R) (2019)]
to calculate the
magnetization  $M ( H , T  )$ and the damping of magnons 
due to classical longitudinal fluctuations of
quantum Heisenberg ferromagnets.
In order to guarantee that for vanishing magnetic field $H \rightarrow 0$,
the magnon spectrum is gapless when the spin rotational invariance
is spontaneously broken, we use a Ward identity to express the  magnon self-energy in terms of the
magnetization.
In two dimensions our approach correctly predicts  
the absence of long-range magnetic order 
for $H=0$ at finite temperature $T$.
The magnon spectrum then exhibits a gap  from which we obtain the transverse correlation length. We also calculate the wave-function renormalization factor of the magnons.
As a mathematical by-product, we derive a 
recursive form of the generalized Wick theorem for spin operators in frequency space
which facilitates the calculation of arbitrary
time-ordered connected correlation functions 
of an isolated  spin in a magnetic field.

\end{abstract}


\maketitle

\tableofcontents

\section{Introduction}

Recently two of us have proposed a new approach to quantum spin systems based on a formally exact renormalization group equation for the
generating functional of the connected spin correlation functions \cite{Krieg19}.
Our method works directly with the physical spin operators, thus  
avoiding  any representation  of  the spin operators in terms of canonical
bosons or fermions acting on a projected Hilbert space.
A similar strategy was adopted half a century ago by
Vaks, Larkin, and Pikin (VLP) \cite{Vaks68,Vaks68b}, who developed an unconventional
diagrammatic approach to quantum spin systems based on a generalized Wick theorem 
for spin operators. A detailed description of this approach
can be found in the textbook by Izyumov and Skryabin \cite{Izyumov88}.
It turns out, however, that the diagrammatic structure of this approach is rather complicated, which is perhaps the reason why this method has not gained wide acceptance.
As pointed out in Ref.~[\onlinecite{Krieg19}], by combining the VLP approach\cite{Vaks68,Vaks68b} with modern
functional renormalization group (FRG) methods \cite{Berges02,Pawlowski07,Kopietz10,Metzner12}, we can
reduce the problem of calculating the spin correlation functions to the problem of solving
the bosonic version of the Wetterich equation \cite{Wetterich93} with a special initial 
condition
determined by the $SU(2)$-spin algebra. Our 
spin functional renormalization group (SFRG) approach
is an extension of the lattice non-perturbative renormalization group approach developed by Machado and 
Dupuis~\cite{Machado10}  for classical spin models, 
and by Ran\c{c}on and Dupuis~\cite{Rancon11a,Rancon11b, Rancon12a,Rancon12b,Rancon14} 
for bosonic quantum lattice models.
Another FRG approach to quantum spin systems is the so-called pseudofermion 
FRG\cite{Reuther10,Reuther11,Reuther11a,Buessen16},
which uses the representation of  spin-$1/2$ operators in terms of Abrikosov 
pseudofermions \cite{Coleman15}
to approximate  the renormalization group flow of spin-$1/2$ quantum spin systems 
by a truncated fermionic FRG flow \cite{Kopietz01,Metzner12}.  
The pseudofermion FRG has been quite successful to map out the phase diagram of various types of
frustrated magnets without long-range magnetic order.\cite{Reuther10,Reuther11,Reuther11a,Buessen16}
In contrast to the pseudofermion FRG, our SFRG can be used for arbitrary spin $S$ because it 
does not rely on the representation of the spin operators in terms of auxiliary
degrees of freedom.

As a first application of our SFRG approach,
in Ref.~[\onlinecite{Tarasevych18}] we have shown how
Anderson's poor man's scaling equations for the anisotropic Kondo model emerge from 
a simple weak coupling truncation of the SFRG flow equations for the irreducible vertices
of this model. Our ultimate goal is to develop the SFRG method into a 
quantitative tool for studying 
quantum spin systems which is valid both in the magnetically ordered and the 
disordered phases.
To realize this program, it is important to test the method for model systems where
the physics is well understood and quantitatively accurate  
results can be obtained by means of other methods.
Here we will use the SFRG to study
the spin-$S$ ferromagnetic quantum Heisenberg model in two dimensions.
At any finite temperature $T > 0$ the magnetization
$M (  H , T > 0)$ of this model is  rigorously known \cite{Mermin66}  to vanish when the
external magnetic field $H$ approaches zero, so that
conventional spin-wave theory breaks down for $H \rightarrow 0$.
If we nevertheless try to calculate
the magnetization for $H =0$ perturbatively, we encounter infrared divergencies
signalling the breakdown of perturbation theory \cite{Hofmann02}.

Theoretical investigations of quantum Heisenberg ferromagnets in two dimensions have a long  
history.
In the 1980s several approximate analytical methods have been developed to calculate
the thermodynamics and the spectrum of renormalized magnons at finite temperature.
For example, in Takahashi's modified spin-wave theory \cite{Takahashi86,Takahashi87}  the 
vanishing of the magnetization for $H =0$ is enforced by adding an effective 
chemical potential to the magnon energies  which regularizes
the infrared divergence encountered in perturbation theory.
Note, however, that for finite magnetic field $H$ 
this strategy is not applicable  because then the magnetization 
does not vanish.  One possibility to generalize Takahashi's modified spin-wave theory
for finite magnetic field is to consider the spin-wave expansion of the 
Gibbs potential for fixed magnetization \cite{Kollar03}.
Another useful  method is based on the 
representation of the spin operators in terms of Schwinger 
bosons \cite{Arovas88}. This representation maps the 
exchange interaction  between spins onto a quartic boson Hamiltonian with
an additional constraint on the physical Hilbert space. 
Often  a simple mean-field treatment of the  
resulting effective boson problem gives already sensible results, so that
Schwinger boson mean-field theory continues to be a popular  analytical method for
studying  spin systems  without long-range magnetic order.
However, going beyond the mean-field approximation
is rather difficult within the Schwinger boson 
approach \cite{Chubukov91,Trumper97,Timm98,Ghioldi18}.
Another analytical approach to quantum ferromagnets is based on the  decoupling
of the equations of motion for the Green functions of the 
spins \cite{Tyablikov67,Junger04,Junger08}.
Of course, the  physical properties  of quantum ferromagnets can  be obtained with high accuracy 
numerically  using Monte Carlo simulations \cite{Kopietz89,Kopietz89b,Henelius00}.

In Ref.~[\onlinecite{Kopietz89}]
the correlation length and the susceptibility of a two-dimensional
quantum Heisenberg ferromagnet have been calculated
via a  special implementation of the 
momentum-shell Wilsonian renormalization group (RG)  technique
using  the Holstein-Primakoff transformation to express the 
spin operators in terms of  bosons.  
At the level of a one-loop approximation the results for the correlation length and the
susceptibility agree
with modified spin-wave theory~\cite{Takahashi86,Takahashi87}
 and Schwinger boson mean-field theory,\cite{Arovas88}
but the  two-loop corrections have been found to modify  
the one-loop results \cite{Kopietz89}.
In the present work, we  will show that the one-loop flow equations
derived in  Ref.~[\onlinecite{Kopietz89}]
can be obtained in a straightforward way within our
 SFRG formalism from a truncated flow
equation for the magnetization.
We then use the SFRG  to derive an improved  flow equation for the magnetization 
which takes into account self-energy and vertex corrections neglected in Ref.~[\onlinecite{Kopietz89}].
Moreover, we will also calculate the damping of spin-waves due to the coupling
to classical longitudinal spin fluctuations. This decay channel of magnons is
neglected in conventional spin-wave theory where the longitudinal fluctuations are not treated as independent degrees of freedom.

The rest of this work is organized as follows.
In Sec.~\ref{sec:exact}, we define  a new hybrid  generating functional 
$\Gamma_{\Lambda} [ \bd{m} , \phi ]$ depending
on the transverse magnetization $\bd{m}$ as well as on a longitudinal exchange field $\phi$
and show that this functional satisfies the Wetterich equation \cite{Wetterich93}.
In Sec.~\ref{sec:vertexexpansion} we give  the general structure of the expansion
of  $\Gamma_{\Lambda} [ \bd{m} , \phi ]$   
in powers of the fields and explicitly write down the
exact flow equations for the magnetization and the two-point vertices.
Section \ref{sec:thermo} is devoted to the explicit calculation of the magnetization
$M ( H , T )$ of a two-dimensional Heisenberg ferromagnet.
We establish the relation of our SFRG approach to
the momentum shell RG of Ref.~[\onlinecite{Kopietz89}]
and go beyond this  work 
by including self-energy and vertex corrections.
In Sec.~\ref{sec:damping}, we calculate the
damping of spin waves  in two-dimensional ferromagnets  due to the coupling to
classical longitudinal fluctuations. 
In Sec.~\ref{sec:conclusions}, we summarize our main results and point out possible 
extensions of our method.

In four appendices, we give  additional technical details. 
In Appendix~A, we derive  the relation between the
connected spin correlation functions and the
irreducible vertices generated by our hybrid functional $\Gamma_{\Lambda} [ \bd{m} , \phi ]$.
In Appendix~B, we give the  time-ordered connected spin correlation functions of a single spin in an external magnetic field  with up to four spins and derive the corresponding irreducible vertices; we also present a simple recursive form
of the generalized Wick theorem for spin operators in frequency space.
In Appendix~C, we write down
equations of motion for the time-ordered connected spin
correlation functions and derive a Ward identity which we use in the main text 
to close the flow equation for the magnetization.
 Finally, in Appendix~D, we derive initial conditions for 
the irreducible vertices in tree approximation where all terms involving loop
integrations over momenta are neglected.

\section{Exact flow equations for quantum spin systems}
 \label{sec:exact}

In this section and in the following  Sec. ~\ref{sec:vertexexpansion}, 
we consider a general anisotropic Heisenberg Hamiltonian of the form 
 \begin{equation}
 {\cal{H}} =  -    H   \sum_i S^z_i   + 
 \frac{1}{2} \sum_{ij} \left[ 
 J^{\bot}_{ij} \bd{S}^{\bot}_i \cdot \bd{S}^{\bot}_j +   J^z_{ij} {{S}}^z_i  {{S}}^z_j    \right],
 \label{eq:hamiltoniananiso}
 \end{equation}
where the external magnetic field $H$
is measured in units of energy,
$\bd{S}^{\bot}_i = ( S^x_i , S^{y}_i )$ 
is the transverse part of the spin operator $\bd{S}_i$,
and $J^{\bot}_{ij}$ and $J^{z}_{ij}$ are transverse and longitudinal exchange couplings.
In Sec.~\ref{sec:thermo},  we will
consider isotropic ferromagnets by specifying  $J^{\bot}_{ij} = J^{z}_{ij} = - V_{ij} < 0$, 
but at this point, we work with general anisotropic exchange couplings.
Following Ref.~[\onlinecite{Krieg19}], we now 
modify the Hamiltonian (\ref{eq:hamiltoniananiso}) by replacing $J^{\bot}_{ij}$ and $J^z_{ij}$ by deformed exchange couplings depending on a continuous parameter~$\Lambda$,
 \begin{eqnarray}
 J^{\bot}_{\Lambda, ij} & = & J^{\bot}_{ij} + R^{\bot}_{ \Lambda, ij},
 \; \; \; \; \; \; 
J^{z}_{\Lambda, ij}  =  J^{z}_{ij} + R^{z}_{\Lambda, ij},
 \end{eqnarray}
where the regulators $R^{\bot}_{ \Lambda, ij} $ and 
$R^{z}_{\Lambda, ij}$ should be chosen such that
for some initial $\Lambda = \Lambda_0$ the deformed model can be solved in a controlled way, and
for some final value of $\Lambda$ the regulators vanish so that we recover our original model. For example, $\Lambda$ can be a momentum scale acting as a cutoff for long-wavelength fluctuations, or simply a dimensionless parameter 
in the interval $[0,1]$ which multiplies the bare interaction.
At this point, it is not necessary to specify the deformation scheme.
Let us write the deformed  Hamiltonian in the form ${\cal{H}}_{\Lambda} =   {\cal{H}}_0 + {\cal{V}}_{\Lambda}$,
where
 \begin{eqnarray}
 {\cal{H}}_0 & = & 
 -    H  \sum_i S^z_i
 \label{eq:Hzdef}
\end{eqnarray}
is the Hamiltonian of isolated spins in a constant magnetic field and
 \begin{eqnarray}
 {\cal{V}}_{\Lambda} & = & 
 \frac{1}{2} \sum_{ij} \left[ 
 J^{\bot}_{\Lambda, ij}  \bd{S}^{\bot}_i \cdot \bd{S}^{\bot}_j +   J^z_{\Lambda, ij} {{S}}^z_i  {{S}}^z_j    \right]
 \label{eq:Vlambdadef}
 \end{eqnarray}
represents the coupling between the spins. 
The generating functional of the deformed Euclidean time-ordered spin correlation functions can then be written as \cite{Krieg19}
  \begin{equation}
 {\cal{G}}_{\Lambda} [ \bd{h} ]
= \ln {\rm Tr} \left[ e^{ - \beta {\cal{H}}_0 } {\cal{T}} e^{ \int_0^{\beta} 
 d \tau    [    \sum_i \bd{h}_i ( \tau ) \cdot {  {\bd{S}}_i ( \tau )  -  
 {\cal{V}}_{\Lambda} ( \tau )   ] } }  \right].
 \label{eq:Gcdef}
 \end{equation}
Here $\beta$ is the inverse temperature, ${\cal{T}}$ denotes time-ordering in imaginary time,
$\bd{h}_i ( \tau )$ are fluctuating source fields, and the time dependence of all operators is in the interaction picture with respect to ${\cal{H}}_0$. 
By simply taking  a derivative of both sides of Eq.~(\ref{eq:Gcdef}) with respect to
$\Lambda$ we can derive an exact functional flow equation for the generating functional
$ {\cal{G}}_{\Lambda} [ \bd{h} ]$. Moreover, the Legendre transform of 
$ {\cal{G}}_{\Lambda} [ \bd{h} ]$ satisfies an exact functional flow equation which is formally identical to the bosonic version of the Wetterich equation \cite{Krieg19,Wetterich93}.
A technical complication of this method is that in a scheme where
initially the exchange couplings are completely switched off the Legendre transform of
$ {\cal{G}}_{\Lambda=0} [ \bd{h} ]$ does not exist \cite{Rancon14,Krieg19} because in this limit the longitudinal spin fluctuations do not have any dynamics.
In Ref.~[\onlinecite{Krieg19}] we have already pointed out that  this problem can be avoided 
by working with the  
generating functional of the amputated connected correlation 
functions. Here we show how this idea is implemented  in practice. Actually, because
only the longitudinal fluctuations are initially static, it is convenient to introduce  a
hybrid functional ${\cal{F}}_{\Lambda} [ \bd{h}^{\bot} , s ]$
which generates connected correlation functions for transverse fluctuations,
but amputates the external legs associated with the longitudinal fluctuations.
Formally, this functional can be defined by
 \begin{eqnarray}
  {\cal{F}}_{\Lambda} [ \bd{h}^{\bot} , s ] & = &  
{\cal{G}}_{\Lambda} \Bigl[ \bd{h}^{\bot}_i, 
 h_i^z = - \sum_j {J}^{z}_{ \Lambda, ij}  s_j \Bigr]
 \nonumber
 \\
& - & 
 \frac{1}{2} \int_0^{\beta} d \tau \sum_{ij} 
 {J}^{z }_{\Lambda, ij} s_i ( \tau ) s_j ( \tau ),
 \hspace{7mm}
 \label{eq:Fdef}
 \end{eqnarray}
where $\bd{h}^{\bot}_i = ( h^x_i , h^y_i )$ is a  transverse 
magnetic source field, and
$s_i ( \tau )$ is a longitudinal source field which can be interpreted as a
fluctuating  magnetic moment in the direction of the external field.
Note that a similar hybrid functional of ``partially amputated connected'' correlation functions has been introduced earlier in Ref.~[\onlinecite{Schuetz06}] to 
derive partially bosonized FRG flow equations for interacting fermions.
By expanding    ${\cal{F}}^{}_{\Lambda} [ \bd{h}^{\bot} , s ]$  in powers of the source fields
$\bd{h}^{\bot}_i ( \tau )$ and
$s_i ( \tau )$, we obtain  connected spin correlation functions with the additional property that
external legs associated
with longitudinal propagators are amputated \cite{Krieg19}.
As a consequence, correlation functions
generated by 
  ${\cal{F}}^{}_{\Lambda} [ \bd{h}^{\bot} , s ]$ with longitudinal external legs
involve powers of the longitudinal interaction $J^z_{\Lambda, ij}$
and therefore vanish for $J^z_{\Lambda, ij} \rightarrow 0$.
For example, the longitudinal two-point function  is given by
 \begin{eqnarray}
 & & F_{\Lambda , ij } ( \tau , \tau^{\prime} )  = 
 \left.
\frac{ \delta^2 {\cal{F}}_{\Lambda} [ 
 \bd{h}^{\bot}_i =0, s ]}{       \delta s_i ( \tau )   \delta s_j ( \tau^{\prime} ) }
 \right|_{ s =0}
  \nonumber
 \\
 &  & = - \delta ( \tau - \tau^{\prime} ) J^z_{\Lambda , ij} +
 \sum_{ k l } J^{z}_{\Lambda , ik } J^{z}_{\Lambda, jl}  G^{zz}_{\Lambda, kl } ( \tau , \tau^{\prime} ),
 \label{eq:FGrelation}
 \end{eqnarray}
where
 \begin{equation}
 G^{zz}_{\Lambda, ij  } ( \tau , \tau^{\prime} ) 
 \left.
\frac{ \delta^2 {\cal{G}}_{\Lambda} [ 
 \bd{h} ]}{    \delta h^z_i ( \tau )  \delta h^z_j ( \tau^{\prime} )  }
 \right|_{ \bd{h} =0}
 \end{equation}
is the longitudinal part of the time-ordered two-spin correlation function. 
The relations between the higher order longitudinal correlation functions is for $n \geq 3$,
 \begin{eqnarray}
 & & F^{z \ldots z}_{\Lambda, i_1 \ldots i_n} (\tau_1, \ldots ,\tau_n) =
\frac{\delta^n \mathcal{F}_\Lambda[ \bd{h}^{\bot} =0 , s ] }{
 \delta s_{i_1} (\tau_1) \ldots \delta s_{i_n}(\tau_n) } \Big|_{ s=0}
\nonumber
\\
 &  & =  (-1)^n \sum_{ j_1 \ldots j_n } J_{\Lambda, i_1 j_1}^z \ldots 
 J_{\Lambda,i_n j_n}^z G_{\Lambda, j_1 \ldots j_n }^{z \ldots z} (\tau_{1},...,\tau_{n}),
 \label{eq:FnGn}
 \hspace{7mm}
\end{eqnarray}
where 
\begin{equation}
G_{\Lambda, i_1 \ldots i_n}^{z \ldots z} (\tau_1,...,\tau_n)=
\frac{\delta^n \mathcal{G}_\Lambda[\bd{h} ] }
{\delta h^z_{i_1} (\tau_1) \ldots \delta h^z_{i_n}(\tau_n)} \Big|_{\bd{h}=0}.
\end{equation}
If we work with a deformation scheme where initially all exchange couplings vanish,
all correlation functions generated by
$  {\cal{F}}_{\Lambda} [ \bd{h}^{\bot} , s ]$ involving longitudinal legs also vanish at the 
initial scale, so that at the first sight this functional does not give rise to a
convenient initial  condition for this cutoff scheme. 
However, the Legendre transform of the functional  ${\cal{F}}^{}_{\Lambda} [ \bd{h}^{\bot} , s ]$  has a well-defined limit for vanishing exchange couplings because the externals legs 
associated with the longitudinal propagators  are removed in the Legendre transform.
As usual, we subtract the regulator terms from the Legendre transform and define the
generating functional of the irreducible vertices as follows,
 \begin{eqnarray}
 & & \Gamma_{\Lambda} [ \bd{m} ,  \phi ]  
  =   
\int_0^{\beta} d \tau 
 \sum_i (  \bd{m}_i  \cdot \bd{h}^{\bot}_i    +  \phi_i s_i  ) 
-   {\cal{F}}_{\Lambda} [ \bd{h}^{\bot} ,  s ]
\nonumber
\\
&  &  -
 \frac{1}{2} \int_0^{\beta} d \tau \sum_{ij} \left( {R}^{ \bot}_{\Lambda, ij} 
 {\bd{m}}_i 
 \cdot {\bd{m}}_j 
    +    {R}^\phi_{\Lambda , ij}  \phi_i  \phi_j  \right),
 \hspace{7mm}
 \label{eq:GammaHMdef}
 \end{eqnarray}
 where the transverse and longitudinal regulators are given by
 \begin{eqnarray} 
 {R}^{ \bot}_{\Lambda, ij} & = &  {J}^{\bot}_{\Lambda , ij} - {J}^{\bot}_{ij} ,
 \\
 {R}^\phi_{\Lambda , ij} & = &  -  [ {\mathbbm{J}}^z_{\Lambda} ]^{-1}_{ij}
+ [  {\mathbbm{J}}^z ]^{-1}_{ij}.
 \end{eqnarray}
 Here ${\mathbbm{J}}^{z}_{ \Lambda}$ is a matrix on the spatial labels 
with matrix elements
 $    [ {\mathbbm{J}}^{z}_{ \Lambda} ]_{ij} =   {J}_{\Lambda, ij}^{z}$.
Note that on the right-hand side of Eq.~(\ref{eq:GammaHMdef})
the sources $\bd{h}_i^{\bot}( \tau ) $ and 
$s_i ( \tau )$ should be expressed in terms of the
transverse magnetization $\bd{m}_i ( \tau )$ and the
longitudinal exchange field $\phi_i ( \tau )$ by inverting  the relations
 \begin{eqnarray}
   \bd{m}_i ( \tau ) & = & 
\frac{ \delta {\cal{F}}_{\Lambda} [ \bd{h}^{\bot} , s ]}{\delta \bd{h}^{\bot}_i ( \tau ) }
 =  \langle {\cal{T}} \bd{S}^\bot_j ( \tau ) \rangle,
 \\
 \phi_i ( \tau )   & = &
\frac{ \delta {\cal{F}}_{\Lambda} [ \bd{h}^{\bot} , s ]}{\delta s_i ( \tau ) } 
 \nonumber
 \\
 & = & -
 \sum_j {J}^z_{\Lambda, ij} \left[ s_j ( \tau )  +  \langle {\cal{T}} S^z_j ( \tau ) \rangle
 \right],
 \label{eq:hexpec}
 \end{eqnarray}
where the expectation values
 $ \langle {\cal{T}} \bd{S}^\bot_j ( \tau ) \rangle $
and
$ \langle {\cal{T}} S^z_j ( \tau ) \rangle $ should be evaluated for finite sources
$\bd{h}^{\bot}_i ( \tau )$ and $s_i ( \tau )$. 
From the last line in Eq.~(\ref{eq:hexpec}), we see that
for  $s_i ( \tau ) =0$ the field $\phi_i ( \tau ) $ can be identified
with the exchange correction to the external magnetic field.
Obviously, if we use  Eq.~\eqref{eq:GammaHMdef}  to express  $s_i ( \tau ) $ as a functional of 
$\phi_i ( \tau )$ we obtain a factor of $   [ {\mathbbm{J}}^{z}_{ \Lambda} ]^{-1}_{ij}$ which cancels
the factors of $J^{z}_{\Lambda , ij}$ in the expansion of the functional
 ${\cal{F}}_{\Lambda} [ \bd{h}^{\bot} , s ]$ in powers of the longitudinal sources $s_i ( \tau )$.
The irreducible vertices generated by 
$\Gamma_\Lambda[\bm{m},\phi]$  are therefore well-defined
even if we use a deformation scheme where initially all exchange couplings vanish. Another way to see this is to explicitly express
 the vertices generated by
$\Gamma_\Lambda[\bm{m},\phi]$
in terms of the connected spin correlation functions, see Appendix~A.

It is now straightforward to derive formally exact FRG flow equations of the two functionals ${\cal{F}}_{\Lambda} [ \bd{h}^{\bot} , s ] $ and $\Gamma_{\Lambda} [ \bd{m} ,  \phi ]$. 
Following the procedure outlined in Ref.~[\onlinecite{Krieg19}], we find that the functional 
${\cal{F}}_{\Lambda} [ \bd{h}^{\bot} , s ] $ satisfies the flow equation
 \begin{widetext}
  \begin{eqnarray}
 & & \partial_{\Lambda} {\cal{F}}_{\Lambda} [ \bd{h}^{\bot} , s ] 
   =  
 -
\frac{1}{2} \int_0^{\beta} d \tau 
 \sum_{ij}  ( \partial_{\Lambda} {J}^{\bot}_{\Lambda, ij} ) \sum_{\alpha =x,y}
 \Biggl[\frac{ \delta^2 {\cal{F}}_{\Lambda} [ \bd{h}^\bot , s  ] }{\delta h_i^{\alpha} ( \tau )   
\delta h_j^{\alpha} ( \tau ) }
 + 
\frac{ \delta {\cal{F}}_{\Lambda} [ \bd{h}^{\bot} , s  ] }{\delta h_i^{\alpha} ( \tau ) }
\frac{ \delta {\cal{F}}_{\Lambda} [ \bd{h}^{\bot} , s  ] }{\delta h_j^{\alpha} ( \tau ) }
  \Biggr]
 \nonumber
 \\
 &  &  +
\frac{1}{2} \int_0^{\beta} d \tau 
 \sum_{ij} \left( \partial_{\Lambda} [ {{\mathbbm{J}}}^{z}_{ \Lambda}]_{ij}^{-1} \right)
 \Biggl[\frac{ \delta^2 {\cal{F}}_{\Lambda} [ \bd{h}^{\bot} , s   ] }{\delta s_i ( \tau )   
\delta s_j  ( \tau ) }
 +
\frac{ \delta {\cal{F}}_{\Lambda} [ \bd{h}^{\bot} , s  ] }{\delta s_i ( \tau ) }
\frac{ \delta {\cal{F}}_{\Lambda} [ \bd{h}^{\bot} , s  ] }{\delta s_j ( \tau ) }
  \Biggr]
  + \frac{1}{2} {\rm Tr} \left[ {\mathbf{J}}^{z}_{ \Lambda} \partial_\Lambda 
 ( {{\mathbf{J}}}^{z}_{\Lambda} )^{-1} \right].
 \label{eq:flowGm}
 \end{eqnarray}
Here
${\mathbf{J}}^{z}_{ \Lambda} $ is a matrix 
in all labels (spin component, lattice site, imaginary time) with matrix elements defined by
 \begin{eqnarray}
 {[} \mathbf{{J}}^z_{\Lambda} ]^{\alpha \alpha^{\prime}}_{ i \tau , j \tau^{\prime} }
  & = &   
 \delta ( \tau - \tau^{\prime} )  \delta_{\alpha z} \delta_{ \alpha^{\prime} z}     
[ {\mathbbm{J}}^{z}_{ \Lambda} ]_{ij}
=
\delta ( \tau - \tau^{\prime} )  
\delta_{\alpha z} \delta_{ \alpha^{\prime} z}    {J}^{z}_{\Lambda, ij}.
 \end{eqnarray}
With this notation
the trace in the last term of Eq.~(\ref{eq:flowGm}) is over all labels;
the formally divergent $\delta$-function $\delta ( \tau =0)$ hidden in the 
last term of Eq.~(\ref{eq:flowGm})
cancels when we combine this term with a similar contribution from the second term. 
Differentiating both sides of Eq.~(\ref{eq:GammaHMdef}) with respect to the deformation parameter $\Lambda$ and using Eq.~\eqref{eq:flowGm}, we obtain 
 \begin{eqnarray}
  & & \partial_{\Lambda} \Gamma_{\Lambda} [ \bd{m} , \phi ]
 = 
  \frac{1}{2} \int_0^{\beta} d \tau \sum_{ij} \biggl[ 
  \sum_{ \alpha = x,y} 
 \frac{ \delta^2 {\cal{F}}_{\Lambda} [ \bd{h}^\bot , s  ] }{\delta h_i^{\alpha} ( \tau )   
\delta h_j^{\alpha} ( \tau ) }   \partial_{\Lambda} R^{ \bot}_{\Lambda, ij} 
 + \frac{ \delta^2 {\cal{F}}_{\Lambda} [ \bd{h}^{\bot} , s  ] }{\delta s_i ( \tau )    \delta s_j  ( \tau ) }  \partial_{\Lambda}  R^\phi_{\Lambda , ij}
 \biggr] 
 - \frac{1}{2} {\rm Tr} \left[ {\mathbf{J}}^{z}_{ \Lambda} \partial_\Lambda 
 ( {{\mathbf{J}}}^{z}_{\Lambda} )^{-1} \right].
 \label{eq:Gammaflow}
 \end{eqnarray}
\end{widetext}
Note that the quadratic subtractions in Eq.~(\ref{eq:GammaHMdef}) eliminate
the terms on the right-hand side of the flow equation (\ref{eq:flowGm}) corresponding to diagrams which are reducible with respect to cutting 
either a transverse propagator line or a longitudinal interaction line.
Finally,
we express the  second derivatives on the right-hand side of Eq.~(\ref{eq:Gammaflow})
in terms of the matrix 
  $\mathbf{\Gamma}^{\prime \prime} _{\Lambda} [ \bd{m} , \phi  ]$
of second functional derivatives of $\Gamma_{\Lambda} [ \bd{m} , \phi ]$ and obtain 
an exact  functional flow equation 
for our quantum spin system which formally resembles the
Wetterich equation \cite{Wetterich93},
 \begin{eqnarray}
& &  \partial_{\Lambda} \Gamma_{\Lambda} [ \bd{m} , \phi  ] 
 \nonumber
 \\
 &  & =
  \frac{1}{2} {\rm Tr} 
 \left\{ \left[
 \left(    \mathbf{\Gamma}^{\prime \prime} _{\Lambda} [ \bd{m} , \phi  ] +
 \mathbf{R}_{\Lambda} \right)^{-1}   + {\mathbf{{J}}}_{\Lambda}^z \right]  \partial_{\Lambda} \mathbf{R}_{\Lambda}    \right\}.
 \label{eq:Wetterichhybrid}
 \hspace{7mm}
 \end{eqnarray}
Here the matrix elements of
$\mathbf{\Gamma}^{\prime \prime} _{\Lambda} [ \bd{m} , \phi ]$
are  given by
 \begin{equation}
 \left(    \mathbf{\Gamma}^{\prime \prime} _{\Lambda} [ \bd{m} , \phi  ]
 \right)_{ i \tau \alpha , j \tau^{\prime} \alpha^{\prime}} =
 \frac{ \delta^2 \Gamma_{\Lambda} [ \bd{m} , \phi  ] }{
 \delta \Phi^{\alpha}_i ( \tau ) \delta \Phi^{\alpha^{\prime}}_j ( \tau^{\prime} )},
  \end{equation}
where we have  combined  the two components of  
$\bd{m}_i ( \tau )$ together with 
$\phi_i ( \tau )$ into a three-component field
 \begin{equation}  
 \left(   \begin{array}{c} \Phi_i^{x} \\ \Phi_i^y 
 \\ \Phi_i^z \end{array} \right)    = \left( \begin{array}{c} {m}^{x}_i  \\
   m^y_i  \\  \phi_i  \end{array}  \right),
 \label{eq:collectivefield}
 \end{equation}
 and
$\mathbf{R}_{\Lambda}$ is diagonal in the field labels  with matrix elements
 \begin{eqnarray}
{ [} \mathbf{R}_{\Lambda} ]^{xx}_{ i \tau , j \tau^{\prime} }
& = & { [} \mathbf{R}_{\Lambda} ]^{yy}_{ i \tau , j \tau^{\prime} }
=    \delta ( \tau - \tau^{\prime} ) 
        {R}^{\bot}_{\Lambda, ij},
 \\
{ [} \mathbf{R}_{\Lambda} ]^{zz}_{ i \tau , j \tau^{\prime} }
 & = &    \delta ( \tau - \tau^{\prime} )   R^\phi_{\Lambda , ij}.
 \end{eqnarray}
For finite external field $H$ or in the presence of a finite spontaneous magnetization
the expectation value $\langle S^z_i ( \tau ) \rangle $ is finite even for 
vanishing sources $s_i ( \tau )$. Then 
the functional $ \Gamma_{\Lambda} [ \bd{m} =0,  \phi ]$ is extremal 
for a finite value $\phi_{\Lambda}$ of the exchange field,
\begin{equation}
 \left.  \frac{ \delta \Gamma_{\Lambda} [ \bd{m} =0, \phi ]}{\delta \phi_i ( \tau ) }
 \right|_{ \phi =  {\phi}_{\Lambda} } =0,
 \label{eq:extremum}
 \end{equation}
where ${\phi}_{\Lambda}$ can be identified with the 
scale-dependent renormalization  of the external magnetic field
due to the exchange interaction.
It  is then convenient to shift the field $\phi_i ( \tau ) = {\phi}_{\Lambda} + \varphi_i ( \tau )$ 
and consider the flow of
 \begin{equation}
 \tilde{\Gamma}_{\Lambda} [ \bd{m} ,  \varphi ]  =
\Gamma_{\Lambda} [ 
 \bd{m} , {\phi}_{\Lambda} + \varphi ] ,
 \label{eq:Gammashift}
 \end{equation}
which is given by
 \begin{eqnarray}
 \partial_{\Lambda} \tilde{\Gamma}_{\Lambda} [ 
 \bd{m},  \varphi ] 
 &     =  &   
\left. \partial_{\Lambda} \Gamma_{\Lambda} [ 
 \bd{m} , \phi ]  \right|_{ \phi  \rightarrow
 {\phi}_{\Lambda} + \varphi }
 \nonumber
 \\
 & + &  
   \int_0^{\beta} d \tau \sum_i 
\frac{ \delta \tilde{\Gamma}_{\Lambda} [ 
 \bd{m} ,  \varphi  ]}{\delta {\varphi}_i ( \tau ) }
 \partial_{\Lambda} {\phi}_\Lambda
 \nonumber
 \\
 &  & \hspace{-22mm} =
 \frac{1}{2} {\rm Tr} 
 \left\{ \left[
 \left( \mathbf{\tilde{\Gamma}}^{\prime \prime}_{\Lambda} [ \bd{m} , \varphi ] +
 \mathbf{R}_{\Lambda} \right)^{-1}   + {\mathbf{J}}_{\Lambda}^z \right]  \partial_{\Lambda} \mathbf{R}_{\Lambda}    \right\}
 \nonumber
  \\
 & +  &  \int_0^{\beta} d \tau \sum_i 
\frac{ \delta \tilde{\Gamma}_{\Lambda} [ 
 \bd{m} ,  \varphi  ]}{\delta {\varphi}_i ( \tau ) }
 \partial_{\Lambda} {\phi}_\Lambda.
 \label{eq:WetterichHeisenberg}
 \end{eqnarray}

\section{Vertex expansion}
\label{sec:vertexexpansion}

\subsection{Exact flow equations}

To construct an approximate solution of the exact flow equation
(\ref{eq:WetterichHeisenberg}), we expand the functional 
$\tilde{\Gamma}_{\Lambda} [  \bd{m},  \varphi ] $ in powers of the transverse magnetization $\bd{m}_i ( \tau )$ and the fluctuation $\varphi_i ( \tau )$ of the longitudinal
exchange field. Then we obtain an infinite hierarchy of flow equations for the
irreducible vertices generated by $\tilde{\Gamma}_{\Lambda} [  \bd{m},  \varphi ] $.
It is convenient to formulate the expansion in momentum-frequency space.
Introducing a  collective label $K = ( \bd{k} , i \omega)$ for momentum 
$\bd{k}$ and Matsubara frequency $i \omega$,
the Fourier expansion of the fields can be written as
 \begin{subequations}
 \label{eq:FTdef}
 \begin{eqnarray}
 \bd{m}_i ( \tau ) & = & 
 \int_K e^{  i ( \bd{k} \cdot \bd{r}_i - \omega \tau )} \bd{m}_{ K },
 \\
 {\varphi}_i ( \tau ) & = & 
 \int_K e^{  i ( \bd{k} \cdot \bd{r}_i - \omega \tau )} {\varphi}_{ K },
 \end{eqnarray}
 \end{subequations}
where  $\int_K = ( \beta N )^{-1} \sum_{\bd{k} , \omega }$.
In terms of  the spherical components of the transverse magnetization,
 \begin{equation}
 m_K^+ = \frac{ m^x_K + i m^y_K }{\sqrt{2}}, \; \; \; 
 m_{K}^- = \frac{ m^x_{K} -  i m^y_{K} }{\sqrt{2}}
 = (m_{-K}^+)^{\ast},
 \label{eq:mpmrelation}
 \end{equation}
the vertex expansion of $\tilde{\Gamma}_{\Lambda} [  \bd{m},  \varphi ] $ up to fourth
order in the fields is of the form
 \begin{widetext}
\begin{eqnarray}
  \tilde{\Gamma}_{\Lambda} [ \bd{m} , \varphi ]  & = &   
  \beta N f_{\Lambda} 
+  \int_K \Gamma^{+-}_{\Lambda} ( K ) m^{-}_{-K } m^+_K
 +  \frac{1}{2!} \int_K \Gamma^{zz}_{\Lambda} ( K ) \varphi_{-K } \varphi_K
 \nonumber
 \\
& + &  \int_{K_1} \int_{K_2}
\int_{K_3} \delta ( K_1 + K_2 + K_3 )  \Gamma^{+-z}_{\Lambda} ( K_1 , K_2 ,  K_3 ) 
 m^-_{K_1} m^+_{K_2} \varphi_{ K_3 } 
 \nonumber
 \\
& + &
  \frac{1}{3!} \int_{K_1} \int_{ K_2} \int_{K_3} 
 \delta ( K_1 + K_2 + K_3 ) 
\Gamma^{zzz}_{\Lambda} ( K_1 , K_2 , K_3  ) \varphi_{K_1 } \varphi_{K_2} \varphi_{ K_3 }
\nonumber
 \\
 &   +  & \int_{K_1} \int_{K_2} \int_{K_3}  \int_{ K_4} 
\delta ( K_1 + K_2 + K_3 + K_4) 
 \biggl\{  \frac{1}{ (2! )^2}
  \Gamma^{++--}_{\Lambda} ( K_1, K_2 ,  K_3 ,  K_4 )  
 m^-_{K_1} m^-_{K_2} m^+_{ K_3 } m^+_{  K_4} 
 \nonumber
 \\
 &  & +
 \frac{1}{ 2! } 
  \Gamma^{+-zz}_{\Lambda} ( K_1, K_2 ,  K_3 ,  K_4 )  
 m^-_{K_1} {m}^+_{K_2} \varphi_{ K_3 } \varphi_{  K_4} 
+  \frac{1}{4!} 
\Gamma^{zzzz}_{\Lambda} ( K_1 , K_2 , K_3 , K_4 ) \varphi_{K_1 } \varphi_{K_2} \varphi_{ K_3 } 
 \varphi_{K_4} \biggr\}
+ \ldots \; , \hspace{7mm}
 \label{eq:Gammahcomplete}
 \end{eqnarray}
\end{widetext}
where $\delta ( K  ) = \beta N \delta_{ \bd{k} , 0 } \delta_{\omega , 0}$ and
we have omitted vertices with five and more external legs.
The superscripts attached to the vertices refer to the field types of the associated external legs. The fact
that in the expansion (\ref{eq:Gammahcomplete})
the field $m^-_K$ is associated with the superscript $^{+}$ in 
$\Gamma_{\Lambda}^{ \cdots + \cdots} ( \cdots K \cdots )$
is related to the fact that  $m^-_K$ is generated by differentiating
${\cal{G}}_{\Lambda} [ \bd{h} ]$ with respect to $h^+_K$, and  
 $m^+_K$  can be obtained by differentiating
${\cal{G}}_{\Lambda} [ \bd{h} ]$ with respect to
 $h^-_K$ (note the alternating superscripts).
The relation between the time-ordered spin correlation functions and the 
irreducible vertices defined via the
expansion (\ref{eq:Gammahcomplete}) is explicitly constructed 
in Appendix~A for vertices with up to four external legs.

\begin{figure}[tb]
 \begin{center}
  \centering
\vspace{7mm}
 \includegraphics[width=0.45\textwidth]{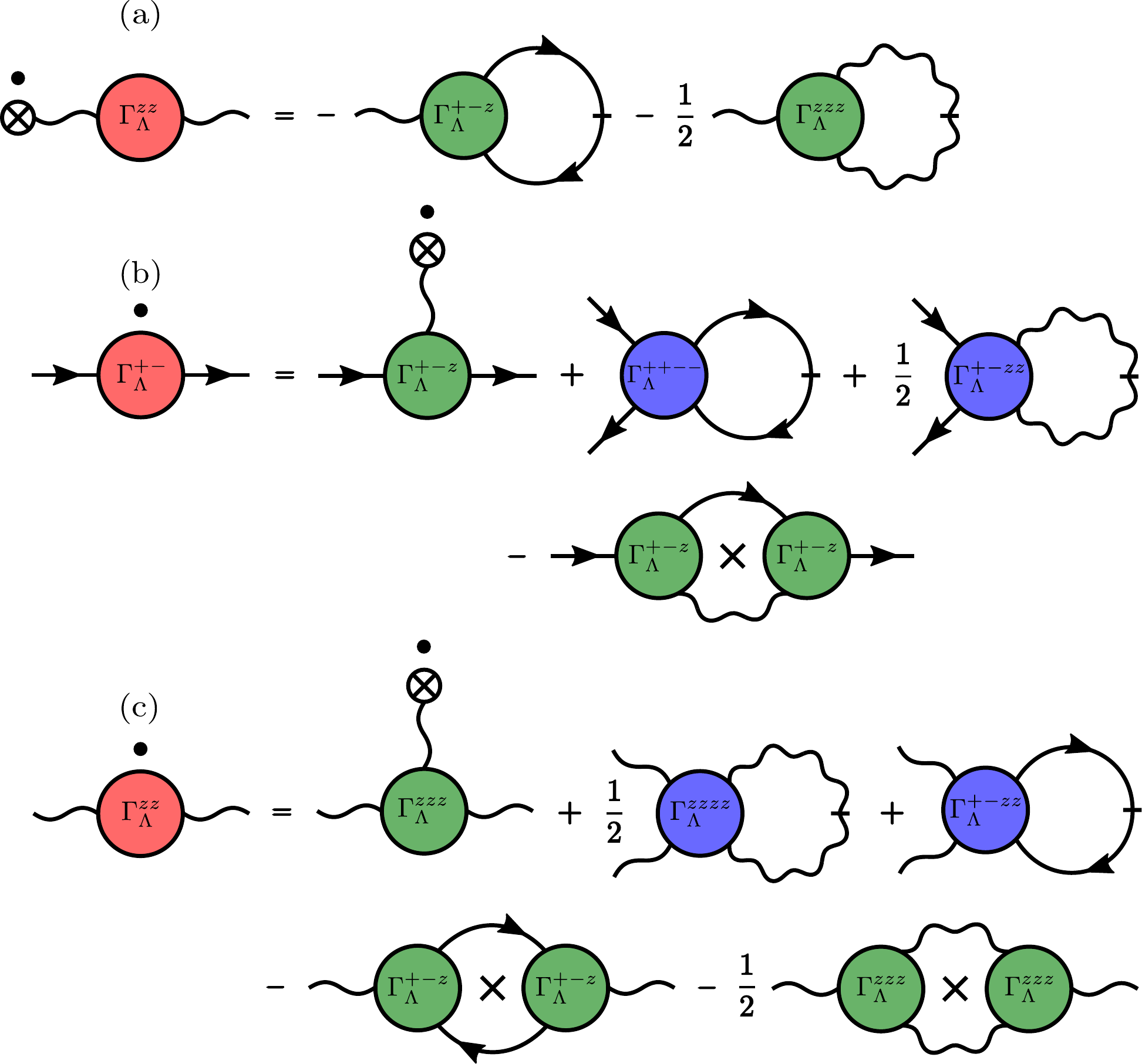}
   \end{center}
  \caption{
Graphical representation of the exact FRG flow equations (\ref{eq:flowh}), 
(\ref{eq:Gammapmflow}) and (\ref{eq:Gammazzflow}) for the exchange field and the two-point vertices.
(a) Flow equation for  the exchange field $\phi_{\Lambda}$; the dotted cross represents the scale derivative $\partial_{\Lambda}
 \phi_{\Lambda}$.
The dot above the circle represents the derivative $\partial_{\Lambda}$ with respect to the
deformation parameter.
(b) Flow equation for the transverse two-point vertex
$\Gamma^{+-}_\Lambda ( K )$.
(c) Flow equation for  the longitudinal two-point vertex
$\Gamma^{zz}_\Lambda ( K )$. 
We use the same color coding as in Ref.~[\onlinecite{Kopietz10}]: two-point vertices are red, three-point vertices are green, and four-point vertices are blue. Outgoing arrows represent the spherical component
 $m^-_{-K}$ of the transverse magnetization field, 
while incoming arrows represent the conjugate field $m^+_K$. 
Wavy lines are associated with the fluctuating part $\varphi$ of the longitudinal exchange field.
Solid arrows and wavy lines represent transverse and longitudinal propagators, 
while lines with an extra slash represent the corresponding single-scale propagators.
A diagram with a cross inside a loop represents the sum of two similar  diagrams
where each of the propagators forming the loop is successively replaced by the
corresponding single-scale propagator (product rule). 
}
\label{fig:flow}
\end{figure}

The exact flow equations for the vertices can be obtained from the functional 
flow equation (\ref{eq:WetterichHeisenberg}) by expanding
both sides in powers of the fields and
comparing coefficients of a given power after proper symmetrization 
\cite{Kopietz10}.
To write down the flow equations, we need the regularized 
transverse and longitudinal propagators
 \begin{eqnarray}
G_\Lambda (K) & = & \frac{1}{\Gamma^{+-}_{\Lambda} ( K )  +  R^{\bot}_{\Lambda} ( \bd{k} )},
 \label{eq:GGamma}
\\
F_\Lambda (K) & = & \frac{1}{\Gamma^{zz}_{\Lambda} ( K )  +  R^{\phi}_{\Lambda} ( \bd{k} )},
\label{eq:Fzzdef}
 \end{eqnarray}
and the corresponding single-scale propagators
 \begin{eqnarray}
 \dot{G}_\Lambda ( K )  &  = &  -
  [  G_{\Lambda} (K) ]^2  \partial_{\Lambda}  R_{\Lambda}^{\bot} ( \bd{k} ),
 \label{eq:Gsinglescale}
 \\
 \dot{F}_\Lambda ( K ) &   = &  -
  [  F_{\Lambda} (K) ]^2  \partial_{\Lambda}  R_{\Lambda}^{\phi} ( \bd{k} ).
\end{eqnarray}
Here the regulator in momentum space are given by
 \begin{align}
 R^{\bot}_{\Lambda} ( \bd{k} ) &= J^{\bot}_{\Lambda} ( \bd{k} ) -  J^{\bot} ( \bd{k} ),
 \label{eq:Rbotmom}
 \\
 R^\phi_{\Lambda} ( \bd{k} ) &=   \frac{1}{J^z ( \bd{k} )} 
 - \frac{1}{J^z_{\Lambda} ( \bd{k} )} ,
 \label{eq:Rphimom}
 \end{align}
where we have used the discrete translational invariance of the lattice to expand the exchange couplings in momentum space,
 \begin{eqnarray}
 J^{\bot}_{ \Lambda , ij } & = & \frac{1}{N} \sum_{\bd{k}} 
e^{ i \bd{k} \cdot ( \bd{r}_i - \bd{r}_j ) }
J^{\bot}_{\Lambda} ( {\bd{k}} ) , 
 \\
J^{z}_{ \Lambda , ij } & = & \frac{1}{N} \sum_{\bd{k}} 
e^{ i \bd{k} \cdot ( \bd{r}_i - \bd{r}_j ) }
J^{z}_{\Lambda} ( {\bd{k}} ) .
 \end{eqnarray}
With this notation  the  flow equation for the constant $f_{\Lambda}$ in Eq.~(\ref{eq:Gammahcomplete}), which
can be identified with the
free energy per lattice site,  can be written as
\begin{align}
 \partial_{\Lambda}  f_{\Lambda} 
 = 
 \int_{ K }& \Big[ G_{\Lambda} ( K ) \partial_{\Lambda} 
 {R}^{\bot}_{\Lambda} ( \bd{k} )    
\nonumber
\\ 
 & \hspace{-6mm} +
 \frac{1}{2}    \left[  F_{\Lambda} ( K ) + J_\Lambda^z ( \bm{k} ) \right] \partial_{\Lambda} 
 {R}^{\phi}_{\Lambda} ( \bd{k} ) \Big] .
 \label{eq:flowF}
\end{align}
Recall that according to Eq.~(\ref{eq:FGrelation}) the longitudinal propagator
$F_{\Lambda} ( K )$ is related to the longitudinal two-spin correlation 
function $G^{zz}_{\Lambda} ( K )$ via
 \begin{equation}
 F_{\Lambda} ( K ) = - J^z_{\Lambda} ( \bd{k} ) + [ J^z_{\Lambda} ( \bd{k} ) ]^2 G^{zz}_{\Lambda} ( K ),
 \label{eq:FGzz}
 \end{equation}
so that the flow equation (\ref{eq:flowF}) can alternatively be written as
\begin{equation}
 \partial_{\Lambda}  f_{\Lambda} 
 = 
 \int_{ K } \left[  \dot{J}^{\bot}_{\Lambda} ( \bd{k} )  G_{\Lambda} ( K )  +
 \frac{1}{2}   \dot{J}^z_{\Lambda} ( \bd{k} )   G^{zz}_{\Lambda} ( K )  \right] ,
 \label{eq:flowF2}
 \end{equation}
where $ \dot{J}^{\bot}_{\Lambda} ( \bd{k} ) = \partial_{\Lambda} 
 {J}^{\bot}_{\Lambda} ( \bd{k} )$ and
 $ \dot{J}^{z}_{\Lambda} ( \bd{k} ) = \partial_{\Lambda} 
 {J}^{z}_{\Lambda} ( \bd{k} )$.
Next, consider the flow equation for the exchange field $\phi_{\Lambda}$,
which  can be obtained from the condition
that the expansion of the generating functional $\tilde{\Gamma}_{\Lambda} [ \bd{m} , \varphi ]$
does not have a term linear in the fluctuation $\varphi$ 
(see Refs.~[\onlinecite{Schuetz06,Kopietz10}]).
This is guaranteed if $\phi_{\Lambda}$ flows according to
 \begin{eqnarray}
 {\Gamma}^{zz}_\Lambda ( 0 ) \partial_{\Lambda}  {\phi}_{\Lambda}  & = &
 -  \int_K \dot{G}_{\Lambda} ( K )  {\Gamma}_{\Lambda}^{+-z} ( - K, K , 0 )
 \nonumber
 \\
 & - & 
  \frac{1}{2} \int_K \dot{F}_{\Lambda} ( K )  {\Gamma}_{\Lambda}^{zzz} ( -K, K , 0 ).
 \hspace{7mm}
 \label{eq:flowh}
 \end{eqnarray}
Finally, let us also write down the  exact  flow equations for the transverse and longitudinal two-point vertices,
 \begin{widetext}
 \begin{eqnarray}
 \partial_{\Lambda} \Gamma^{+-}_{\Lambda} ( K ) & = &  {\Gamma}^{+-z}_{\Lambda} (- K , K , 0) 
 \partial_{\Lambda} \phi_{\Lambda}
+ \int_Q \dot{G}_{\Lambda} ( Q ) 
  {\Gamma}^{++--}_{\Lambda} ( - K, - Q , Q , K )
 + \frac{1}{2} \int_Q \dot{F}_{\Lambda} ( Q ) 
  {\Gamma}^{+-zz}_{\Lambda} ( - K, K , - Q , Q )
 \nonumber
 \\
& - &  \int_Q [ \dot{G}_{\Lambda} ( Q ) F_{\Lambda} (  Q-K )  +  
 {G}_{\Lambda} ( Q ) \dot{F}_{\Lambda} (  Q - K ) ]
 {\Gamma}^{+-z}_{\Lambda} ( - K,  Q , K-Q) 
 {\Gamma}^{+-z}_{\Lambda} ( - Q,  K , Q-K ) ,
 \label{eq:Gammapmflow}
\\
 &  & 
\nonumber
 \\
 \partial_{\Lambda} \Gamma^{zz}_{\Lambda} ( K ) & = &   
 {\Gamma}^{zzz}_{\Lambda} (- K , K , 0) \partial_{\Lambda}   \phi_{\Lambda}
 + \int_Q \dot{G}_{\Lambda} ( Q ) 
  {\Gamma}^{+-zz}_{\Lambda} ( - Q, Q, -K , K )
 + \frac{1}{2} \int_Q \dot{F}_{\Lambda} ( Q ) 
  {\Gamma}^{zzzz}_{\Lambda} ( -Q, Q , - K , K )
 \nonumber
 \\
& - &  \int_Q [ \dot{G}_{\Lambda} ( Q ) G_{\Lambda} ( Q- K )  +  
 G_{\Lambda} ( Q ) \dot{G}_{\Lambda} ( Q- K ) ]
 {\Gamma}^{+-z}_{\Lambda} ( - Q, Q -K ,  K) 
 {\Gamma}^{+-z}_{\Lambda} ( K -Q  ,  Q , -K )
 \nonumber
 \\ 
 & - & \frac{1}{2} \int_Q [ \dot{F}_{\Lambda} ( Q ) F_{\Lambda} ( Q- K )  +  
 {F}_{\Lambda} ( Q ) \dot{F}_{\Lambda} ( Q- K ) ]
 {\Gamma}^{zzz}_{\Lambda} ( -Q, Q-K , K ) 
 {\Gamma}^{zzz}_{\Lambda} ( K - Q, Q, -K ) . 
\label{eq:Gammazzflow}
 \end{eqnarray}
\end{widetext}
Graphical representations of
the exact flow equations (\ref{eq:flowh})--(\ref{eq:Gammazzflow}) are 
shown  in Fig.~\ref{fig:flow}.
Following VLP \cite{Vaks68}, it is convenient to parametrize the longitudinal
spin correlation function $G^{zz}_{\Lambda} ( K )$ 
in terms of the interaction-irreducible polarization $\Pi_{\Lambda} ( K )$ by setting
 \begin{equation}
 G^{zz}_{\Lambda} ( K ) = \frac{ \Pi_{\Lambda} ( K ) }{ 1 + J^z_{\Lambda} ( \bd{k} ) 
 \Pi_{\Lambda} ( K )}.
 \label{eq:GzzPi}
 \end{equation}
Using the relation (\ref{eq:FGzz}) this implies that the 
longitudinal two-point function $F_{\Lambda} ( K )$ generated by
our functional ${\cal{F}}_{\Lambda} [ \bd{h}_{\bot} , s ]$ can be written as
  \begin{equation}
 F_{\Lambda} ( K ) = - \frac{ J_{\Lambda}^{z} ( \bd{k} ) }{ 1 + J^z_{\Lambda} ( \bd{k} ) 
 \Pi_{\Lambda} ( K )}.
 \end{equation}
Hence, the longitudinal propagator  $F_{\Lambda} ( K )$ can be identified with an effective screened
interaction between longitudinal spin fluctuations.
Finally, using Eq.~(\ref{eq:Fzzdef}) to express $F_{\Lambda} ( K )$ in terms of the longitudinal 
two-point vertex
$\Gamma^{zz}_{\Lambda} ( K )$ which is 
generated by our hybrid functional $\tilde{\Gamma}_{\Lambda}
 [ \bd{m} , \varphi ]$ defined via
Eqs.~(\ref{eq:GammaHMdef}) and (\ref{eq:Gammashift}), we obtain
\begin{equation}
 \Gamma^{zz}_{\Lambda} ( K ) = - \frac{1}{J^{z} ( \bd{k} ) } - \Pi_{\Lambda} ( K ).
 \label{eq:GammaPi}
 \end{equation}
We conclude that, up to a minus sign, the right-hand side 
of the flow equation (\ref{eq:Gammazzflow}) for 
 $ \Gamma^{zz}_{\Lambda} ( K )$
can be identified with the flow equation for the irreducible longitudinal 
polarization $\Pi_{\Lambda} ( K )$
defined via Eq.~(\ref{eq:GzzPi}).

\subsection{Deformation scheme: switching off the transverse interaction}
\label{sec:initial}

Let us now specify our deformation scheme. In general, the deformed exchange
interactions $J_{\Lambda_0 , ij}^{\bot}$ and $J^z_{\Lambda_0 , ij }$ at the initial value
$\Lambda  = \Lambda_0$ of the deformation parameter should be chosen such that the correlation 
functions of the deformed model can be calculated in a controlled way.\cite{Krieg19}
The simplest possibility is to completely switch off the deformed exchange couplings,
$J_{\Lambda_0 , ij}^{\bot} = J^z_{\Lambda_0 , ij } =0$, so that the initial system consists of
non-interacting spins subject to an external magnetic field. Note, however, that even in this case 
 the time-ordered spin correlation functions and the corresponding irreducible vertex functions are 
rather complicated and reflect the non-trivial on-site spin correlations between different spin components implied by the $SU(2)$-spin algebra. In Appendix~B we explicitly derive 
the corresponding initial values  
for the correlation functions
and the vertices  with up to four external legs.

In order to establish the relation of our SFRG approach and the previous
momentum-shell RG calculation for quantum Heisenberg ferromagnets \cite{Kopietz89}, 
we choose in this work  a different deformation scheme, where  only the 
transverse interaction is initially switched off
while the longitudinal interaction is not deformed at all,
\begin{equation}
 J^{z}_{\Lambda} ( \bd{k} )  =  J^{z}( \bd{k} ) .
 \label{eq:Jbotmomentum}
\end{equation}
For our purpose it is sufficient to regularize the long-wavelength modes of the transverse
 interaction
via a sharp cutoff, 
\begin{eqnarray}
 J^{\bot}_{\Lambda} ( \bd{k} ) & = & \Theta (  k - \Lambda  ) J^{\bot} ( \bd{k} ),
 \label{eq:Jbotsharp}
 \end{eqnarray}
which amounts to the following choice of the transverse regulator:
 \begin{equation}
 R^{\bot}_{\Lambda} ( \bd{k} ) =  J^{\bot}_{\Lambda} ( \bd{k} ) - J^{\bot} ( \bd{k} ) =
 - \Theta (  \Lambda - k  ) J^{\bot} ( \bd{k} ).
 \end{equation}
The initial value $\Lambda_0$  of the deformation parameter $\Lambda$
(which has units of momentum) should be chosen 
of the order of the inverse lattice spacing.
Since $\partial_{\Lambda} J^z_{\Lambda} ( \bd{k} ) =0$ in this cutoff scheme,
all terms involving the longitudinal single-scale propagator $\dot{F}_{\Lambda} ( K)$
in the flow equations (\ref{eq:flowh}--\ref{eq:Gammazzflow}) 
shown graphically in  Fig.~\ref{fig:flow} can be  omitted.
Our deformed model at the initial scale $\Lambda = \Lambda_0$ is then an Ising model, which is still nontrivial.

\subsection{Initial conditions}
 \label{sec:initial}

For the deformation scheme discussed above, the initial conditions for the SFRG flow are determined by the magnetization and the correlation functions of a spin S Ising model, which cannot be calculated exactly.
However, in this work we are not interested in the critical regime of the Ising model, but we focus on the low-temperature regime $T \ll | J^z (0 ) |$  where a perturbative calculation of the correlation functions is possible. 
In order to calculate the initial conditions in a controlled way,  we follow VLP and assume that the range $r_0$ of the exchange interaction is large compared with the lattice spacing.
In this case, momentum integrals are controlled by the small parameter $1/r_0$, and the initial values of the correlation functions can be calculated by solving the corresponding hierarchy of equations of motion perturbatively in powers of loop integrations, see Appendix C. 
To leading order, i.e., in the so called tree approximation, one may simply neglect all loops.
As discussed in Appendix~D, then
the infinite hierarchy of equations of motion decouples.
The resulting initial magnetization $M_0 = \langle S_i^{z} \rangle$ is given by the self-consistent mean-field approximation, obtained from the solution of
\begin{equation}
M_0 = b\left( \beta \left( H + \phi_0 \right) \right),
 \label{eq:M0def}
\end{equation}
where
\begin{equation}
\phi_0= -J^z (0) M_0
\end{equation}
is the initial value of the exchange field introduced in Eq.~\eqref{eq:extremum} and
 \begin{equation} 
  b(y) = \bigl( S + \frac{1}{2} \bigr) \coth  \left[ \bigl( S + \frac{1}{2} \bigr) y \right] - \frac{1}{2} \coth \left[ \frac{y}{2} \right]
 \label{eq:brillouin}
 \end{equation}
is the spin-$S$ Brillouin function.
The transverse two-spin correlation function is then simply given by
\begin{eqnarray}
 G_0 ( K ) & =& \frac{M_0}{ H+\phi_0 - i \omega} ,
 \label{eq:G0def}
 \end{eqnarray}
while the longitudinal two-spin correlation function is
 \begin{eqnarray}
 G^{zz}_0 ( K ) & =& \frac{\beta \delta_{ \omega , 0} b^{\prime} ( \beta (H+\phi_0) )}{ 1 + J^z ( \bd{k} )
 \beta \delta_{ \omega , 0} b^{\prime} ( \beta (H+\phi_0) ) }.
 \end{eqnarray}
Here $b^{\prime} ( y )$ is the derivative of the Brillouin function.
Using Eq.~(\ref{eq:GGamma}) we find that in this case the initial value of the transverse two-point vertex is
\begin{eqnarray}
 \Gamma^{+-}_0 ( K ) & = &    J^{\bot} ( \bd{k} ) +   \frac{ H + \phi_0 - i \omega}{M_0} 
 \nonumber
 \\
 & =  &   \frac{ H + M_0 [ J^{\bot} ( \bd{k} ) - J^z (0) ]  - i \omega}{M_0} ,
 \label{eq:Gammapmtree}
 \end{eqnarray} 
while from Eq.~(\ref{eq:GzzPi})
we conclude that the intial value for the polarization is
 \begin{equation}
 \Pi_0 ( K ) = \beta \delta_{\omega,0} b^{\prime} ( \beta ( H + \phi_0 ))  ,
 \end{equation}
which according to Eq.~(\ref{eq:GammaPi}) implies for the initial value of the
longitudinal two-point vertex,
\begin{eqnarray}
  \Gamma^{zz}_0 ( K ) & = & - \frac{1}{ J^{z} ( \bd{k} )}     -   \beta \delta_{ \omega , 0} b^{\prime} ( \beta ( H + \phi_0 ))  ,
 \label{eq:Gammazztree}
 \end{eqnarray}
The initial values of the higher-order spin correlation functions and corresponding irreducible
vertices are derived  in Appendix D.

\section{Magnetization of two-dimensional ferromagnets}
 \label{sec:thermo}

In this section we  calculate the magnetic equation  of state $M ( H , T )$ of an isotropic Heisenberg ferromagnet in two dimensions at finite temperature.  
For $H=0$ the magnetization vanishes at any finite temperature~\cite{Mermin66},
and a naive spin-wave expansion gives a divergent result for $ M (  H=0, T )$.
Non-perturbative methods are thus necessary to obtain reliable results for $M ( H, T )$ for sufficiently small magnetic fields \cite{Hofmann02}.
To simplify our notation, we set from now on
 \begin{equation}
 J^{\bot} ( \bd{k} ) = J^z ( \bd{k} ) = - V_{\bd{k}},
 \end{equation}
where for a ferromagnet the Fourier transform
$V_{\bd{k} }$ of the  
exchange couplings satisfies 
$V_0 \equiv V_{\bd{k} =0} > 0$. 
At low temperatures only long-wavelength 
modes are thermally excited, so that we may expand~\cite{footnoteJ} 
\begin{equation}
 V_{\bd{k}} = V_0 - \frac{\rho_0}{M_0} {k}^2 + {\cal{O}} ( k^4 ),
 \label{eq:Vexpansion}
 \end{equation}
where the value $\rho_0$ of the bare spin-stiffness depends on the range of the interaction.
For nearest-neighbor ferromagnetic exchange $J >0$ on a $D$-dimensional hypercubic lattice with lattice spacing $a$, we have $V_0 = 2 D J$ and $\rho_0 = M_0 J a^2$. 
However, the initial conditions discussed in Sec.~\ref{sec:initial} are only valid if the exchange interaction is long-range;
for short-range interaction, the tree approximaton for the vertices at the initial scale is uncontrolled.
Nevertheless, the comparision of our FRG results with Monte Carlo simulations for two-dimensional ferromagnets with nearest-neighbor interaction shows that even in this case our truncation works rather well, see Sec.~\ref{subsec:magcurves} below.

  \subsection{One-loop approximation with sharp momentum cutoff}

As already mentioned in Sec.~\ref{sec:initial},
to establish the precise relation between our SFRG approach and an old
momentum-shell renormalization group calculation  
for quantum ferromagnets~\cite{Kopietz89}, 
we use the sharp momentum cutoff (\ref{eq:Jbotsharp}) for the transverse exchange couplings and no cutoff at all for the longitudinal couplings.
In this case, all terms involving the
longitudinal single-scale propagator $\dot{F}_\Lambda ( K )$ in our flow equations should be simply omitted, so that the flow equation (\ref{eq:flowh})
for the exchange field $\phi_{\Lambda} = - J^z (0) M_{\Lambda} = V_0 M_{\Lambda}$ reduces to
 \begin{eqnarray}
 {\Gamma}^{zz}_\Lambda ( 0 ) \partial_{\Lambda}  {\phi}_{\Lambda}  & = &
 -  \int_K \dot{G}_{\Lambda} ( K )  {\Gamma}_{\Lambda}^{+-z} ( K, K , 0 ),
 \label{eq:flowh2}
 \end{eqnarray}
where the transverse single-scale propagator  defined in Eq.~(\ref{eq:Gsinglescale})
is with our cutoff scheme given by
 \begin{eqnarray}
 \dot{G}_\Lambda ( K )  
 &   = & - \frac{ \partial_{\Lambda} R^{\bot}_{\Lambda} ( \bd{k} ) }{
[  \Gamma^{+-}_{\Lambda} ( K ) +    R^{\bot}_{\Lambda} ( \bd{k} ) ]^2 }
  \nonumber
 \\
 & = & 
-\frac{ \delta (k -  \Lambda  ) {V}_{\bd{k}}  }{ [ \Gamma^{+-}_{\Lambda}
 ( K ) + \Theta (  \Lambda - k ) {V}_{\bd{k}}   ]^{2} }.
 \end{eqnarray}
A technical complication of the sharp momentum cutoff scheme is that it leads to expressions involving both $\delta$ and $\Theta$ functions, which should be carefully defined using the identity~\cite{Morris94,Kopietz10}
 \begin{equation}
 \delta (x ) f ( \Theta (x ) ) = \delta (x) \int_0^1 dt f ( t ).
 \end{equation}
Taking this into account the single-scale propagator reads
 \begin{equation}
\dot{G}_\Lambda ( K )   =   \delta ( k - \Lambda )  \left[ 
\frac{1}{  \Gamma^{+-}_{\Lambda}  ( K ) +  {V}_{\bd{k}}   } 
-
\frac{1}{  \Gamma^{+-}_{\Lambda} ( K ) } 
   \right]     .
 \label{eq:Gpmsharp}
 \end{equation}

\begin{figure}[tb]
 \begin{center}
  \centering
\vspace{7mm}
\hspace*{-3mm}
 \includegraphics[width=0.5\textwidth]{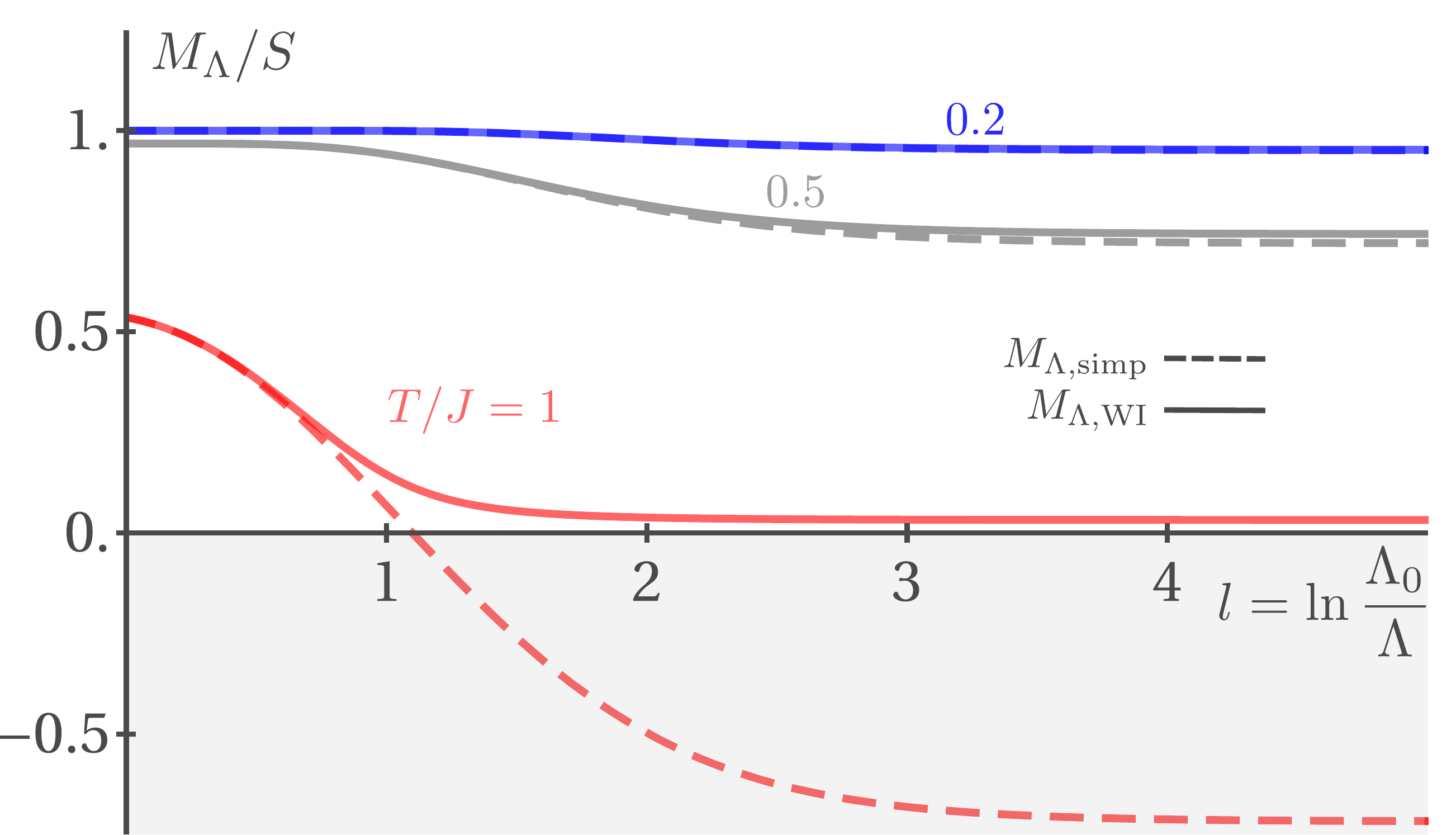}
   \end{center}
  \vspace{-2mm}
  \caption{
RG flow of the magnetization $M_\Lambda$ as a function of the logarithmic flow parameter
$l = \ln ( \Lambda_0 / \Lambda )$ for  $H/J=0.125$
and
$T / J =1,0.5,0.2$ (red, gray, blue; the curves are labeled by the corresponding values of $T/J$)
. The dashed lines 
represent  the solution of the simplest truncation  (\ref{eq:Mflow1})
of the flow equation for the magnetization 
($M_{\Lambda,\text{simp}}$) which completely neglects self-energy and vertex corrections. The solid lines show the solution of the improved flow equation (\ref{eq:Mflow2}) 
($M_{ \Lambda,\text{WI}}$) which takes into account the RG flow of the magnon self-energy via the Ward 
identity (\ref{eq:WIsigma}). Note that $M_{\Lambda,\text{WI}}$, 
in contrast to $M_{\Lambda,\text{simp}}$, does not flow to unphysical negative values for high temperatures.
}
\label{fig:Mflow1}
\end{figure}

In the simplest approximation, the flow of the
vertices $\Gamma_{\Lambda}^{zz} (0)$ and
$ {\Gamma}_{\Lambda}^{+-z} ( K, K , 0 )$ is neglected, so that these vertices 
are approximated by their initial values at scale $\Lambda = \Lambda_0$ where the 
deformed transverse exchange interaction vanishes.  Moreover, at low temperatures, the derivatives of the Brillouin function are exponentially small and can be neglected. In this case,
we may approximate $\Gamma_0^{zz} (0) \approx 1/V_0$ and
${\Gamma}_{0}^{+-z} ( K, K , 0 ) \approx 1/ M_0$ [see Eq.~(\ref{eq:threetree}) 
in Appendix D], 
so that Eq.~(\ref{eq:flowh2})
reduces to the following flow equation for the scale-dependent magnetization: 
 \begin{equation}
 \partial_{\Lambda} M_{\Lambda} = - \frac{1}{M_0}  
 \int_K \dot{G}_{\Lambda} ( K ) .
 \label{eq:Mflowsimp}
 \end{equation}
Neglecting self-energy corrections to the transverse propagator, we may approximate
 \begin{eqnarray}
 \dot{G}_{\Lambda} ( K ) & \approx&
 - \frac{ \delta ( k - \Lambda )}{ \Gamma^{+-}_0 ( K ) }
= - \frac{ \delta ( k - \Lambda ) M_0 }{ H + E_{ {\bd{k}} }  - i \omega },
 \nonumber
 \\
 & &
 \label{eq:singlescale}
 \end{eqnarray}
where we have used the initial value (\ref{eq:Gammapmtree})
for the transverse two-point function and 
introduced the magnon dispersion
 \begin{equation}
 E_{ \bd{k}  } 
= M_0 ( V_0 - V_{ \bd{k} } )
\approx \rho_0 k^2.
\label{eq:Disp}
 \end{equation} 
Recall that the bare spin stiffness $\rho_0$ is defined via the expansion
(\ref{eq:Vexpansion})
of the Fourier transform $V_{\bd{k}}$ of the exchange coupling.
Substituting Eq.~(\ref{eq:singlescale})  
into Eq.~(\ref{eq:Mflowsimp}) and performing the 
integrations and the Matsubara sum,
we obtain in $D $ dimensions
 \begin{equation}
 \partial_{\Lambda} M_{\Lambda} = K_D   \frac{a^D \Lambda^{D-1}}{ e^{ \beta (H + \rho_0 \Lambda^2 )}-1},
 \label{eq:Mflow1}
 \end{equation}
where $K_D$ is the surface area of the $D$-dimensional unit sphere divided by $(2 \pi )^D$.
In Fig.~\ref{fig:Mflow1}, the result of the numerical integration of the flow equation (\ref{eq:Mflow1}) 
for $D=2$ is represented by dashed lines. 
Obviously, for sufficiently small magnetic field
and high temperatures the magnetization flows to unphysical negative values
within this approximation. This is clearly an artefact 
of the simple approximations used to derive Eq.~(\ref{eq:Mflow1}).
In Sec.~\ref{sec:vertex}, we show how this problem can be cured 
by taking into account a Ward identity relating the magnon self-energy to the magnetization.
From Eq.~(\ref{eq:Mflow1}), it is furthermore straightforward to
recover the RG flow equations for 
quantum ferromagnets derived 30 years ago in Ref.~[\onlinecite{Kopietz89}]
(see also Ref.~[\onlinecite{Kopietz89b}])
using the  momentum-shell RG technique. 
In this approach, the low-temperature behavior of quantum ferromagnets in $D$ dimensions is encoded in the RG flow of the following three dimensionless rescaled coupling constants:
 \begin{subequations}
\begin{eqnarray}
t & = & \frac{T  (a \Lambda)^{D-2}  }{ JS M_{\Lambda}  },
 \label{eq:tPK}
 \\
g & = & \frac{(a \Lambda )^D}{M_{\Lambda}},
 \label{eq:gPK}
 \\
 h & = & \frac{H }{JS} (a \Lambda )^{-2},
 \label{eq:hPK}
 \end{eqnarray}
\end{subequations} 
where  $\Lambda = \Lambda_0 e^{-l}$ is the running momentum cutoff
and the energy $J$ is defined by\cite{footnoteJ}   
 \begin{equation}
J = \rho_0/ ( M_0 a^2).
 \label{eq:Jdef}
 \end{equation}
 Using the flow equation (\ref{eq:Mflow1})  for the scale-dependent magnetization $M_{\Lambda}$, 
we find that the above
rescaled couplings satisfy the system of flow equations:
 \begin{subequations}
 \label{eq:1loopflow}
  \begin{eqnarray}
 \partial_l t & = & (2-D) t + K_D \frac{ gt}{e^{g (1+h)/t }-1 },
 \\
 \partial_l g & = & - D g + K_D \frac{ g^2}{e^{g (1+h)/t }-1 },
 \label{eq:g1loop}
 \\
 \partial_l h & = & 2 h,
 \label{eq:h1loop}
 \end{eqnarray}
\end{subequations}
in agreement with the flow equations given  in Refs.~[\onlinecite{Kopietz89,Kopietz89b}].
By integrating these flow equations from $l =0$ up to some finite $l = l_{\ast}$ where the renormalized dimensionless temperature $t_{\ast}$ is of the order of unity, we can estimate the correlation length $\xi$ in units of the lattice spacing $a$.  For $H=0$, the result is
 \begin{equation}
  \frac{\xi}{a}  \propto \sqrt{ \frac{JS}{T}} e^{  2 \pi J S^2 /T},
 \end{equation}
where the precise prefactor cannot be determined with this method.
Note also that flow equations similar to Eqs.~(\ref{eq:1loopflow}) for 
quantum antiferromagnets have been obtained by
Chakravarty, Halperin, and Nelson \cite{Chakravarty88,Chakravarty89} 
by applying  the conventional momentum-shell RG technique to
the quantum nonlinear sigma model
which is believed to describe 
the low-energy and long-wavelength physics of
quantum Heisenberg antiferromagnets in the renormalized classical regime (see Ref. [\onlinecite{Rancon13}] for a  derivation within the FRG).

\subsection{Self-energy and vertex corrections}
\label{sec:vertex}
From Fig.~\ref{fig:Mflow1},  we see that
in two dimensions the flow equation (\ref{eq:Mflow1})
implies that at sufficiently small magnetic field and large temperature 
the flowing magnetization $M_{\Lambda} ( H , T )$  
becomes negative at some finite scale $\Lambda_{c} = e^{ - l_c }$, so that
we cannot integrate the flow all the way down to $\Lambda =0$.
This is of course an unphysical feature of our truncation. Within our SFRG approach we can
construct a better truncation by including the flow of the magnon self-energy
$\Sigma_{\Lambda} ( K )$ as well as vertex corrections. 
However, in the presence of a finite spontaneous magnetization,  the magnon spectrum
must be gapless, so that the self-energy $\Sigma_{\Lambda} ( K=0 )$ must vanish for $H \rightarrow 0$. To implement this condition without fine-tuning the initial condition,
it is crucial to truncate the infinite hierarchy of flow equations such that the Ward identity
 \begin{equation}
 \chi_{\bot} \equiv G ( \bd{k} =0, i \omega =0 ) = \frac{ M}{H} 
 \label{eq:WI1}
 \end{equation}
is satisfied at least at the end of the flow. This  identity relating the exact
transverse uniform susceptibility $\chi_{\bot}$ to the exact magnetization $M$ 
has been discussed long time ago by
Patashinskii and Prokrovskii \cite{Patashinskii73}. 
In Appendix~C, we give a rigorous  derivation
of this identity using the Heisenberg equations of motion.

\subsubsection{Magnon self-energy from the Ward identity}

We define the flowing magnon self-energy $\Sigma_{\Lambda} ( K ) $
by writing the transverse two-point vertex $\Gamma^{+-}_{\Lambda} ( K )$
in the form
 \begin{equation}
 \Gamma^{+-}_{\Lambda} ( K ) = \frac{ H + E_{\bd{k}} - i \omega}{M_0} +
 \Sigma_{\Lambda} ( K ),
 \end{equation}
so that the regularized transverse propagator is
 \begin{equation}
 G_{\Lambda} ( K )  =  \frac{M_0}{ H + E_{\bd{k}} - i \omega + M_0 R_{\Lambda}^{\bot} ( \bd{k} ) + M_0
 \Sigma_{\Lambda} ( K ) }.
 \end{equation}
Obviously,  the Ward identity (\ref{eq:WI1}) can be implemented
by demanding that the flowing transverse two-point 
vertex at vanishing momentum and frequency 
is related to the flowing magnetization via
 \begin{equation}
 \Gamma^{+-}_{\Lambda} ( 0 ) = \frac{H}{M_{\Lambda}},
 \end{equation}
which implies that  the magnon self-energy $\Sigma_{\Lambda} (0)$
at vanishing momentum and frequency 
is related to the magnetization via
 \begin{equation}
 \Sigma_{\Lambda} (0) = \frac{H}{M_{\Lambda}} - \frac{H}{M_0}.
 \label{eq:WIsigma}
 \end{equation}
Neglecting the momentum- and frequency-dependence  of the self-energy,
we then obtain for
the transverse single-scale propagator instead of Eq.~(\ref{eq:singlescale}),
\begin{eqnarray}
 \dot{G}_{\Lambda} ( K ) & = &  - \frac{ \delta ( k - \Lambda )}{
 \frac{ H + E_{ \bd{k} }  - i \omega}{M_0} + \Sigma_{\Lambda} (0 ) }
 \nonumber
 \\
 & = &  - \frac{ \delta ( k - \Lambda ) M_0}{ 
 \Delta_{\Lambda}
 + E_{ \bd{k} }  - i \omega } 
 \label{eq:Gsinglescale}
 \end{eqnarray} 
where
 \begin{equation}
 \Delta_{\Lambda} = \frac{ H M_0}{ M_{\Lambda} }
 \label{eq:Deltadef}
 \end{equation}
can be interpreted as a scale-dependent gap of the magnon dispersion.
The resulting modified flow equation for the magnetization is
 \begin{equation}
 \partial_{\Lambda} M_{\Lambda} = K_D   \frac{a^D \Lambda^{D-1}}{ e^{ \beta 
 ( \Delta_{\Lambda} + \rho_0 \Lambda^2 ) }-1}.
 \label{eq:Mflow2}
 \end{equation}
A numerical solution of this  modified flow equation
is shown by the solid lines in Fig.~\ref{fig:Mflow1}.
Note that now the magnetization never flows to unphysical negative values.
The definition (\ref{eq:hPK}) of the 
rescaled dimensionless magnetic field should then be replaced by
 \begin{equation}
 h  =  \frac{H }{J M_{\Lambda}} (a \Lambda )^{-2},
 \end{equation}
and the modified system of flow equations for the rescaled couplings $t$, $g$, and $h$ reads
 \begin{subequations}
 \label{eq:1loop2}
  \begin{eqnarray}
 \partial_l t & = & (2-D) t + K_D \frac{ gt}{e^{g (1+h)/t }-1 },
 \\
 \partial_l g & = & - D g + K_D \frac{ g^2}{e^{g (1+h)/t }-1 },
 \label{eq:g1loop2}
 \\
 \partial_l h & = & 2 h + K_D \frac{ g h}{e^{g (1+h)/t }-1 }.
 \end{eqnarray}
\end{subequations}
In contrast to our earlier  flow equation (\ref{eq:h1loop}), the
renormalized magnetic field now has a nontrivial fluctuation correction.

In order to quantify the effects of additional higher order vertex corrections on the magnetization, we investigate in the following two more advanced  truncations.

\subsubsection{Vertex correction from the Katanin substitution}
A simple approximate method to take vertex corrections into account is the so-called Katanin substitution, which amounts to
replacing the single-scale propagator in the flow equations by a total 
derivative, \cite{Katanin04}
 \begin{equation}
  \dot{G}_{\Lambda} ( K ) \rightarrow \partial_{\Lambda} G_{\Lambda} ( K ).
 \end{equation}
This substitution, known from the FRG formulation of interacting 
Fermi systems \cite{Katanin04,Metzner12}, has been found to be essential to obtain meaningful results in the pseudofermion FRG approach to quantum spin 
systems \cite{Reuther10,Reuther11,Reuther11a,Buessen16}.
The effect of higher-order vertices is thereby partially taken into account in a weak coupling truncation. In this way, the violation of Ward identities is shifted to a higher order in the vertex expansion.
Within our simple truncation the Katanin substitution amounts to replacing the single-scale propagator in Eq.~(\ref{eq:Gsinglescale}) by
\begin{eqnarray}
 & & \dot{G}_{\Lambda} ( K )  \rightarrow     \partial_{\Lambda}
 \frac{ \Theta ( k - \Lambda )}{
 \frac{ H + E_{\bd{k} }  - i \omega}{M_0} + \Sigma_{\Lambda} (0 ) }
 \nonumber
 \\
 &  &  =  - \frac{ \delta ( k - \Lambda )}{
 \frac{ H + E_{\bd{k} }  - i \omega}{M_0} + \Sigma_{\Lambda} (0 ) }
 - \frac{  \Theta ( k - \Lambda ) \partial_{\Lambda} \Sigma_{\Lambda} (0)}{
 \bigl[   \frac{H + E_{ \bd{k} }  - i \omega}{M_0} + \Sigma_{\Lambda} (0 ) \bigr]^2 }.
 \nonumber
 \\
 & &
 \label{eq:GKatanin}
 \end{eqnarray}
With this replacement the right-hand side of 
our truncated flow equation (\ref{eq:Mflowsimp}) for the magnetization
becomes a total derivative. Using again the Ward identity (\ref{eq:WIsigma})
to express $\Sigma (0 )$ at the end of the flow in terms of the magnetization $M$, 
we obtain
 \begin{equation}
 M = M_0  - \frac{1}{N} \sum_{\bd{k}} \frac{1}{ 
 e^{  \beta (H M_0 /M  + E_{ \bd{k} }) }  -1}    .
 \label{eq:Mselfcon}
 \end{equation}
In one and two dimensions, the zero-field susceptibility
$\chi = \lim_{ H \rightarrow 0} M / H$ is expected to be finite for any non-zero temperature, so that the gap  $\Delta_{\Lambda}$ approaches a finite limit for $H \rightarrow 0$.
In this limit Eq.~(\ref{eq:Mselfcon})
reduces to the following equation for the zero-field susceptibility:
 \begin{equation}
 M_0 = \frac{1}{N} \sum_{\bd{k}} \frac{1}{ 
 e^{  \beta (M_0 / \chi  + E_{\bd{k}} ) }  -1}    .
 \end{equation}
At low temperatures ($ T \ll V_0)$, we can solve this equation by
approximating
 \begin{equation}
 \frac{1}{ 
 e^{  \beta (M_0 / \chi  + E_{ \bd{k} })  }  -1} \approx \frac{T}{  
 M_0 / \chi  + \rho_0 k^2 }
 \end{equation}
and imposing an ultraviolet cutoff $\Lambda_0$ on the momentum-integration such that
 $ \beta \rho_0 \Lambda_0^2 = 1$. 
Defining the energy scale\cite{footnoteJ}
$J$ in terms of  $\rho_0$ as in Eq.~(\ref{eq:Jdef})  and using the fact that at low temperatures
 $M_0 \approx S$, we obtain for the susceptibility at vanishing magnetic field,
 \begin{equation}
 \chi  =  \frac{M_0}{T}   e^{ 4 \pi J S^2 / T }.
 \label{eq:chi1loop}
 \end{equation}
The transverse correlation length $\xi ( H, T )$ can be
defined by writing the
magnon dispersion for small momenta as
 \begin{equation}
    H M_0 / M + E_{\bd{k}}   \approx
\rho_0 (  \xi^{-2} + k^2).
\label{eq:corr}
 \end{equation}
For $ H \rightarrow 0$, 
this leads to the identification
 \begin{equation}
 \xi = \sqrt{ \rho_0 \chi / M_0 }.
 \end{equation}
With $\chi$ given by Eq.~(\ref{eq:chi1loop}),  this yields
 \begin{equation}
  \frac{\xi}{a} = \sqrt{ \frac{JS}{T}} e^{  2 \pi J S^2 /T},
 \end{equation}
which agrees (up to a numerical prefactor of the order of unity) 
with previous one-loop calculations based on
modified spin-wave theory \cite{Takahashi86}, Schwinger-Boson mean-field theory \cite{Arovas88}, and a one-loop momentum shell renormalization group calculation \cite{Kopietz89}.
As shown in Ref.~[\onlinecite{Kopietz89}], however, a more accurate two-loop calculation gives an addition factor of $T /(JS^2)$ in front of the 
exponential, i.e., $ \xi \propto ( T / JS )^{1/2} e^{  2 \pi J S^2 /T}$. 
Similarly, the temperature dependence of the prefactor 
in the expression (\ref{eq:chi1loop}) for the susceptibility is 
known to be modified by two-loop corrections \cite{Kopietz89}.
Moreover, the one-loop approximation  for the zero-field 
susceptibility given in Refs.~[\onlinecite{Takahashi86,Arovas88,Kopietz89}]  is a factor of $T /(JS)$ smaller than the result
(\ref{eq:chi1loop}). A possible reason  for  this discrepancy 
is that in our truncation we have neglected the frequency and momentum dependence of
the magnon self-energy.

\begin{figure}[tb]
 \begin{center}
\vspace{0.5mm}
\hspace*{-2mm}
 \includegraphics[width=0.5\textwidth]{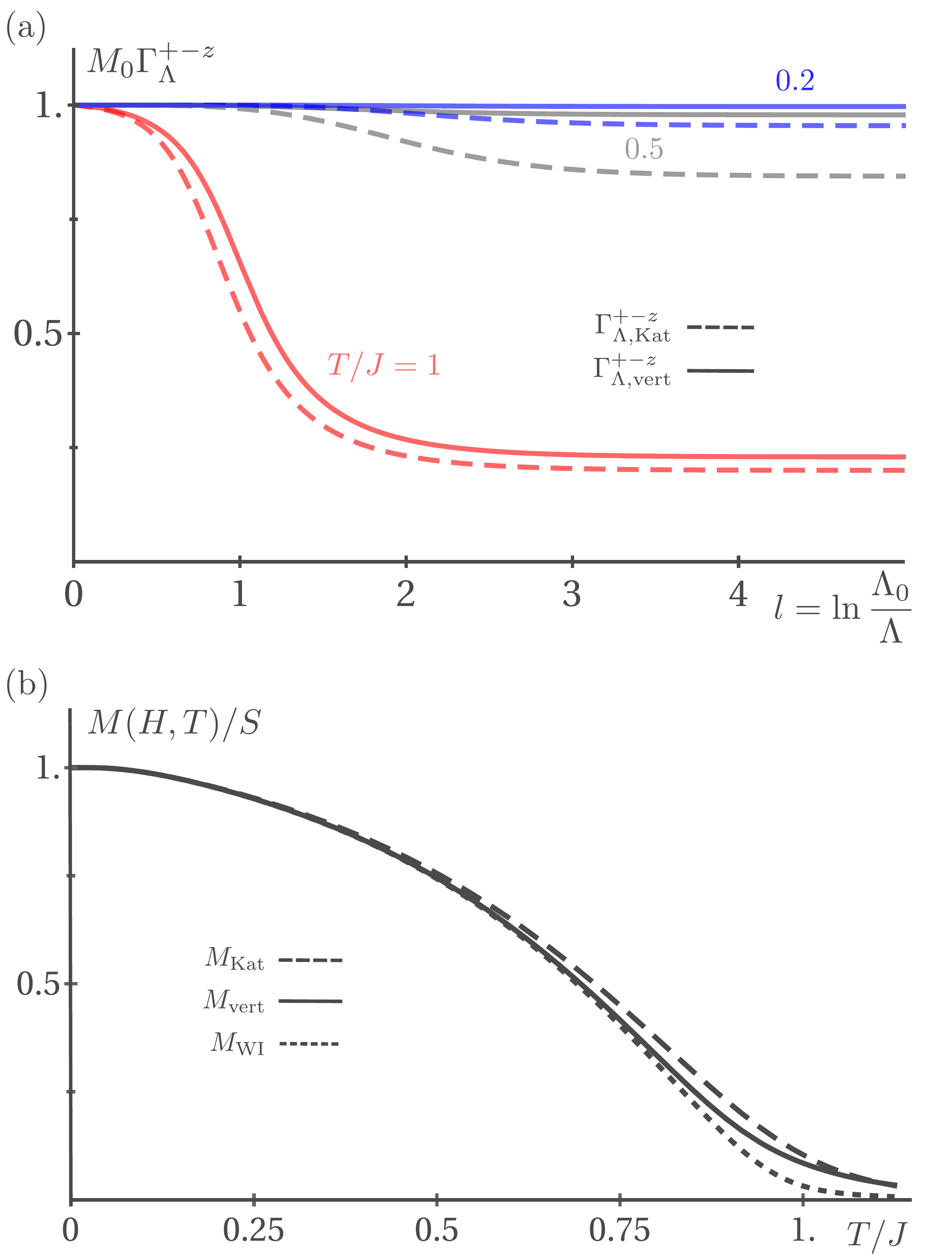}
   \end{center}
  \vspace{-2mm}
  \caption{
  (a) RG flow of the vertex $\Gamma_\Lambda^{+-z}$  as a function of the logarithmic flow parameter $l = \ln ( \Lambda_0 / \Lambda )$ for $T / J =1,0.5,0.2$ (red, gray, blue) and $H/J=0.125$, calculated either 
using the Katanin substitution 
($\Gamma_{\Lambda,\text{Kat}}^{+-z}$, see
Eqs. (\ref{eq:Mflowapproxvert}) and (\ref{eq:gammaKat})), 
or the approximate solution of the 
SFRG flow equations ($\Gamma_{\Lambda,\text{vert}}^{+-z}$, see 
Eqs.~(\ref{eq:Mflowapproxvert2}) and (\ref{eq:vertflow})).
The qualitative behavior of $\Gamma_{ \Lambda,\text{vert}}^{+-z}$ 
and $\Gamma_{ \Lambda,\text{Kat}}^{+-z}$ is similar throughout the temperature range considered.  
(b) Resulting magnetization $M(H,T)=M_{ \Lambda \rightarrow 0}(H,T)$ as a function of temperature. The dotted line represents the solution of the flow equation
(\ref{eq:Mflow2})
without vertex correction where $M_0\Gamma_\Lambda^{+-z} = 1$.
The different magnetization curves coincide for high and low temperatures and show only small deviations in the intermediate range.
}
\label{fig:vertexkat}
\end{figure}

Within our approach, it is possible to quantify the vertex correction which is implicitly taken
into account via the Katanin substitution. Therefore we use the 
Ward identity (\ref{eq:WIsigma}) to express the scale-derivative of the  self-energy on the right-hand side  of
Eq.~(\ref{eq:GKatanin}) in terms of the derivative of the magnetization,
 \begin{equation}
 \partial_{\Lambda} 
 \Sigma_{\Lambda} (0) = - \frac{H}{ M_{\Lambda}^2 } \partial_{\Lambda} M_{\Lambda}.
 \label{eq:wardflow}
 \end{equation}
Our truncated flow equation (\ref{eq:Mflowsimp}) can then be written as
  \begin{eqnarray}
 \partial_{\Lambda} M_{\Lambda} 
& = &  \frac{M_0    }{\beta N } \sum_{\bd{k} , \omega} 
  \frac{ \Gamma^{+-z}_{\Lambda, {\rm Kat} }  \delta ( k - \Lambda )}{
 \Delta_{\Lambda} + E_{\bd{k} }  - i \omega  },
 \label{eq:Mflowapproxvert}
 \end{eqnarray}
where the scale-dependent mixed three-point vertex is given by
 \begin{eqnarray} 
M_0 \Gamma^{+-z}_{\Lambda, {\rm Kat} } = \frac{1}{ 1+ \frac{ \Delta_{\Lambda}}{ M_{\Lambda} }
   \int_K
 \frac{   \Theta ( \Lambda_0 - k ) \Theta ( k - \Lambda) }{ [ \Delta_{\Lambda}  + E_{ \bd{k} }
    - i \omega  ]^2 }  }.
 \hspace{7mm}
 \label{eq:gammaKat}
 \end{eqnarray}
We conclude that the Katanin substitution amounts to the assumption that
the mixed three-legged  vertex $\Gamma^{+-z}_{\Lambda} ( - K ,  K , 0)$
in the exact FRG flow equation (\ref{eq:flowh}) for the exchange field $\phi_{\Lambda}$
can be approximated
by a  momentum- and frequency-independent constant
$ \Gamma^{+-z}_{\Lambda, {\rm Kat}}  $ which is linked to the flowing magnetization
$M_{\Lambda}$ via Eq.~(\ref{eq:gammaKat}).  In Fig.~\ref{fig:vertexkat} we show the flow of 
$ \Gamma^{+-z}_{\Lambda, {\rm Kat}}$ 
in two dimensions for $T > 0$ and finite magnetic field.
Although the vertex correction
becomes important when the magnetization is significantly reduced from its initial value,
the flow of the magnetization is very similar to the flow
obtained without vertex correction.

\subsubsection{Vertex correction from flow equations}

We have shown in the previous subsection 
that  the Katanin substitution (\ref{eq:GKatanin}) amounts to
replacing  the three-legged vertex $\Gamma^{+-z}_{\Lambda} ( - K , K , 0 )$
in the flow equation  (\ref{eq:flowh2}) for the expectation value
$\phi_{\Lambda}$ of the exchange field
by a momentum- and frequency-independent coupling
$ \Gamma^{+-z}_{\Lambda, {\rm Kat}}$ given
by Eq.~(\ref{eq:gammaKat}). To check the validity of this substitution,
we now give an independent calculation of the vertex $\Gamma^{+-z}_{\Lambda}$.
In principle, we could write down the corresponding exact flow equation, which depends on various higher-order vertices. Alternatively, we can use the
Ward identity (\ref{eq:WIsigma}) to determine
 $\Gamma^{+-z}_{\Lambda}$ from the requirement that
the flow equations for $\partial_{\Lambda} M_{\Lambda}$ and $\partial_{\Lambda}
 \Sigma_{\Lambda} (0)$ give consistent results.  
Within our cutoff scheme where only the transverse part of the exchange interaction
is deformed, the exact flow equation (\ref{eq:Gammapmflow}) for the
transverse two-point vertex reduces to 
 \begin{eqnarray}
 & & \partial_{\Lambda} \Sigma_{\Lambda} (0)  =  
 \Gamma^{+-z}_{\Lambda}(0,0,0)  V_0 \partial_{\Lambda} M_{\Lambda} 
 \nonumber
 \\ 
&  &  +  \int_Q \dot{G}_{\Lambda} ( Q) \Gamma^{++--}_{\Lambda} ( 0, -Q, Q , 0 )
 \nonumber
 \\
 &  & - \int_Q \dot{G}_{\Lambda} ( Q) F_{\Lambda} ( Q) 
  \Gamma^{+-z}_{\Lambda}  (0, Q , -Q )  \Gamma^{+-z}_{\Lambda}  (-Q, 0 , Q )    .
 \hspace{7mm}
 \nonumber
 \\
 & &
 \label{eq:sigmaflowapprox}
 \end{eqnarray}
To simplify the calculation, 
we replace the vertices in the second and third lines of Eq.~(\ref{eq:sigmaflowapprox})
by their initial values given by the tree approximation discussed in Appendix~D,
see Eq.~(\ref{eq:Gammappmmtree}).
This amounts to neglecting the contribution involving the three-legged vertices
in the last line of 
Eq.~(\ref{eq:sigmaflowapprox}), and replacing in the second line
  \begin{equation}
  \int_Q \dot{G}_{\Lambda} ( Q) \Gamma^{++--}_{\Lambda} ( 0, -Q, Q , 0 ) 
 \rightarrow U_0  \int_Q \dot{G}_{\Lambda} ( Q),
 \end{equation}
where the coupling constant $U_0$ is given by
 \begin{equation}
 U_0  \approx \frac{ H + V_0 M_0}{ M_0^3 }.
 \label{eq:U0value}
 \end{equation}
Using the Ward identity in the form (\ref{eq:wardflow}) to express $\partial_{\Lambda}
 \Sigma_{\Lambda} (0)$ in terms of $\partial_{\Lambda} M_{\Lambda}$, 
we finally obtain
  \begin{eqnarray}
 \partial_{\Lambda} M_{\Lambda} 
& = &  - \frac{U_0}{ H / M_{\Lambda}^2 + V_0 \Gamma^{+-z}_{\Lambda} }
 \int_K \dot{G}_{\Lambda} ( K ).
 \end{eqnarray}
On the other hand,  for a momentum-independent three-legged vertex, 
the flow equation (\ref{eq:flowh2}) for $M_{\Lambda}$
is of the form
 \begin{equation}
\partial_{\Lambda} M_{\Lambda} 
 =   -\Gamma^{+-z}_{\Lambda} 
 \int_K \dot{G}_{\Lambda} ( K ),
 \end{equation}
so that in this approximation
the vertex $ \Gamma^{+-z}_{\Lambda } $ satisfies the compatibility condition
 \begin{equation}
 \Gamma^{+-z}_{\Lambda} = \frac{ U_0 }{ H / M_{\Lambda}^2 + V_0 \Gamma^{+-z}_{\Lambda} }.
 \end{equation}
Solving for $\Gamma^{+-z}_{\Lambda}$ we obtain
 \begin{eqnarray}
 & & \Gamma^{+-z}_{\Lambda} =   
 \left[ \left( \frac{ H }{2 V_0 M_{\Lambda}^2 }
 \right)^2 + \frac{  U_0}{ V_0   } \right]^{1/2}
  - \frac{ H }{2  V_0  M_{\Lambda}^2 }.
 \label{eq:vertflow} 
 \end{eqnarray} 
Keeping in mind that $  U_0 / V_0    = 1/M_0^2 + H /( V_0 M^3_0 )$,
we see that at the initial scale $\Lambda = \Lambda_0$ where $M_{\Lambda_0} = M_0$
the right-hand side of Eq.~(\ref{eq:vertflow}) reduces to $1/M_0$, while for small
$M_{\Lambda}$ the vertex vanishes as
 \begin{equation} 
 \Gamma^{+-z}_{\Lambda} \sim \frac{ H + V_0 M_0}{H} \frac{ M_{\Lambda}^2}{ 
 M^2_0 }.
 \end{equation}
With a sharp momentum cutoff the
flow of the magnetization is therefore given by
\begin{eqnarray}
 \partial_{\Lambda} M_{\Lambda} =
\frac{M_0  }{\beta N } \sum_{\bd{k} , \omega} 
  \frac{ \Gamma^{+-z}_{\Lambda }  \delta ( k - \Lambda )}{
 \Delta_{\Lambda} + E_{\bd{k} }  - i \omega  }.
 \label{eq:Mflowapproxvert2}
 \end{eqnarray}
In Fig.~\ref{fig:vertexkat} we compare the flow of the vertex  (\ref{eq:vertflow})
with the corresponding vertex flow implied by the Katanin substitution, see
Eq.~(\ref{eq:gammaKat}). The qualitative behavior is similar,
so that the Katanin substitution at least qualitatively takes the vertex correction 
due to the flow of the mixed three-point vertex into account.  
Figure~\ref{fig:vertexkat}(b) reveals furthermore that the resulting magnetization curves 
of the three different approximations used in this section do not differ significantly, indicating that the 
fulfilment of the Ward identity~(\ref{eq:WI1}) is the essential feature of any of these truncations.

\subsubsection{Wave-function renormalization}

Finally, let us include the wave-function renormalization factor $Z_{\Lambda}$
of the magnons which is related to 
the frequency-dependence of the magnon self-energy via 
 \begin{equation}
 \Sigma_{\Lambda} ( \bd{k} , i \omega ) = \Sigma_{\Lambda} (0 ) + M_{0}^{-1}
 \left( 1 - Z_{\Lambda}^{-1} \right) i \omega + {\cal{O}} ( \omega^2, k^2 ).
 \end{equation}
For our purpose it is sufficient to approximate the vertices in the exact flow equation
(\ref{eq:Gammapmflow}) for the transverse two-point vertex by their initial values
given in Appendix D. Then we obtain
\begin{equation}
 \partial_{\Lambda} \Sigma_{\Lambda} ( K ) = 
 \partial_{\Lambda} \Sigma_{\Lambda}^a ( K ) + 
 \partial_{\Lambda} \Sigma_{\Lambda}^b ( K ),
 \end{equation}
with
\begin{eqnarray}
  \partial_{\Lambda} \Sigma^{a}_{\Lambda} ( \bd{k} , i \omega ) 
 & = &   \frac{1}{\beta N M_0^2} \sum_{ \bd{k}^{\prime} ,  \omega^{\prime} }
 \dot{G}_{\Lambda} ( \bd{k}^{\prime} , i  \omega^{\prime} )  
 \nonumber
 \\
 &  \times & \Bigl[  G_1^{-1} (  \omega )     +  G_1^{-1} (   \omega^{\prime} )     - V_{ \bd{k} - \bd{k}^{\prime}} - V_0 \Bigr], 
 \nonumber
 \\
 & &
 \\
 \partial_{\Lambda} \Sigma^{b}_{\Lambda} ( \bd{k} , i \omega ) 
 &  = &  - \frac{1}{\beta N M_0^2} \sum_{ \bd{k}^{\prime} ,  \omega^{\prime} }
 \dot{G}_{\Lambda} ( \bd{k}^{\prime} , i  \omega^{\prime} ) 
 \frac{ \beta \delta_{ \omega , \omega^{\prime}} b^{\prime}}{ 1 - \beta b^{\prime} V_{ 
 \bd{k} - \bd{k}^{\prime}} }
 \nonumber
 \\
 &  \times &
 [ G_1^{-1} (   \omega )
 - V_{ \bd{k} - \bd{k}^{\prime}} ]
[ G_1^{-1} (   \omega^{\prime} ) 
 - V_{ \bd{k} - \bd{k}^{\prime}} ],
 \nonumber
 \\
 & &
 \label{eq:SigmaI}
 \end{eqnarray}
where with our cutoff scheme
 \begin{equation}
 G_1^{-1} (  \omega ) = \frac{ H + V_0 M_0 - i \omega}{M_0 },
 \end{equation}
see Eq.~(\ref{eq:G0def}).
At low temperatures, the derivatives of the Brillouin function are exponentially small, so that the second contribution $\Sigma^{b}_{\Lambda} ( \bd{k} , i \omega ) $ can be neglected.
From the frequency-dependence of $\Sigma^{a}_{\Lambda} ( \bd{k} , i \omega ) $, we obtain for the flowing wave-function renormalization
 \begin{eqnarray}
\partial_{\Lambda} Z_{\Lambda} & = &
 - \frac{ Z_{\Lambda}^2}{ M_0^2} \int_K \dot{G}_{\Lambda} ( K )
 \nonumber
 \\
 & = & K_D \frac{ ( a \Lambda )^{D-1}}{M_0} \frac{ Z_{\Lambda}^3}{
 e^{ \beta  Z_{\Lambda} ( \Delta_{\Lambda} + \rho_0 \Lambda^2 ) } -1 }.
 \label{eq:Zflow}
 \end{eqnarray} 
In the following section, we discuss the resulting magnetization and wave-function renormalization as functions of temperature and magnetic field strength.
\begin{figure}[tb]
 \begin{center}
\vspace{0.5mm}
\hspace*{-2mm}
 \includegraphics[width=0.5\textwidth]{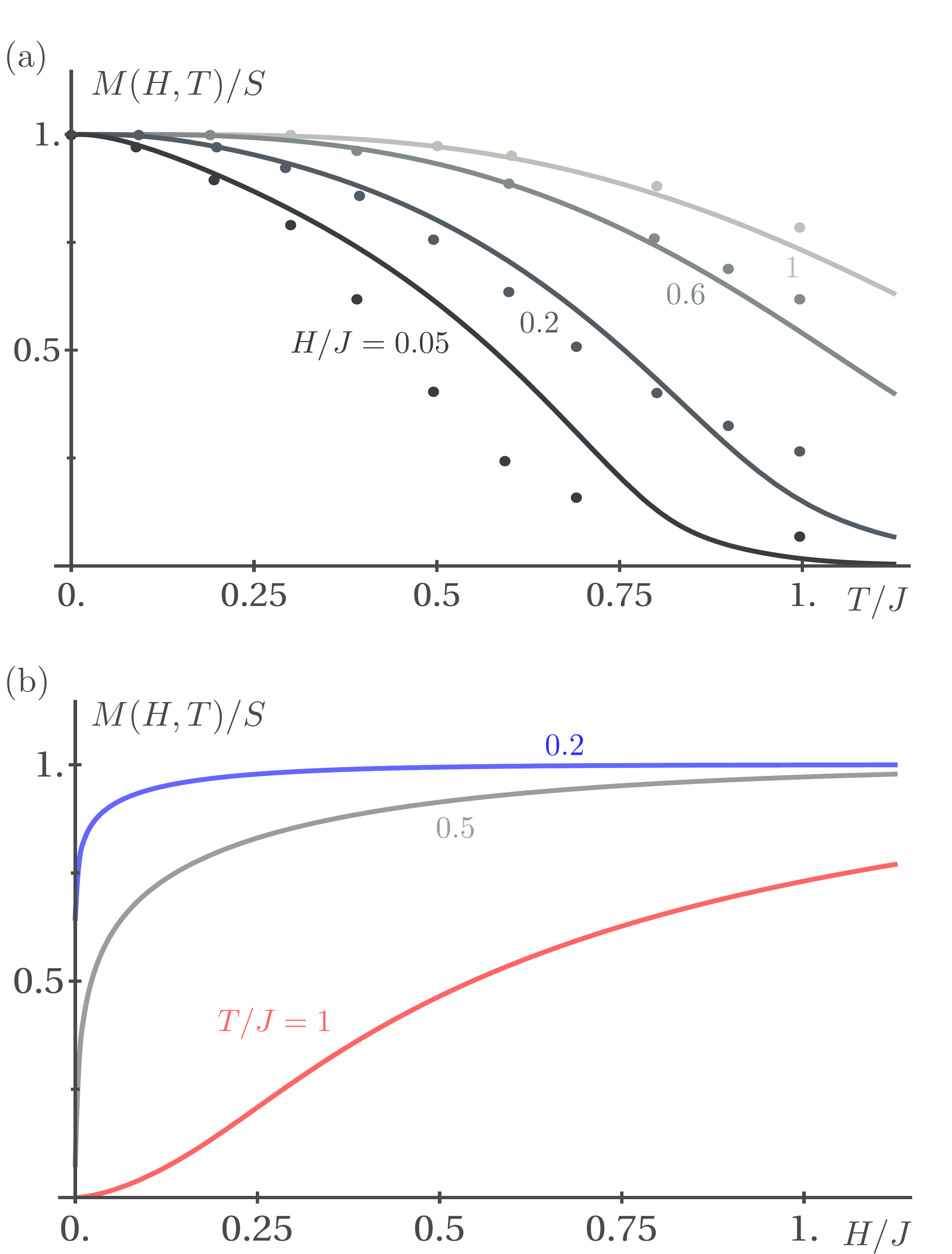}
   \end{center}
  \vspace{-2mm}
  \caption{
  (a) Magnetization curve $M (T,H)$ in $D = 2$ as a function of temperature obtained from the numerical integration of Eqs.~  \eqref{eq:Zflow} and \eqref{eq:Mflow3} for $H/J=1.0,0.6,0.2,0.05$ (top to bottom; the values of $H/J$ are written next to the curves).
Here $J = \rho_0 /( M_0 a^2)$ is defined in terms of the bare spin-stiffness $\rho_0$, 
see Eqs.~(\ref{eq:Vexpansion}) and~(\ref{eq:Jdef}).
The dots are the Monte Carlo results of Ref.~[\onlinecite{Henelius00}]
for a spin $S=1/2$ Heisenberg ferromagnet with nearest neighbor exchange $J$.
  (b) $M (H,T)$ as a function of the magnetic field for temperatures $T/J=1,0.5,0.2$ (red, gray, blue; the values of $T/J$ are written next to the curves).
}
\label{fig:magcurve}
\end{figure}
\begin{figure}[tb]
 \begin{center}
\vspace{0.5mm}
\hspace*{-2mm}
 \includegraphics[width=0.5\textwidth]{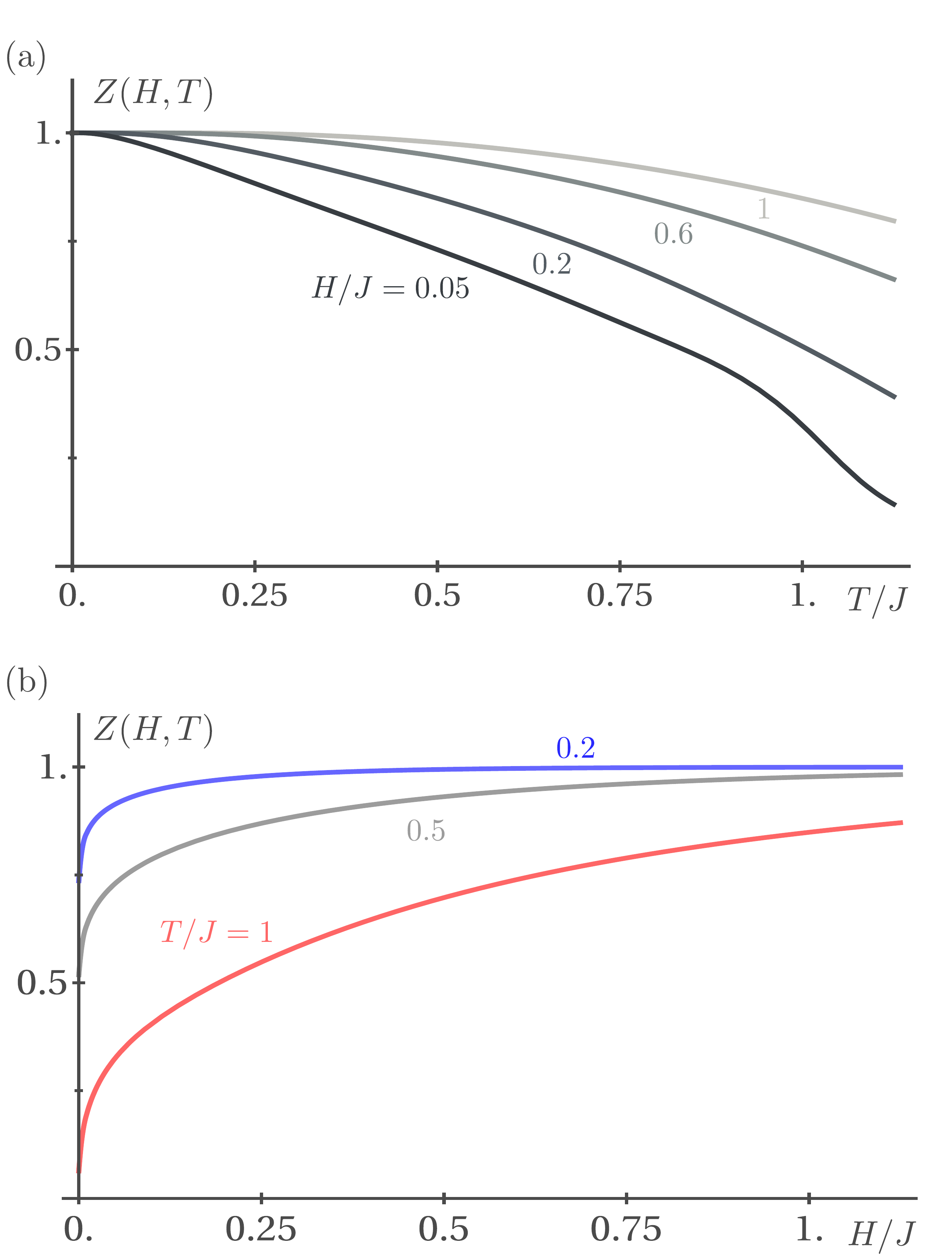}
 \end{center}
   \vspace{-2mm}
  \caption{
  (a)  Wave-function renormalization factor $Z ( H , T )$ in $D = 2$ as a function of temperature obtained from the numerical integration of Eqs.~\eqref{eq:Zflow} and ~\eqref{eq:Mflow3} for $H/J=1.0,0.6,0.2,0.05$ (top to bottom). (b) $Z ( H , T )$ as a function of magnetic field for temperatures $T/J=1,0.5,0.2$ (red, gray, blue).
  }
\label{fig:Zres}
\end{figure}
\subsection{Magnetization curves}
\label{subsec:magcurves}

Taking into account the wave-function renormalization factor $Z_{\Lambda}$ and vertex correction given in
Eq.~(\ref{eq:vertflow}), our modified flow equation for the magnetization
becomes 
\begin{equation}
 \Lambda \partial_{\Lambda} M_{\Lambda} = M_0 K_D   \frac{   Z_{\Lambda} \Gamma^{+-z}_{\Lambda}   
( a \Lambda)^{D}    }{ e^{ \beta Z_{\Lambda}
 ( \Delta_{\Lambda} + \rho_0 \Lambda^2 ) }-1},
 \label{eq:Mflow3}
 \end{equation}
which should be solved simultaneously with the flow equation  \eqref{eq:Zflow}
for the wave-function renormalization factor.
Note  that the vertex $\Gamma^{+-z}_{\Lambda}$ and the gap $\Delta_{\Lambda} = H M_0 / M_{\Lambda}$
are functions of the flowing magnetization $M_{\Lambda}$, see Eqs.~(\ref{eq:vertflow}) and
(\ref{eq:Deltadef}).
By integrating the flow equations (\ref{eq:Mflow3}) and (\ref{eq:Zflow})  from some large initial scale $\Lambda_0$ 
of the order of the inverse lattice spacing down to $\Lambda =0$, we obtain
nonperturbative expressions for the magnetic equation of state
$M ( H , T )$ and the corresponding wave-function renormalization factor $Z( H, T)$,
 which remain well-defined in low dimensions even in the limit of vanishing
magnetic field, see Figs.~\ref{fig:magcurve} and~\ref{fig:Zres}.
Recall that in Eq.~(\ref{eq:Jdef}) we have defined the  
energy scale $J = \rho_0 /(M_0 a^2 ) $  in terms of the bare spin  stiffness \cite{footnoteJ}.
For low temperatures and finite magnetic field, both $M/S$ and $Z$ 
remain close to unity, indicating that magnons are well-defined quasiparticles. 
When $T/J$ is not small the magnetization and the wave-function renormalization factor
both decrease.
The identity (\ref{eq:corr}) implies that then also the correlation length  decreases.
For not too small magnetic field, the quantitative agreement of our SFRG calculation
with Monte Carlo simulations \cite{Henelius00} 
extends to temperatures of order $J$, while
for $H  \ll J$, our SFRG result for $M ( H , T )$  agrees with the Monte Carlo results only
for temperatures $T \lesssim 0.2 J$. This is due to the fact that
in our truncation we have neglected all terms involving the derivatives of the
Brillouin function, which for $H \ll J$ is only  justified for temperatures
$T \ll J$.
From Fig.~\ref{fig:Zres} (b), we also see that
for any finite magnetic field $H$ the wave-function renormalization factor
$Z$ remains finite. On the other hand, $Z ( H, T )$ vanishes
if we take the 
limit $H\rightarrow 0$ for $T > 0$; the corresponding steepening of the slope $\partial M(H,T) / \partial H \Huge|_{H=0}$ reflects the exponential behavior of the susceptibility calculated in Eq.~\eqref{eq:chi1loop}.
The fact that the Monte Carlo calculations have been performed for a 
nearest neighbor Heisenberg model whereas
our truncation of the hierarchy of the SFRG flow equations is controlled 
only for long-range interactions~\cite{footnoteJ}
does not affect the quantitative accuracy of our SFRG calculation at low temperatures because
in this regime the relevant energy scale is set by the spin stiffness $\rho_0$ 
defined in terms of the small-momentum expansion (\ref{eq:Disp}) of the magnon dispersion.
We conclude that at low temperatures
the magnetization curves predicted by our SFRG approach agree with controlled
Monte Carlo simulations for two-dimensional quantum Heisenberg 
ferromagnets~\cite{Henelius00}.
Similar results have been obtained with a $1/N$-expansion~\cite{Timm98}, with
Green function methods~\cite{Junger04},  and with an exact 
diagonalization calculation~\cite{Junger08}.

\section{Magnon damping due to classical longitudinal fluctuations}
\label{sec:damping}

\begin{figure}[tb]
 \begin{center}
 \vspace*{-2mm}
 \hspace*{-2mm}
 \includegraphics[width=0.5\textwidth]{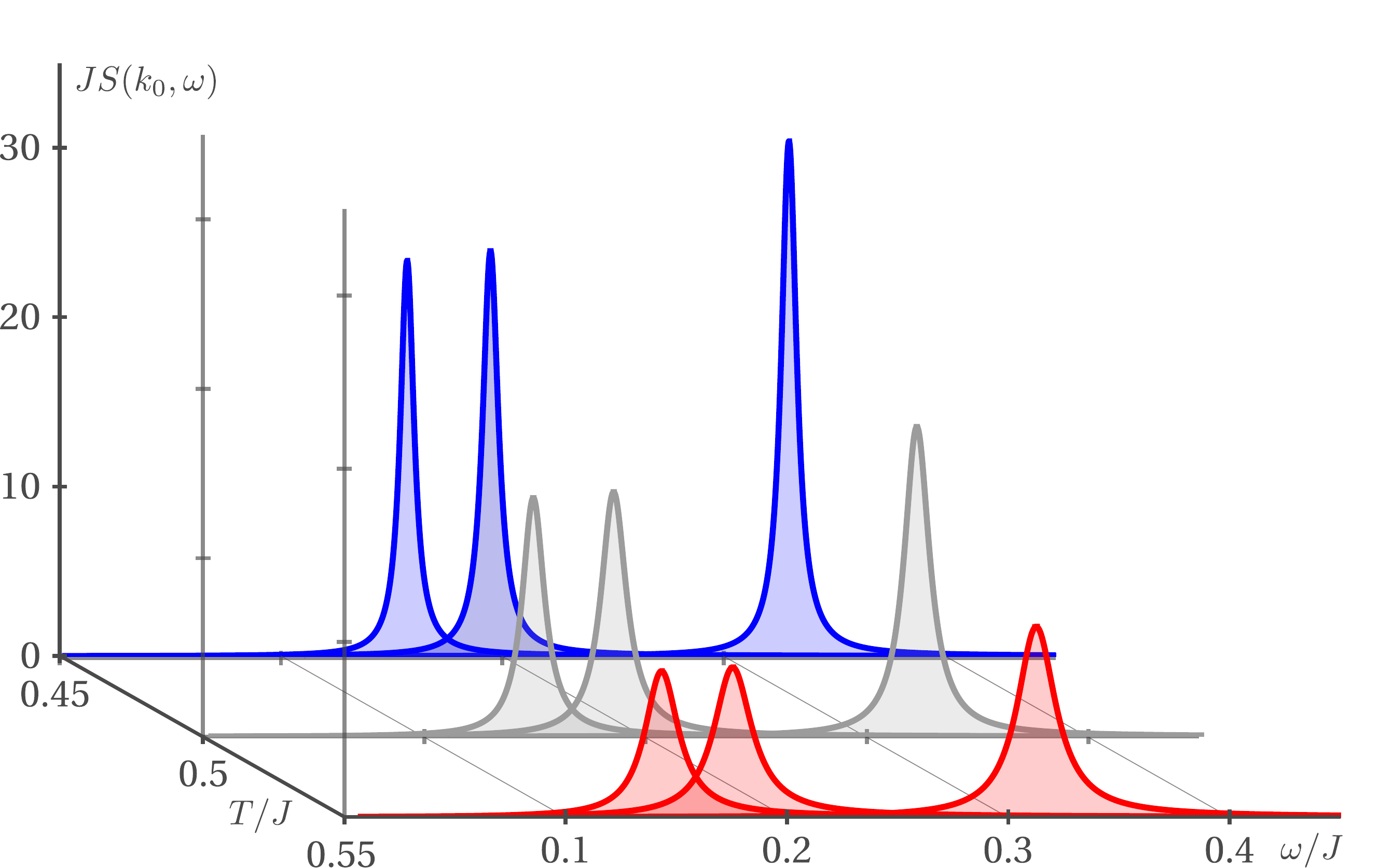}
   \end{center}
  \vspace{-2mm}
  \caption{
Spectral density function $S( \bd{k}_0,\omega)$ 
as a function of frequency and temperature for $k_0a=0.75$ and different magnetic fields $H/J 
=0.001,0.01,0.1$ (from left to right). The quasiparticle peaks sharpen with increasing magnetic field strength and decreasing temperature. Note that the lineshapes of the peaks are not symmetrical around $\omega= Z\Delta+Z E_{\bd{k}_0}$ but show a slight asymmetry due to the $\omega$-dependence of $\gamma(\bd{k}_0,\omega)$.
}
\label{fig:damp}
\end{figure}

Let us now calculate the damping of magnons
due to the coupling to classical longitudinal spin fluctuations at intermediate temperatures.
This decay channel  is not properly taken into account
 in  the usual spin-wave expansion where fluctuations of the length of the
magnetic moments are not treated  as independent degrees of freedom \cite{Hofmann02}.
 For a three-dimensional ferromagnet, 
the leading perturbative contribution to this decay process has already been 
studied
by VLP \cite{Vaks68b} using their spin-diagrammatic approach.
To obtain their result for the decay rate within our SFRG approach, 
we can neglect self-energy corrections to the single-scale propagator
on the right-hand side of the flow equation for the self-energy
$ \Sigma^{b}_{\Lambda} ( \bd{k} , i \omega ) $ in
Eq.~(\ref{eq:SigmaI}). The right-hand side of the flow equation is then 
a total derivative which can easily be integrated over the flow parameter.
After analytic continuation to real frequencies we then obtain 
the perturbative result for the damping given by VLP \cite{Vaks68b},
 \begin{eqnarray}
\gamma ( \bd{k} , \omega ) & = & - M_0 {\rm Im} 
 \Sigma^b_{\Lambda} ( \bd{k} , \omega + i 0 )
 \nonumber
 \\
& = & \frac{\pi b^{\prime}   }{N  } \sum_{\bd{k}^{\prime}} 
  \delta ( H + E_{\bd{k}^{\prime}} - \omega  )
 \frac{     ( V_{\bd{k}^{\prime}} - V_{ \bd{k} - \bd{k}^{\prime} } )^2    }{ 1 - \beta b^{\prime} V_{ 
 \bd{k} - \bd{k}^{\prime}} }.
 \hspace{7mm}
 \label{eq:damp1int}
 \end{eqnarray}
In two dimensions, this expression is not valid for small 
magnetic field, because we know that  self-energy corrections generate a gap in the magnon spectrum and therefore 
cannot be neglected.
Unfortunately, the inclusion of self-energy corrections
requires also a consistent renormalization of the higher-order interaction vertices which is 
beyond the scope of this work. A simple
phenomenological way to take self-energy effects into account is to
replace the bare magnetic field $H$ 
in Eq.~(\ref{eq:damp1int}) by the gap
 \begin{equation}
 \Delta  = \frac{H M_0}{M (H,T) }
 \end{equation} 
and divide the external frequency by the wave-function 
renormalization factor $Z$, both of which are calculated by means of our SFRG approach. 
This leads to the following expression for the magnon damping,
\begin{eqnarray}
\gamma ( \bd{k} , \omega )  
& = & \frac{\pi b^{\prime}   }{N  } \sum_{\bd{k}^{\prime}} 
  \delta ( \Delta  + E_{\bd{k}^{\prime}} - \omega /Z  )
 \frac{     ( V_{\bd{k}^{\prime}} - V_{ \bd{k} - \bd{k}^{\prime} } )^2    }{ 1 - \beta b^{\prime} V_{ 
 \bd{k} - \bd{k}^{\prime}} }.
 \nonumber
 \\
 & &
 \label{eq:dampres}
 \end{eqnarray}
Formally, the renormalizations described by  $\Delta$ and $Z$ can be 
generated by replacing the mean-field inverse propagators $G_1^{-1} ( i \omega )$
in the second line of Eq.~(\ref{eq:SigmaI})
by appropriate renormalized propagators, which amounts to assuming that
the structure of renormalized vertex resembles the bare one.
Evaluating $\gamma(\bm{k},\omega)$ within a small-momentum expansion, we  plot in Fig. \ref{fig:damp} the
resulting spectral density function
\begin{align}
S(\bm{k},\omega)&= \frac{1}{\pi} \text{Im } G(\bm{k},i\omega\rightarrow \omega+i0)
\nonumber
\\
&=
\frac{1}{\pi}
 \frac{M_0 \gamma(\bm{k},\omega)}{\left[\Delta+E_{k}-\omega/Z\right]^2+ \gamma(\bm{k},\omega)^2 }
\end{align}
as a function of temperature and magnetic field for a generic  momentum $\bd{k}_0$. Note that the threshold for the spectral weight 
is controlled by the $\delta$ function in Eq.~(\ref{eq:dampres}), which
sets 
$ S(\bm{k},\omega)=0$ for $\omega<Z (\Delta+E_{\bd{k}_0})$.
In the limit $H \rightarrow 0$, this implies that the spectral weight 
$S(\bm{k},\omega)$ vanishes for $\omega<ZM_0/\chi(T) +E_{\bd{k}_0}$, 
where $\chi ( T) $ is the susceptibility for vanishing field.
With increasing magnetic field and decreasing temperature the resulting quasiparticle-peaks grow and sharpen, 
so that the magnons are stabilized in this regime.
The same holds true when the magnitude of the  
magnon momentum $k_0$ is lowered (not shown in Fig.~\ref{fig:damp}).
The momentum and frequency dependence of $\gamma(\bm{k},\omega)$ furthermore leads to a  non-Lorentzian asymmetry of the spectral lineshape, 
which increases when  the quasiparticle-peaks broaden.  
It should be emphasized that 
at low temperatures ($T \ll J$) the derivative $b^{\prime}$ of the Brillouin function and hence also the
magnon damping in Eq.~(\ref{eq:dampres})  are exponentially small. In this regime we expect that 
the magnon damping is dominated by two-body scattering of renormalized magnons. The proper treatment of these type of processes,
taking into account that at finite temperature magnons cannot propagate over distances larger than 
the correlation length, is beyond the scope of this work.

\section{Summary and outlook}
 \label{sec:conclusions}

In this work we have used the  spin functional renormalization group approach (SFRG)
recently proposed by two of us \cite{Krieg19} to study the
thermodynamics and dynamics
of two-dimensional quantum Heisenberg ferromagnets at low temperatures.
We have established the precise relation between the SFRG and an old 
momentum-shell RG-calculation\cite{Kopietz89}, and have derived the temperature dependence of the susceptibility and transverse correlation length,
which agree up to a prefactor in the susceptibility with Takahashi's modified spin-wave theory \cite{Takahashi86} and Schwinger-Boson calculations,\cite{Arovas88} respectively.
Our SFRG results for the magnetic equation of state in two dimensions agrees 
quite well with numerical simulations.
Moreover, we have also shown how the damping of transverse spin-waves due to the
coupling to longitudinal spin fluctuations can be obtained
within our SFRG approach. 
At intermediate temperatures, this damping can be substantial in two dimensions and 
generates an asymmetry in the spectral  line shape.

This work also contains several technical advances 
which will be helpful  for future applications of the SFRG to quantum spin systems.
Of particular importance are the following three points.

\begin{enumerate}

\item The construction of the 
hybrid functional $\Gamma_{\Lambda} [ \bd{m} , \phi ]$ 
in Eq.~(\ref{eq:GammaHMdef}) which generates vertices which are irreducible
in the transverse propagator line and in the longitudinal interaction line.
This functional  satisfies the Wetterich equation (\ref{eq:Wetterichhybrid})
and has a well-defined initial condition even if we start 
from a deformed Hamiltonian where
the exchange interactions are completely switched off. The 
two-point functions generated by this functional
can be identified physically with the self-energy $\Sigma_{\Lambda} ( K )$ 
of the transverse magnons and the irreducible polarization $\Pi_{\Lambda} ( K )$ of longitudinal spin fluctuations. This establishes the precise relation between our SFRG approach and the diagrammatic method for quantum spin systems developed by 
VLP \cite{Vaks68,Vaks68b} and by others \cite{Izyumov88,Izyumov02}.

\item The use of the Ward identity \eqref{eq:WI1} to obtain a closed flow equation for 
the magnetization. This is crucial in the regime where the spin-rotational symmetry is spontaneously broken and the magnon spectrum is gapless. The Ward identity guarantees in this case that the 
gap in the magnon spectrum vanishes in the entire symmetry broken phase without
fine-tuning the initial conditions. Note also that, in contrast
to Takahashi's modified spin-wave theory \cite{Takahashi86,Takahashi87},
 in our approach, we do not assume a priori that for $H=0$ the magnetization  vanishes in two dimensions.

\item In a deformation scheme where initially the 
exchange interactions are completely switched off, the initial condition 
for the renormalization group flow 
is determined by the connected imaginary-time ordered correlation functions
of a single spin in a magnetic field. The diagrammatic calculation of these correlation function using the
generalized Wick theorem for spin operators derived by VLP \cite{Vaks68,Vaks68b} are rather cumbersome. 
We have derived an  explicit recursive form of the
generalized Wick theorem for spin operators in frequency space, 
see Eq.~(\ref{eq:recursion}), which 
provides us with an efficient method to calculate the higher-order connected
spin correlation functions.

\end{enumerate}

Another challenging problem where our
SFRG approach promises to be useful is the calculation of the longitudinal 
part of the dynamic structure factor of an ordered ferromagnet.
VLP \cite{Vaks68b} suggested that for sufficiently small wave-vectors
the longitudinal structure factor of an ordered ferromagnet should exhibit a diffusive peak 
at vanishing frequency and two inelastic peaks at frequencies corresponding to the magnon energies.
However, the  diagrammatic resummation  within the framework of the 
spin-diagrammatic approach to confirm this scenario has not been found \cite{Vaks68b,Maleev73}.
Recently, it has been shown by means of a kinetic equation approach
that in the absence of momentum-relaxing scattering processes (such as umklapp processes or scattering by disorder) the 
 longitudinal structure factor of a two-dimensional ferromagnet in a magnetic field
exhibits a linearly dispersing hydrodynamic sound mode at low frequencies, which is  induced by fluctuations of the magnon density \cite{Rodriguez18}. In three dimensions,
such a mode was also found by 
Izyumov {\it{et al.}} \cite{Izyumov02} by means of the
spin-diagram technique \cite{Vaks68,Vaks68b,Izyumov88}.  It would be interesting to see whether such 
a sound-like magnon mode can also be obtained within our SFRG approach.

Finally, let us point out that
our SFRG approach can be extended in many directions.
Our goal is to develop our  method to become a useful tool for studying 
various types of frustrated spin systems in the regime without 
long-range magnetic order. 
Note that in the past few years these types of systems have
been studied using the so-called pseudofermion FRG
\cite{Reuther10,Reuther11,Reuther11a,Buessen16}, which relies on the represention of the spin-$1/2$ operators in terms of fermionic operators \cite{Coleman15}. 
The unphysical states which are introduced by this representation have to be projected out which in practice
can only be achieved approximately.
This problem does not arise within our SFRG approach where we directly work with the
physical spin operators.

\section*{Acknowledgments}
This work was financially supported by the Deutsche Forschungsgemeinschaft (DFG)
through project KO 1442/10-1.

\begin{appendix}

\section*{APPENDIX A:  Relations between irreducible vertices and connected correlation functions}
\setcounter{equation}{0}
\renewcommand{\theequation}{A \arabic{equation}}
 \label{sec:tree}

In this appendix we work out the precise relation between the
irreducible vertices generated by our  hybrid functional 
$\tilde{\Gamma}^{}_{\Lambda} [ \bd{m}, \varphi ]=
 \Gamma^{}_{\Lambda} [ \bd{m},\phi_\Lambda+\varphi ]$ 
defined via Eqs.~(\ref{eq:Gammashift}) and (\ref{eq:Gammahcomplete})
on the one hand, 
and the connected spin correlation functions generated by $\mathcal{G}_{\Lambda} [\bd{h}]$
defined in Eq.~(\ref{eq:Gcdef}).
With a slight variation of the notation introduced in Eq.~(\ref{eq:collectivefield}),
we define the three-component field $\bd{\Phi}_K$ with spherical transverse components, 
 \begin{equation}
 \bd{\Phi}_K = 
 \left( \begin{array}{c}
 \Phi^{+}_K  \\ \Phi^-_K \\ \Phi^z_K 
 \end{array} \right)
 = \left( \begin{array}{c}
 m^+_K  \\ m^-_K  \\ \phi_K  \end{array} \right),
 \end{equation}
where the spherical Fourier components $m^{\pm}_K$  are defined 
as in Eq.~(\ref{eq:FTdef}) implying
$( m_K^+ )^{\ast} = m_{-K}^-$, see Eq.~(\ref{eq:mpmrelation}).
Note that in this appendix
the superscript $\alpha$ assumes the values $+,-,z$.
We also introduce the conjugate three-flavor source field 
 \begin{equation}
 \bd{j}_K  = \left( \begin{array}{c}
 j^{+}_K  \\ j^-_K \\ j^z_K 
 \end{array} \right)
 = \left( \begin{array}{c}
 h^+_K  \\ h^-_K  \\ s_K  \end{array} \right).
 \end{equation}
The generating functional $   \Gamma_{\Lambda} [ \bd{\Phi} ] =
\Gamma_{\Lambda} [ \bd{m} , \phi ]$
defined in Eq.~(\ref{eq:GammaHMdef})  can then be written as
  \begin{equation}
 \Gamma_{\Lambda} [ \bd{\Phi} ]  
  =  \int_K
 \bd{\Phi}_{K}^{\dagger}  \bd{j}_K
-   {\cal{F}}_{\Lambda} [ \bd{j} ]
- \frac{1}{2} \int_K \bd{\Phi}_K^\dagger \mathbf{R}_{\Lambda} ( \bd{k} )  
 \bd{\Phi}_{K} ,
 \end{equation}
where the hybrid functional ${\cal{F}}_{\Lambda} [ \bd{j} ] =
{\cal{F}}_{\Lambda} [ h^+, h^- , s ]$ is defined in Eq.~(\ref{eq:Fdef}), i.e.,
 \begin{eqnarray}
  {\cal{F}}^{}_{\Lambda} [ \bd{j} ] &  = &  
{\cal{G}}_{\Lambda} \Bigl[ {h}^{+}_K, h^-_K , h^z_K = - J^z_{\Lambda} ( \bd{k} ) s_K \Bigr]
 \nonumber
 \\
 & & - \frac{1}{2} \int_K
 {J}^{z }_{\Lambda}(\bd{k}) s_{-K}  s_{K}.
 \end{eqnarray}
The regulator matrix $\mathbf{R}_{\Lambda} ( \bd{k} )$
is diagonal in flavor space,
\begin{align}
\mathbf{R}_{\Lambda} ( \bd{k} )
=
\begin{pmatrix}
R^\perp_\Lambda(\bd{k}) & 0& 0
\\
0 & R^\perp_\Lambda(\bd{k}) & 0
\\
0 & 0 & R^\phi_\Lambda(\bd{k})
\end{pmatrix},
\end{align}
where the transverse and longitudinal regulators  
$R^{\bot}_{\Lambda} ( \bd{k} )$ and $R^{\phi}_{\Lambda} ( \bd{k} )$ in momentum space  
are given in Eqs.~(\ref{eq:Rbotmom}) and (\ref{eq:Rphimom}).
To take into account that the third component $\phi_K$ of $\bd{\Phi}_K$
can have a finite expectation value, we set  $\phi_K = \delta (K ) \phi_{\Lambda}
 + \varphi_K$ and define [see Eq.~(\ref{eq:Gammashift}]
 \begin{equation}
 \tilde{\Gamma}_{\Lambda} [ \bd{\tilde{\Phi}} ] = \Gamma_{\Lambda} [ 
 m^+_K , m^-_K , \phi_K = \delta ( K ) \phi_{\Lambda} + \varphi_K ],
 \end{equation}
where the third component of
\begin{equation}
 \bd{\tilde{\Phi}}_K 
 = \left( \begin{array}{c}
 m^+_K  \\ m^-_K  \\ \varphi_K  \end{array} \right)
 \end{equation}
contains the fluctuating part $\varphi_K$ 
of the longitudinal exchange field $\phi_K$.
The expansion of
 $\tilde{\Gamma}_{\Lambda} [ \bd{\tilde{\Phi}} ]$ in powers of 
$\bd{\tilde{\Phi}}$  defines the
irreducible hybrid vertices, 
see Eq.~(\ref{eq:Gammahcomplete}).
Similarly, we may expand the functional ${\cal{F}}_{\Lambda} [ \bd{j} ]$
in powers of the sources $\bd{j}$,
 \begin{widetext}
\begin{eqnarray}
{\cal{F}}_{\Lambda} [ \bd{j}  ] &   =   &
  {\cal{F}}_{\Lambda} [ 0 ] 
+  \int_K G^{+-}_{\Lambda} ( K ) h^{-}_{-K } h^+_K
 +  \frac{1}{2!} \int_K F_{\Lambda} ( K ) s_{-K } s_K
 \nonumber
 \\
  & + &    \int_{K_1} \int_{K_2}
\int_{K_3} \delta ( K_1 + K_2 + K_3 )  F^{+-z}_{\Lambda} ( K_1 , K_2 ,  K_3 ) 
h^-_{K_1} h^+_{K_2} s_{ K_3 } 
\nonumber
\\
 &+ &
  \frac{1}{3!} \int_{K_1} \int_{ K_2} \int_{K_3} 
 \delta ( K_1 + K_2 + K_3 ) 
F^{zzz}_{\Lambda} ( K_1 , K_2 , K_3  ) s_{K_1 } s_{K_2} s_{ K_3 }
+ ... \quad.
\end{eqnarray}
\end{widetext}
The relations between the correlation functions  generated by 
${\cal{F}}_{\Lambda} [ \bd{j} ]$ and the vertices generated by 
its (subtracted) Legendre transform
$\tilde{\Gamma}_{\Lambda} [ \bd{\tilde{\Phi}} ]$ can be obtained by successive differentiation of the
field-dependent relation of the Hessian matrices,
\begin{equation}
\left( \bd{\tilde{\Gamma}}_\Lambda''[ \bd{\tilde{\Phi}} ]
+
\mathbf{R}_{\Lambda}
 \right)_{a a^{\prime}}  =
\Bigl( \mathbf{{F}}_\Lambda''[\bd{j} ] \Bigr)^{-1}_{a a^{\prime}},
\label{eq:Hessian}
\end{equation}
where the collective labels  $a = ( \alpha , K )$ and $a^{\prime} = ( \alpha^{\prime} , K^{\prime} )$ 
represent  the flavour index $\alpha = +,-,z$ in combination with the momentum-frequency  label $K =
( \bd{k} , i \omega )$, and the double-primes represent the second functional derivatives with respect to the corresponding fields,
\begin{eqnarray}
\left( \bd{\tilde{\Gamma}}_\Lambda''[  \bd{\tilde{\Phi}} ] \right)_{aa'} 
&= & \frac{\delta^2 \tilde{\Gamma}_\Lambda[\bd{\tilde{\Phi}}] }{\delta 
\tilde{\Phi}_{K}^{\alpha} \delta \tilde{\Phi}_{K'}^{\alpha'} },
 \\
\Bigl(  \mathbf{F}_\Lambda''[\bd{j} ] \Bigr)_{aa'} &= & \frac{\delta^2 
\mathcal{F}_\Lambda[\bd{j}] }{\delta j_{K}^{\alpha} \delta j_{K'}^{\alpha'} }.
\end{eqnarray}
By taking additional derivatives of Eq.~\eqref{eq:Hessian} 
with respect to $\tilde{\Phi}_K^\alpha$ 
we find for the third- and  fourth-order derivative tensors,
 \begin{widetext}
\begin{eqnarray}
\left( \bd{\tilde{\Gamma}}_\Lambda^{(3)}[  \bd{\tilde{\Phi}} ] \right)_{a_1a_2 a_3} 
&= &
-
\sum_{a_1^{\prime} a_2^{\prime} a_3^{\prime}}
\left[ \prod_{i=1}^3
\left( \mathbf{F}_\Lambda''[\bd{j}  ]\right)^{-1}_{a_i a_i^{\prime}} \right]
\left(  \mathbf{F}_\Lambda^{(3)}[\bd{j} ] \right)_{a_1^{\prime} a_2^{\prime} a_3^{\prime}} ,
\label{eq:Treegamma3}
\\
\left( \bd{\tilde{\Gamma}}_\Lambda^{(4)}[ \bd{\tilde{\Phi}} ] \right)_{a_1 a_2 a_3 a_4} 
& =&
-
\sum_{a_1^{\prime} a_2^{\prime} a_3^{\prime} a_4^{\prime}}
\left[ \prod_{i=1}^4
\left(\mathbf{F}_\Lambda''[\bd{j} ]\right)^{-1}_{a_i a_i^{\prime}}  \right]
\left( \mathbf{F}_\Lambda^{(4)}[\bd{j} ] \right)_{a_1^{\prime} a_2^{\prime} a_3^{\prime} a_4^{\prime}} 
\nonumber
\\
&+ & \mathcal{S}_{a_1, a_2; a_3,a_4}
\frac{1}{2} 
\sum_{a_1^{\prime} a_2^{\prime}} 
\left( 
\bd{\tilde{\Gamma}}_\Lambda^{(3)}[ \bd{\tilde{\Phi}} ] \right)_{a_1 a_2 a_1^{\prime}}
\left(\mathbf{F}_\Lambda''[\bd{j} ]\right)^{-1}_{a_1^{\prime} a_2^{\prime}}
\left( \bd{\tilde{\Gamma}}_\Lambda^{(3)}[ \bd{\tilde{\Phi}}] \right)_{a_2^{\prime} a_3 a_4}.
\label{eq:Treegamma4}
\end{eqnarray}
Here, the operator
$ \mathcal{S}_{a_1, a_2; a_3,a_4}$ symmetrizes 
the expression to its  right
with respect to the exchange of all labels \cite{Kopietz10}, and
the summation over the internal labels 
is defined by $\sum_{a}= \int_K \sum_{\alpha=+,-,z}$.
By setting $\tilde{\bd{\Phi}} =0$ and $\bd{j} =0$
in Eqs.~(\ref{eq:Treegamma3}) and (\ref{eq:Treegamma4}) we obtain the desired
 expansion of the three-point and four-point vertices generated by our hybrid functional
 $\tilde{\Gamma}_{\Lambda} [ \bd{\tilde{\Phi}} ]$ 
in powers of the correlation functions generated by ${\cal{F}}_{\Lambda} [ \bd{j} ]$.
Using the fact that for vanishing sources
 \begin{subequations}
\begin{align}
&\left( \mathbf{F}_{\Lambda}''[0] \right)_{KK'}^{+-}=\delta(K+K')G^{-+}_\Lambda(K)=\delta(K+K')G_\Lambda(-K),
\\
&\left( \mathbf{F}_{\Lambda}''[0] \right)_{KK'}^{-+}=\delta(K+K')G^{+-}_\Lambda(K)=\delta(K+K')G_\Lambda(K),
\\
&\left( \mathbf{F}_{\Lambda}''[0] \right)_{KK'}^{zz}=\delta(K+K')F_\Lambda(K),
\end{align}
\end{subequations}
where the scale-dependent transverse  propagator interaction $G_{\Lambda} ( K )$
is given in Eq.~(\ref{eq:GGamma}) and the longitudinal effective interaction
$F_{\Lambda} ( K ) $ is given in Eq.~(\ref{eq:Fzzdef}),
we obtain from Eq.~(\ref{eq:Treegamma3}) for the three-point vertices
defined via the vertex expansion  (\ref{eq:Gammahcomplete})
in momentum-frequency space,
\begin{align}
 \Gamma_\Lambda^{+-z} ( K_1 , K_2 , K_3  )  = & - G_\Lambda^{-1}  (-K_1 ) G_\Lambda^{-1} (K_2 ) 
 F_\Lambda^{-1} ( K_3 )    F_\Lambda^{+-z} ( - K_1 , -K_2 , -K_3  ),
 \label{eq:F3general1}
 \\
 \Gamma_\Lambda^{zzz} ( K_1 , K_2 , K_3  )  = & - 
 \left[ \prod_{i=1}^3 
 F_\Lambda^{-1}  (K_i )  \right]  F_\Lambda^{zzz} ( - K_1 ,- K_2 , -K_3  ).
 \label{eq:F3general2}
 \end{align}
Using the relation (\ref{eq:Fdef}) between the hybrid functional ${\cal{F}}_{\Lambda} [ \bd{j} ]$
and the generating functional ${\cal{G}}_{\Lambda} [ \bd{h} ]$ of the
connected spin correlation functions, we can express the three-point functions
$F_\Lambda^{+-z} ( - K_1 , -K_2 , -K_3  )$
and  $F_\Lambda^{zzz} ( - K_1 ,- K_2 , -K_3  )$ on the right-hand side of
Eqs.~(\ref{eq:F3general1}) and (\ref{eq:F3general2})
in terms of the connected spin correlation functions. In momentum-frequency space we obtain
\begin{eqnarray}
F_\Lambda^{-1}(K_3) F_\Lambda^{+-z} ( - K_1 , -K_2 ,- K_3  )
& = &   F_\Lambda^{-1}(K_3) \left(-J^z_{\Lambda,\bd{k_3}} \right) 
G_\Lambda^{+-z} ( - K_1 , -K_2 ,- K_3  )
 \nonumber
 \\
 & = &
\left[ 1+J^z_{\Lambda,\bd{k_3}}\Pi_\Lambda(K_3) \right] 
G_\Lambda^{+-z} ( - K_1 , -K_2 , -K_3  ),
 \label{eq:GFpmz}
\\
 \left[
\prod_{i=1}^3
F_\Lambda^{-1}  (K_i ) \right]
F_\Lambda^{zzz} ( - K_1 , -K_2 , -K_3  ) 
& = &  
 \left[
\prod_{i=1}^3 
\left( 1+J^z_{\Lambda,\bd{k}_i}\Pi_\Lambda(K_i) \right)
 \right]
G_\Lambda^{zzz} ( - K_1 , - K_2 , - K_3  ).
 \label{eq:GFzzz}
\end{eqnarray}
Note that the relation (\ref{eq:GFzzz}) between longitudinal three-point functions 
follows also directly from Eq.~(\ref{eq:FnGn}).
Substituting Eqs.~(\ref{eq:GFpmz}) and (\ref{eq:GFzzz})
into Eqs.\eqref{eq:F3general1} and \eqref{eq:F3general2}, we obtain the expansion of the three-point vertices in terms of the connected spin correlation functions,
\begin{align}
 \Gamma_\Lambda^{+-z} ( K_1 , K_2 , K_3  )  = & - G_\Lambda^{-1}  (-K_1 ) G_\Lambda^{-1} (K_2 ) 
\left[ 1+J^z_{\Lambda,\bd{k_3}}\Pi_\Lambda(K_3) \right] 
   G^{+-z}_{\Lambda} ( - K_1 , -K_2 , -K_3  ),
 \label{eq:G3general1}
 \\
 \Gamma_\Lambda^{zzz} ( K_1 , K_2 , K_3  )  = & - 
 \left[ \prod_{i=1}^3 \left( 1+J^z_{\Lambda,\bd{k}_i}\Pi_\Lambda(K_i) \right) \right]
   G_\Lambda^{zzz} ( - K_1 , -K_2 , -K_3  ).
 \label{eq:G3general2}
 \end{align}
Analogously,  using Eq.\eqref{eq:Treegamma4}
we find for the four-point vertices,
\begin{align}
 \Gamma_\Lambda^{++--} ( K_1 , K_2   ,  K_3 , K_4 )  = &
 - G_\Lambda^{-1}  (-K_1 ) G_\Lambda^{-1} (- K_2 ) 
 G_\Lambda^{-1}  (K_3 ) G_\Lambda^{-1} (K_4 ) G_\Lambda^{++--} ( - K_1 , - K_2 , -  K_3 , - K_4 ) 
 \nonumber
 \\
  & + \Bigl\{ \Gamma_\Lambda^{+-z} ( K_1 , K_3 , - K_1 - K_3 ) F ( - K_1 -  K_3 )  
 \Gamma_\Lambda^{+-z} ( K_2  , K_4 , - K_2  - K_4 )
   + ( K_3 \leftrightarrow K_4 ) \Bigr\} ,
   \label{eq:G4general1}
 \nonumber
 \\
  &
 \\
  \Gamma_\Lambda^{+-zz} ( K_1 ,  K_2 , K_3 , K_4 )  = & - G_\Lambda^{-1} (-K_1 )
  G_\Lambda^{-1} (  K_2 ) [ 1 + J^z_{\Lambda,\bd{k}_3} \Pi ( K_3 ) ]  [ 1 + J^z_{\Lambda,\bd{k}_4} \Pi ( K_4 ) ]
 G_\Lambda^{+-zz} ( - K_1 , - K_2 , - K_3 , - K_4 )
 \nonumber
 \\
  &   + \Bigl\{ \Gamma_\Lambda^{+-z} ( K_1,  - K_1 - K_3, K_3  ) 
 G_\Lambda (  - K_1 -  K_3)  \Gamma_\Lambda^{+-z} ( - K_2 -   K_4 , K_2,  K_4 )
 + ( K_3 \leftrightarrow K_4 ) \Bigr\} 
 \nonumber
 \\
  &   +  \Gamma_\Lambda^{zzz} ( K_3,   K_4, - K_3 - K_4  ) 
 F_\Lambda (  - K_3 -  K_4)  \Gamma_\Lambda^{+-z} ( K_1, K_2 , - K_1 - K_2 )  .
 \label{eq:G4general2}
 \\
 \Gamma_\Lambda^{zzzz} ( K_1 , K_2   ,  K_3 , K_4 )  = &
 -  \left[ \prod_{i=1}^4 
 \left(  1 + J^z_{\Lambda,\bd{k}_i} \Pi ( K_i )  \right) \right]
 G_\Lambda^{zzzz} ( - K_1 , - K_2 , -  K_3 , - K_4 ) 
 \nonumber
 \\
  & + \Bigl\{ \Gamma_\Lambda^{zzz} ( K_1 , K_3 , - K_1 - K_3 )   
  [ 1 + J^z_{\Lambda,-\bd{k}_1-\bd{k}_3} \Pi ( -K_1-K_3 ) ]
 \Gamma_\Lambda^{zzz} ( K_2  , K_4 , - K_2  - K_4 )\Bigr\} ,
  \label{eq:G4general3}
\end{align} 
where we have introduced  the notation $\left\{ f(.., K_3,  ..) +  (K_3 \leftrightarrow K_4) \right\} = \left\{  f(..,K_3,..)+  f(..,K_4,..) \right\}$ with an arbitrary function  $f$.
\end{widetext}

\section*{APPENDIX B: Generalized blocks and Wick theorem for spin operators}
\setcounter{equation}{0}
\renewcommand{\theequation}{B \arabic{equation}}
 \label{sec:tree}

For vanishing exchange interaction 
all spins  are completely decoupled so that all spin correlation functions are diagonal in the site index.
The on-site time-ordered connected spin correlation functions can
then be expanded in frequency space  as follows,
\begin{eqnarray}
 & & G^{\alpha_1 \ldots \alpha_n }_0 ( \tau_1 , \ldots, \tau_n )  \equiv 
 \langle {\cal{T}} [ S^{\alpha_1} ( \tau_1 ) \ldots S^{\alpha_n} ( \tau_n ) ] \rangle_{\rm connected}
 \nonumber
 \\
 &  & = \frac{1}{\beta^n } \sum_{ \omega_1 \ldots \omega_n } e^{ - i ( \omega_1 \tau_1 + \ldots
 + \omega_n \tau_n )} \tilde{G}^{\alpha_1 \ldots \alpha_n }_0 (  \omega_1 , \ldots ,  \omega_n ).
 \nonumber
 \\
 & &
 \label{eq:FT2}
 \end{eqnarray}
Translational invariance in imaginary time implies that
we can factor out a frequency-conserving $\delta$-function,
 \begin{eqnarray}
 & &   \tilde{G}^{\alpha_1 \ldots \alpha_n }_0 (  \omega_1 , \ldots ,  \omega_n )
 \nonumber
 \\
 & = & \delta ( \omega_1 + \cdots + \omega_n )
 {G}^{\alpha_1 \ldots \alpha_n }_0 (  \omega_1 , \ldots ,  \omega_n ),
 \end{eqnarray}
where we have introduced the discretized $\delta$-function in frequency space,
 \begin{equation}
\delta ( \omega ) = \beta \delta_{\omega,0}.
 \end{equation}
In the book by Izyumov and Skryabin \cite{Izyumov88}
time-ordered connected spin correlation functions
${G}^{\alpha_1 \ldots \alpha_n }_0 (  \omega_1 , \ldots ,  \omega_n )$ 
are called {\it{generalized blocks}}.
In the local limit the longitudinal correlations are purely static and
can be expressed in terms of the derivatives of the Brillouin function, 
 \begin{eqnarray}
  \tilde{G}^{zz}_0 (  \omega_1 , \omega_2  ) & = & 
 \delta ( \omega_1 ) \delta ( \omega_2 )    b^{\prime} ,
   \label{eq:AppBG0zz}
 \\
  \tilde{G}^{zzz}_0 (  \omega_1 ,  \omega_2 , \omega_3 ) & = & 
 \delta ( \omega_1  )  \delta (  \omega_2  ) \delta ( \omega_3 )   b^{\prime \prime} ,
 \\
 \tilde{G}_0^{zzzz} (  \omega_1 ,  \omega_2 , \omega_3 , \omega_4) & = & 
 \delta ( \omega_1  )  \delta (  \omega_2  ) \delta ( \omega_3 ) \delta ( \omega_4 )   b^{\prime \prime \prime}  .
 \hspace{7mm}
 \end{eqnarray}
The transverse two-spin correlation functions are
 \begin{eqnarray}
 \tilde{G}^{+-}_0 (  \omega , \omega^{\prime}   ) & = &  
 \delta ( \omega + \omega^{\prime} )
 G_0 ( \omega ) , 
  \label{eq:AppBG0}
 \\
 \tilde{G}^{-+}_0 (  \omega , \omega^{\prime}  ) & = &  
\delta ( \omega + \omega^{\prime} )
 G_0 ( - \omega ) ,
 \end{eqnarray}
where 
 \begin{equation}
 G_0 ( \omega ) = \frac{ b }{H - i \omega},
 \end{equation}
and the spin-$S$ Brillouin function $b = b ( \beta H)$ is  given
in Eq.~(\ref{eq:brillouin}).
The mixed three-spin correlation function is
  \begin{eqnarray}
 & &b  {G}^{+-z}_0 (  \omega_1 ,  \omega_2 ,  \omega_3 )  = 
 -   G_0 (  \omega_1 ) G_0 (    - \omega_2 )  
 +        G_0 (  \omega_1 ) \delta ( \omega_3 )  b^{\prime}  ,
 \nonumber
 \\
 & &
 \label{eq:G3loc}
 \end{eqnarray}
and the purely transverse connected four-spin correlation function is
 \begin{eqnarray}
  & & b^2   {G}^{++--}_0 (  \omega_1 ,  \omega_2 ,  \omega_3 ,  \omega_4)  =
  \nonumber
 \\
 &   & 
  -  G_0 (  \omega_1 ) G_0 (  \omega_2 ) 
 [ G_0 (  -  \omega_3 ) + G_0 (  -   \omega_4 ) ]
 \nonumber
 \\
 & & +  G_0 (  \omega_1 ) G_0
 (  \omega_2 ) [ \delta ( \omega_1 + \omega_3 )
 + \delta ( \omega_1 +   \omega_4 ) ]  b^{\prime}  .
 \hspace{7mm}
 \label{eq:G4loc1}
 \end{eqnarray}
There is also a mixed four-spin correlation function involving two transverse and
two longitudinal spin components,
 \begin{eqnarray}
& &  b^2  {G}^{+-zz}_0 (  \omega_1 ,   \omega_2 ,  \omega_3 ,  \omega_4)  =  
 \nonumber
 \\
& &    G_0 (  \omega_1 ) G_0 (   -  \omega_2 ) [ G_0 (  \omega_1 +   \omega_3 ) 
+ G_0 (  \omega_1 +   \omega_4 ) ]
 \nonumber
 \\
 & & -  G_0 (  \omega_1 ) G_0 (    - \omega_2 ) [ \delta ( \omega_3 ) + \delta ( \omega_4 ) ]  b^{\prime}
 \nonumber
 \\
 & & +  G_0  (  \omega_1 ) \delta ( \omega_3 ) \delta ( \omega_4 )  
b b^{\prime \prime} .
 \label{eq:G4loc2}
 \end{eqnarray}
 The above expressions for the connected spin correlation functions up to fourth order
have have first been derived by VLP\cite{Vaks68}, see also  Ref.[\onlinecite{Izyumov88}].
The  higher-order connected spin correlation functions can in principle be calculated diagrammatically using the generalized Wick theorem for spin operators derived by VLP \cite{Vaks68}. Since the diagrammatic approach is rather tedious,
we have developed an alternative method to generate the
higher-order connected spin correlation functions in frequency space
based on the hierarchy of equations of motion given in Eq.~(\ref{eq:eommaster}).
Noting that in the limit of vanishing exchange couplings 
only the first two terms on the right-hand side of Eq.~(\ref{eq:eommaster}) survive,
and denoting the resulting local limit of the
connected correlation function by
 \begin{widetext}
 \begin{eqnarray}
 &  & G^{(n,n,m)}_0 ( \omega_1 \ldots \omega_n; \omega_1^{\prime} \ldots
 \omega_n^{\prime} ; \omega_1^{\prime \prime} \ldots \omega_m^{\prime \prime} )
 =    G_0^{ \overbrace{+ \cdots +}^{n}   \overbrace{- \cdots -}^n 
\overbrace{z \cdots z}^m }
 ( \omega_1 \ldots \omega_n; \omega_1^{\prime} \ldots
 \omega_n^{\prime} ; \omega_1^{\prime \prime} \ldots \omega_m^{\prime \prime} )
 \end{eqnarray}
we obtain from Eq.~(\ref{eq:eommaster}) the following expression relating
the correlation function involving
$2n$ transverse and $m$ longitudinal spin components
to a linear combination of connected correlation functions involving  at 
most $2n+m -1$ spins,
 \begin{eqnarray}
 b G_0^{(n,n,m)} ( \omega_1 \ldots \omega_n; 
 \omega_1^{\prime} \ldots \omega_n^{\prime} ; \omega_1^{\prime \prime} \ldots \omega_m^{\prime \prime} )
 & = & - G_0 ( \omega_1 ) \sum_{ \nu =1}^m G_0^{(n,n,m-1)} ( \omega_1 
+ \omega_{\nu}^{\prime \prime} , \omega_2 \ldots \omega_n; 
 \omega_1^{\prime} \ldots \omega_n^{\prime} ; 
 \omega_1^{\prime \prime} \ldots 
 {\slashed{\omega}}_{\nu}^{\prime \prime}
 \ldots   \omega_m^{\prime \prime} )
 \nonumber
 \\
 &  & \hspace{-40mm} + G_0 ( \omega_1 ) \sum_{ \nu =1}^n
 G_0^{(n-1,n-1, m+1)} ( \omega_2  \ldots \omega_n ; \omega_1^{\prime}   
 \ldots \slashed{\omega}_{\nu}^{\prime} \ldots \omega_n^\prime ; 
 \omega_1^{\prime \prime} \ldots \omega_m^{\prime \prime}    ,   
 \omega_1 + \omega_{\nu}^{\prime}  ),
 \label{eq:recursion}
 \end{eqnarray}
\end{widetext} 
where the slashed symbol ${\slashed{\omega}}_{\nu}^{\prime \prime}$  in the list
$\omega_1^{\prime \prime} \ldots 
 {\slashed{\omega}}_{\nu}^{\prime \prime}
 \ldots   \omega_m^{\prime \prime} $ means that
${\omega}_{\nu}^{\prime \prime}$ should be omitted.
The recursion relation (\ref{eq:recursion}) can be viewed as an algebraic form of the
generalized Wick theorem for spin operators.
By iterating this relation, we can express 
$G_0^{(n,n,m)} ( \omega_1 \ldots \omega_n; 
 \omega_1^{\prime} \ldots \omega_n^{\prime} ; \omega_1^{\prime \prime} \ldots \omega_m^{\prime \prime} )$ as a linear combination of terms involving products of the transverse propagators $G_0 ( \omega )$ and
purely longitudinal blocks 
 \begin{equation}
 G_0^{  \overbrace{z \cdots z}^k } ( \omega_1 \ldots \omega_k ) = \left( \prod_{ i=1}^{k-1} \delta ( \omega_i ) 
 \right) b^{(k-1)}
 \end{equation}
up to $k = n + m$, where $b^{(k)}$ denotes the $k$-th derivative of the Brillouin function.

The corresponding irreducible vertices can be obtained using the
relations between irreducible vertices and connected spin correlation functions
derived in Appendix~A, see Eqs.~(\ref{eq:G3general1}--\ref{eq:G4general3}).
We find that the longitudinal vertices 
are simply given by the negative of the corresponding generalized blocks,
 \begin{eqnarray}
 \Gamma^{zz}_0 ( \omega ) & = & - \delta ( \omega ) b^{\prime},
 \\
  \Gamma^{zzz}_0 ( \omega_1 , \omega_2 , \omega_3 ) & = &  - \delta ( \omega_1  ) 
 \delta ( \omega_2 ) b^{\prime \prime},
 \label{eq:blockzzz}
 \\
 \Gamma^{zzzz}_0 ( \omega_1 , \omega_2 , \omega_3 , \omega_4) & = &  
 - \delta ( \omega_1  ) 
 \delta ( \omega_2 ) \delta ( \omega_3 ) b^{\prime \prime \prime}.
 \hspace{7mm}
 \end{eqnarray}
The transverse two-point vertex is 
 \begin{equation}
 \Gamma_0^{+-} ( \omega ) = G_0^{-1} ( \omega )=\frac{ H - i \omega }{b} ,
 \end{equation}
and the  mixed three-point vertex is related to the mixed three-spin correlation function via
 \begin{eqnarray}
  & &  \Gamma_0^{+-z} ( \omega_1 , \omega_2 , \omega_3 ) =
 \nonumber
 \\
  &  &   - G_0^{-1} ( - \omega_1 ) G_0^{-1} ( \omega_2 ) G_0^{+-z} ( - \omega_1 , - \omega_2 , - \omega_3 ), \hspace{7mm}
 \label{eq:Gammapmzlocal}
 \end{eqnarray}
 which gives
 \begin{equation}
b  \Gamma_0^{+-z} ( \omega_1 , \omega_2 , \omega_3 )
 =  1 -     G^{-1}_0 ( \omega_2 )  \delta ( \omega_3 ) b^{\prime}    .
 \label{eq:block3}
 \end{equation}
The transverse and the mixed four-point vertices are given by
 \begin{widetext}
 \begin{eqnarray}
 \Gamma_0^{++--} ( \omega_1 , \omega_2 , \omega_3 , \omega_4 )  &  = &
  - G_0^{-1}( - \omega_1 ) G_0^{-1} ( - \omega_2 )  
G_0^{-1} ( \omega_3 )
  G_0^{- 1} ( \omega_4 )
 G^{++--}_0 ( - \omega_1 , - \omega_2 , - \omega_3 , - \omega_4 ),
 \\
 & & 
 \nonumber
 \\
 \Gamma_0^{+-zz} ( \omega_1 , \omega_2 , \omega_3 , \omega_4 )  &  = &
  - G_0^{-1}( - \omega_1 ) G_0^{-1} ( - \omega_2 )
   G^{+-zz}_0 ( - \omega_1 , - \omega_2 , - \omega_3 , - \omega_4 )
 \nonumber
 \\
 & + &  \left\{ \Gamma_0^{+-z} ( \omega_1 , - \omega_1 - \omega_3,  \omega_3 )
 G_0 ( - \omega_1 - \omega_3 )
 \Gamma_0^{+-z} (  - \omega_2 - \omega_4,  \omega_2 , \omega_4 )
 + ( \omega_3 \leftrightarrow \omega_4 ) \right\},
 \end{eqnarray}
 \end{widetext}
which gives
 \begin{eqnarray}
 & & b^2 \Gamma_0^{++--} ( \omega_1 , \omega_2 , \omega_3 , \omega_4 )  = 
  G_0^{-1} ( \omega_3 ) + G_0^{-1} ( \omega_4 ) 
 \nonumber
 \\
 &  &-  [ \delta ( \omega_1 + \omega_3 ) + \delta ( \omega_1 + \omega_4 )]  b^{\prime}  
 G_0^{-1} ( \omega_3 ) G_0^{-1} ( \omega_4 ) ,
 \hspace{7mm}
 \nonumber
 \\
 & &
 \label{eq:Gammappmmlocal}
 \end{eqnarray}
and
 \begin{eqnarray}
 & & b^2 \Gamma^{+-zz}_0 ( \omega_1 , \omega_2 , \omega_3 , \omega_4 )
  =   -  [ \delta ( \omega_3 )   +  \delta ( \omega_4 ) ]  b^{\prime}
 \nonumber
 \\
 & & \hspace{10mm} + 
    \delta ( \omega_3 ) \delta ( \omega_4 )   G_0^{-1} ( \omega_2 )  [ 2 (b^{\prime} )^2 - b b^{\prime 
 \prime} ]    .
 \label{eq:Gammapmzzlocal}
 \end{eqnarray}
In a cutoff scheme where initially 
the exchange interaction is completely switched off, the above 
expressions for the vertices define the initial condition
for the SFRG flow equations.

\section*{APPENDIX C: Ward identity and hierarchy of equations of motion}
\setcounter{equation}{0}
\renewcommand{\theequation}{C\arabic{equation}}
 \label{sec:ward}

In the regime where the spin-rotational invariance is spontaneously broken,
Goldstone's theorem guarantees that the energy dispersion of the spin-waves is gapless
for vanishing external field $ H \rightarrow 0$.
To construct a truncation of the SFRG flow equations which does not violate this, 
it is crucial to  take into account an exact Ward identity forcing the vanishing of the spin-wave gap
when $M$ remains finite  for $H \rightarrow 0$.
To derive this  Ward identity, consider the imaginary time Heisenberg equation of motion 
for the spin operator $\bd{S}_i ( \tau )$ with deformed Hamiltonian 
${\cal{H}}_{\Lambda} = {\cal{H}}_0 + {\cal{V}}_{\Lambda}$
given by Eqs.~(\ref{eq:Hzdef}) and (\ref{eq:Vlambdadef}) under the influence of an additional fluctuating source field $\bd{h}_i ( \tau )$. Writing for simplicity
$J^\bot_{ij} = J^{\bot}_{ \Lambda, ij} $ and
$J^z_{ij} = J^z_{ \Lambda, ij}$, and choosing
our coordinate system such that
the uniform magnetic field $\bd{H} = H \bd{e}_z$ points in  $z$-direction $\bd{e}_z$,
the Heisenberg equation of motion for the spin operators 
$\bd{S}_i =    S^z_i \bd{e}_z  + \bd{S}^{\bot}_i    =      \bd{S}^\parallel_i  + \bd{S}^{\bot}_i$
in imaginary time can be written as 
 \begin{eqnarray}
 i \partial_{\tau} \bd{S}_i ( \tau )  =  \bd{S}_i ( \tau ) & \times &  \biggl[
 \bd{H} + \bd{h}_i ( \tau ) 
 \nonumber
 \\
 &  &    - \sum_j \left( J^z_{ij} \bd{S}^\parallel_j 
 + J^{\bot}_{ij}
\bd{S}^{\bot}_j ( \tau ) \right) \biggr].
 \hspace{7mm}
 \label{eq:eom}
 \end{eqnarray}
Taking the source-dependent average  of both sides of Eq.~(\ref{eq:eom}) we obtain
 \begin{widetext}
 \begin{eqnarray}
 i \partial_{\tau} \langle \bd{S}_i ( \tau ) \rangle_{ \bd{h} } & = & 
  \langle \bd{S}_i ( \tau ) \rangle_{\bd{h}}  \times \biggl[
 \bd{H} + \bd{h}_i ( \tau ) 
 -  \sum_j \left( J^z_{ij}   \langle \bd{S}^{\parallel}_j ( \tau ) \rangle_{\bd{h}} 
  + J^{\bot}_{ij}   \langle \bd{S}^{\bot}_j ( \tau ) \rangle_{\bd{h}} 
\right) \biggr]
 \nonumber
  \\
 &- &  \sum_j \left( J^{z}_{ij} 
   \langle \delta \bd{S}_i ( \tau ) \times \delta \bd{S}^{\parallel}_j ( \tau ) \rangle_{\bd{h}}
   + J^{\bot}_{ij} 
   \langle \delta \bd{S}_i ( \tau ) \times \delta \bd{S}^{\bot}_j ( \tau ) \rangle_{\bd{h}}
 \right) ,
 \label{eq:eomaverage}
 \end{eqnarray}
where $\delta \bd{S}_i ( \tau )  = \bd{S}_i ( \tau ) - \langle \bd{S}_i ( \tau ) \rangle_{\bd{h}}$ and the
average symbol is defined as follows,
 \begin{equation}
 \langle \ldots \rangle_{\bd{h}} =
 \frac{ {\rm Tr} \left[ e^{ - \beta {\cal{H}}_0 }
 {\cal{T}} e^{ \int_0^{\beta} d \tau [ \sum_i \bd{h}_i ( \tau )  \cdot {\bd{S}}_i ( \tau ) - {\cal{V}}_{\Lambda} ( \tau ) ] } \ldots \right] }{
{\rm Tr} \left[ e^{ - \beta {\cal{H}}_0 }
 {\cal{T}} e^{ \int_0^{\beta} d \tau [ \sum_i \bd{h}_i ( \tau )  \cdot {\bd{S}}_i ( \tau ) - {\cal{V}}_{\Lambda} ( \tau ) ] }  \right]}.
 \end{equation}
By definition, the averages in Eq.~(\ref{eq:eomaverage}) can be expressed in terms of the derivatives of the
generating functional ${\cal{G}}_{\Lambda} [ \bd{h} ]$ of the connected
imaginary-time spin correlation functions defined in Eq.~(\ref{eq:Gcdef}).
It is convenient to express the part of the source field $\bd{h}_i ( \tau )$ which is perpendicular
to the external magnetic field in terms of the spherical components
 \begin{equation}
 {h}^{\pm}_i ( \tau ) = \frac{1}{\sqrt{2}} \left[ h^x_i ( \tau ) \pm i h^y_i ( \tau ) \right].
 \end{equation}
Then Eq.~(\ref{eq:eomaverage}) reduces to the following three identities:
 \begin{subequations} 
 \begin{eqnarray}
   \partial_{\tau} \left[
   \frac{ \delta {\cal{G}}_{\Lambda} }{\delta h_i^{z} ( \tau ) } \right]  & = &  
    \frac{ \delta {\cal{G}}_{\Lambda} }{\delta h_i^{-} ( \tau ) }  h_i^- ( \tau ) -   
 \frac{ \delta {\cal{G}}_{\Lambda} }{\delta h_i^{+} ( \tau ) }  h_i^+ ( \tau )
 \nonumber
 \\
 & - &    \sum_j J^{\bot}_{ij} \Biggl[
  \frac{ \delta {\cal{G}}_{\Lambda} }{\delta h_i^{-} ( \tau ) }
 \frac{ \delta {\cal{G}}_{\Lambda} }{\delta h_j^{+} ( \tau ) } 
 -    \frac{ \delta {\cal{G}}_{\Lambda} }{\delta h_i^{+} ( \tau ) }
 \frac{ \delta {\cal{G}}_{\Lambda} }{\delta h_j^{-} ( \tau ) } 
+  \frac{ \delta^2 {\cal{G}}_{\Lambda} }{\delta h_i^{-} ( \tau )   \delta h_j^{+} ( \tau ) }
-  \frac{ \delta^2 {\cal{G}}_{\Lambda} }{\delta h_i^{+} ( \tau )   \delta h_j^{-} ( \tau ) }
 \Biggr],
 \label{eq:Ward1}
 \\
  \partial_{\tau} \left[
   \frac{ \delta {\cal{G}}_{\Lambda} }{\delta h_i^{+} ( \tau ) } \right]  &  = & 
    \frac{ \delta {\cal{G}}_{\Lambda} }{\delta h_i^{+} ( \tau ) }  [ H + h_i^z ( \tau ) ] -   
 \frac{ \delta {\cal{G}}_{\Lambda} }{\delta h_i^{z} ( \tau ) }  h_i^- ( \tau )
 \nonumber
 \\
 & -  &    \sum_j \Biggl[J^z_{ij}
  \frac{ \delta {\cal{G}}_{\Lambda} }{\delta h_i^{+} ( \tau ) }
 \frac{ \delta {\cal{G}}_{\Lambda} }{\delta h_j^{z} ( \tau ) } 
 -  J^{\bot}_{ij}  \frac{ \delta {\cal{G}}_{\Lambda} }{\delta h_i^{z} ( \tau ) }
 \frac{ \delta {\cal{G}}_{\Lambda} }{\delta h_j^{+} ( \tau ) } 
+  J^z_{ij}\frac{ \delta^2 {\cal{G}}_{\Lambda} }{\delta h_i^{+} ( \tau )   \delta h_j^{z} ( \tau ) }
-  J^{\bot}_{ij} \frac{ \delta^2 {\cal{G}}_{\Lambda} }{\delta h_i^{z} ( \tau )   \delta h_j^{+} ( \tau ) }
 \Biggr],
 \\
 \partial_{\tau} \left[
   \frac{ \delta {\cal{G}}_{\Lambda} }{\delta h_i^{-} ( \tau ) } \right]  & = & 
  -   \frac{ \delta {\cal{G}}_{\Lambda} }{\delta h_i^{-} ( \tau ) }  [ H + h_i^z ( \tau ) ] +
 \frac{ \delta {\cal{G}}_{\Lambda} }{\delta h_i^{z} ( \tau ) }  h_i^+ ( \tau )
 \nonumber
 \\
 & + &  \sum_j \Biggl[
  J^z_{ij}  \frac{ \delta {\cal{G}}_{\Lambda} }{\delta h_i^{-} ( \tau ) }
 \frac{ \delta {\cal{G}}_{\Lambda} }{\delta h_j^{z} ( \tau ) } 
 -    J^{\bot}_{ij} \frac{ \delta {\cal{G}}_{\Lambda} }{\delta h_i^{z} ( \tau ) }
 \frac{ \delta {\cal{G}}_{\Lambda} }{\delta h_j^{-} ( \tau ) } 
+ J^{z}_{ij} \frac{ \delta^2 {\cal{G}}_{\Lambda} }{\delta h_i^{-} ( \tau )   \delta h_j^{z} ( \tau ) }
-  J^{\bot}_{ij} \frac{ \delta^2 {\cal{G}}_{\Lambda} }{\delta h_i^{z} ( \tau )   \delta h_j^{-} ( \tau ) }
 \Biggr].
 \label{eq:Ward3}
 \end{eqnarray}
\end{subequations}
\end{widetext}
To derive the Ward identity (\ref{eq:WI1}), we
take  another functional derivative $\frac{\delta }{ \delta h^+_n ( \tau^{\prime} )}$
of Eq.~(\ref{eq:Ward3}) and then set all sources equal to zero.
For an isotropic ferromagnet with $J^{z}_{ij} = J^{\bot}_{ij} = - V_{ij}$
the resulting
equation of motion for the transverse two-spin correlation function
in real space and imaginary time can be written as
 \begin{eqnarray}
 \partial_{\tau} G^{+-}_{ i n} ( \tau - \tau^{\prime} )
 & = & - H G^{+-}_{ i n} ( \tau - \tau^{\prime} ) + \delta_{in} \delta ( \tau -
 \tau^{\prime} ) M
 \nonumber
  \\
 &  & \hspace{-28mm} + \sum_j V_{ij} \Bigl[  G^{+-}_{ in} ( \tau - \tau^{\prime} ) M
 -   G^{+-}_{ jn} ( \tau - \tau^{\prime} )    M
 \nonumber
 \\
 &  & \hspace{-15mm} +  
  G^{+- z }_{ inj} ( \tau , \tau^{\prime} , \tau )
 -   G^{+-z}_{ jni} ( \tau , \tau^{\prime} , \tau  )   \Bigr],
 \end{eqnarray}
where $M = \langle S^z_i ( \tau ) \rangle$
is the local moment  for the 
system with deformed Hamiltonian ${\cal{H}}_{\Lambda}$ in the absence of sources, i.e. for  $\bd{h}_i ( \tau ) =0$.
Summing both sides of this equation over the site label $i$ and integrating
$\int_0^{\beta} d \tau $
we see that the last term on the right-hand side vanishes due to the antisymmetry of the terms in the square braces
with respect to $i \leftrightarrow j$, while the left-hand side vanishes due to the
periodicity of the imaginary time spin correlation functions.
Noting that the uniform transverse susceptibility is given by
 \begin{equation}
 \chi_{\bot} = G^{+-} (K=0)  =  \int_0^{\beta} d \tau \sum_i G^{+-}_{ in} ( \tau -
 \tau^{\prime} ),
 \end{equation}
we finally arrive at the Ward identity \cite{Patashinskii73} 
$\chi_{\bot}= M/H$, see  Eq.~(\ref{eq:WI1}) of the main 
text.

By successively taking higher-order derivatives of the functional relations 
(\ref{eq:Ward1})--(\ref{eq:Ward3}), we can derive equations of motion  for all
higher-order connected spin correlation functions. For example,
starting from the third  relation (\ref{eq:Ward3}) we can obtain the equations of motion for the
connected spin correlation functions
 \begin{widetext}
 \begin{equation} 
   G^{(n,n,m)} ( X_1 , \cdots , X_n ; X_1^{\prime} , \ldots , X_n^{\prime} ;
 X_1^{\prime \prime} , \ldots , X_m^{\prime \prime} )  =
 G^{ \overbrace{+ \cdots +}^{n}   \overbrace{- \cdots -}^n 
\overbrace{z \cdots z}^m }
 (    X_1 , \cdots , X_n ; X_1^{\prime} , \ldots , X_n^{\prime} ;
 X_1^{\prime \prime} , \ldots , X_m^{\prime \prime} )    
 \label{eq:Gnnmdef}
 \end{equation} 
involving  $2n \geq 2$ transverse and $m$ longitudinal spin components
by taking $n$ derivatives with respect to the $h^+$-sources, 
$n-1$ derivatives with respect to
the $h^-$-sources, and $m$ derivatives with respect to 
the $h^z$-sources. In Eq.~(\ref{eq:Gnnmdef}) the symbols 
$X_i = ( {\bd{r}}_i , \tau_i ) $  are collective labels 
representing the lattice sites ${\bd{r}}_i$ and the imaginary time $\tau_i$.
Transforming all objects to momentum-frequency space,
 \begin{eqnarray}
  & & G^{(n,n,m)} ( X_1 , \cdots , X_n ; X_1^{\prime} , \ldots , X_n^{\prime} ;
 X_1^{\prime \prime} , \ldots , X_m^{\prime \prime} )    =  
 \int_{ K_1 } \ldots \int_{ K_n }
 \int_{ K_1^{\prime} } \ldots \int_{ K_n^{\prime} }
 \int_{ K_1^{\prime \prime} } \ldots \int_{ K_m^{\prime \prime} }
 e^{ i      \sum\nolimits_{i=1}^n ( K_i X_i + K_i^{\prime} X_i^{\prime}) + 
  i \sum\nolimits_{ i=1}^m K_i^{\prime \prime} X_i^{\prime \prime}  }
 \nonumber
 \\
 & & \times \delta \left( \sum\nolimits_{i=1}^n ( K_i + K_i^{\prime} ) + 
 \sum\nolimits_{ i=1}^m K_i^{\prime \prime} \right)
 G^{(n,n,m)} ( K_1 , \cdots , K_n ; K_1^{\prime} , \ldots , K_n^{\prime} ;
 K_1^{\prime \prime} , \ldots , K_m^{\prime \prime} ) ,
\end{eqnarray}
where $K_i X_i = \bd{k}_i \cdot \bd{r}_i - \omega_i \tau_i $,
 we obtain the following infinite hierarchy of equations,
 \begin{eqnarray}
 & & ( H - i \omega_1 )  G^{(n,n,m)} ( K_1, K_2 \ldots K_{n}; K_1^{\prime} \ldots K_n^{\prime} ; K_1^{\prime \prime} \ldots K_m^{\prime \prime} ) =
 \nonumber
 \\
 &  & -  \sum_{ \nu =1}^m G^{(n,n,m-1)} ( K_1 
+ K_{\nu}^{\prime \prime} , K_2 \ldots K_{n}; 
 K_1^{\prime} \ldots K_n^{\prime} ; 
 K_1^{\prime \prime} \ldots 
 {\slashed{K}}_{\nu}^{\prime \prime}
 \ldots   K_m^{\prime \prime} )
 \nonumber
 \\
 &  & + \sum_{ \nu =1}^n
 G^{(n-1,n-1, m+1)} ( K_2  \ldots K_n ; K_1^{\prime}   
 \ldots \slashed{K}_{\nu}^{\prime} \ldots K_n^\prime ; 
 K_1^{\prime \prime} \ldots K_m^{\prime \prime}    ,   
 K_1 + K_{\nu}^{\prime}  )
 \nonumber
 \\
 & & + \sum_{\nu =0}^{n-1} \sum_{ \mu =0}^m 
{\cal{S}}^{(+)}_{ K_2 \ldots K_{\nu+1} ; K_{ \nu+2} \ldots K_{n} }
{\cal{S}}^{(-)}_{ K_1^{\prime} \ldots K_{\nu}^{\prime} ; K^{\prime}_{ \nu+1} \ldots K^{\prime}_{n} }
{\cal{S}}^{(z)}_{ K_1^{\prime \prime} \ldots K_{\mu}^{\prime \prime} ; 
 K^{\prime \prime }_{ \mu+1} \ldots K^{\prime \prime}_{m} }
 \Biggl\{
 \nonumber
 \\
 & &   \hspace{14mm} 
 \biggl[ J^z \Bigl( \sum\nolimits_{ i =1}^\nu (\bd{k}_{i+1} + \bd{k}^{\prime}_i ) + 
  \sum\nolimits_{i =1}^\mu   \bd{k}_i^{\prime \prime} \Bigr)     
 - J^\bot \Bigl( \bd{k}_1 + \sum\nolimits_{ i =1}^\nu (\bd{k}_{i+1} + \bd{k}^{\prime}_i ) + 
 \sum\nolimits_{i =1}^\mu \bd{k}_i^{\prime \prime} \Bigr)     
\biggr]
 \nonumber
 \\
 & &  \hspace{12mm} \times
 G^{(\nu , \nu , \mu +1)} \Bigl( K_2 \ldots K_{\nu +1}; K_1^{\prime} \ldots K_{\nu}^{\prime};  K_1^{\prime \prime}  \ldots K_{\mu}^{\prime \prime} , - \sum\nolimits_{ i=1}^{\nu} ( K_{i+1} + K^{\prime}_i ) -
 \sum\nolimits_{ i=1}^{\mu} K^{\prime \prime}_{i} \Bigr)
 \nonumber
 \\
 & & \hspace{12mm} \times
G^{(n- \nu , n- \nu , m - \mu )} \Bigl(  K_1  
 +  \sum\nolimits_{ i=1}^{\nu} ( K_{i+1} + K^{\prime}_i ) +
 \sum\nolimits_{ i=1}^{\mu} K^{\prime \prime}_{i} ,
 K_{ \nu +2} \ldots K_{n}  ; 
K^{\prime}_{ \nu +1} \ldots K^{\prime}_n  ;
 K^{\prime \prime}_{\mu +1 } \ldots K^{\prime \prime}_m \Bigr)
\Biggl\}
 \nonumber
 \\
 &  &  + \int_Q [ J^z ( \bd{q} ) - J^{\bot} (  \bd{k}_1 + \bd{q} ) ]
  G^{ ( n,n ,  m+1)} \bigl( K_1 +  Q, K_2  \ldots K_n;  K_1^{\prime} \ldots K_n^{\prime} ;
 K_1^{\prime \prime} \ldots K_m^{\prime \prime} ,  - Q  \bigr),
 \label{eq:eommaster}
 \end{eqnarray}
\end{widetext} 
where the slashed symbol ${\slashed{K}}_{\nu}^{\prime \prime}$  in the list
$K_1^{\prime \prime} \ldots 
 {\slashed{K}}_{\nu}^{\prime \prime}
 \ldots   K_m^{\prime \prime} $ means that
${K}_{\nu}^{\prime \prime}$ should be deleted from the list, and the operators 
${\cal{S}}^{( \alpha )}$ (where $\alpha =+,-,z$) symmetrize all expressions with respect to permutations of the
$K$-labels belonging to the same spin component, see Ref.~[\onlinecite{Kopietz10}] for an 
explicit definition of these operators.
To obtain the equations of motion of  purely longitudinal correlation functions, we take 
functional derivatives of the
first equation (\ref{eq:Ward1}) with respect to the longitudinal sources and then set all sources
equal to zero. Note that only the terms involving the second functional derivative
in the second line of Eq.~(\ref{eq:Ward1}) survive in this limit, which in Fourier space
involve a loop integral.
If we deform our model such that the loop integrations are small (this is the case
in the presence of a strong external magnetic field, or if the exchange interaction is long-ranged),
then all terms involving loops can be dropped. We call this the tree approximation.
 In this limit we can solve the simplified hierarchy
of flow equations recursively, because  the right-hand side of the hierarchy (\ref{eq:eommaster})
involves correlation functions of the same order as the left-hand side with the same arguments or 
of lower order. In Appendix~D we explicitly give the tree approximation for 
the vertices with up to four external legs.

\section*{APPENDIX D: Tree approximation for correlation functions and vertices}
\setcounter{equation}{0}
\renewcommand{\theequation}{D \arabic{equation}}
 \label{sec:tree}

For the approximate  solution of the SFRG flow equations 
it is convenient to  include the exchange couplings between different spins
at the initial scale within the tree approximation, which graphically amounts to neglecting
all diagrams involving closed loops.
In the spirit of VLP \cite{Vaks68}, we therefore assume that
the exchange interaction is 
long-ranged so that its Fourier transform is dominated by
momenta $ | \bd{k} | \lesssim  k_0 \ll 1/a$, where $a$ is the lattice spacing. 
Loop integrations over momenta are then suppressed by powers of $ k_0 a$, so that
perturbation theory in powers of loops is controlled by the
small parameter $k_0 a$. To leading order, we may simply 
neglect all loops (tree approximation).
As already pointed out  at the end of Appendix C, 
in this limit the infinite hierarchy of equations of motion decouples. 
To simplify our notation, let us  rename here 
the Fourier transform of the exchange couplings as follows,
 \begin{equation}
 J^{z} ( \bd{k} ) = - V^z_{\bd{k}}, \; \; \; \; 
J^{\bot} ( \bd{k} ) = - V^{\bot}_{\bd{k} }.
 \end{equation}
The tree approximation for the transverse propagator is
 \begin{equation}
 G_0 ( K ) = G_0 ( \bd{k} , i \omega  ) = \frac{M_0}{ H + E_{\bd{k}} - i \omega },
 \label{eq:appcg0}
 \end{equation}
where the magnetization $M_0$ in self-consistent mean-field approximation
is defined in Eq.~(\ref{eq:M0def}), and $E_{\bd{k}} = M_0 ( V^z_0 - V^{\bot}_{\bd{k}})$ is the magnon dispersion.
The longitudinal effective interaction
is in tree approximation given by
 \begin{equation}
 F_0 ( K ) = F_0 ( \bd{k} , i \omega ) = \frac{ V^z_{\bd{k}}}{ 1 -
 \delta ( \omega ) b^{\prime}  V^z_{\bd{k}}     },
 \end{equation}
while the tree approximation for 
the mixed three-spin correlation function can be written as
\begin{widetext}
 \begin{eqnarray}
  & & M_0 G^{+-z}_{\rm tree} ( K_1 , K_2 , K_3 )  = 
 - G_0 (K_1 ) G_0 ( - K_2 )
 \left\{ 1 -  \bigl[     G^{-1}_0 ( -K_2 )  +  V^\bot_{\bd{k}_2 } - V^z_{\bd{k}_3 } 
 \bigr]
 \frac{    \delta ( \omega_3 )  b^{\prime}}{ 1 - \beta b^{\prime} 
 V^z_{\bd{k}_3}}  \right\}  .
 \label{eq:G3tree}
 \end{eqnarray}
Moreover, from the equation of motion (\ref{eq:eommaster})
we obtain for
transverse four-spin correlation function in tree approximation
 \begin{eqnarray}
 & & M_0^2 G^{++--}_{\rm tree} ( K_1 , K_2 , K_3 , K_4 ) =
  -  G_0 ( K_1) G_0 ( K_2 )  G_0 ( - K_3 ) G_0 ( - K_4 )
 \Biggl\{
G^{-1}_0 (- K_3 )  + V^{\bot}_{\bd{k}_3} - V^z_{\bd{k}_1 + \bd{k}_3 }
 \nonumber
 \\
 &  & -  \frac{ \delta ( \omega_1 + \omega_3 ) b^{\prime}}{ 1 -
 \beta V^z_{\bd{k}_1 + \bd{k}_3 } b^{\prime} }
\left[  G^{-1}_0 ( -K_3)  +  V^{\bot}_{\bd{k}_3 } -  V^z_{\bd{k}_1 + \bd{k}_3 }  \right]
\left[  G^{-1}_0 ( -K_4)  +  V^\bot_{\bd{k}_4 } - V^z_{\bd{k}_1 + \bd{k}_3 }  \right]
 + ( K_3 \leftrightarrow K_4 ) \Biggr\}.
 \label{eq:G4tree1}
 \end{eqnarray}
 Finally, the tree approximation for the mixed four-spin correlation function can be written as
 \begin{eqnarray}
& & M_0^2 G^{+-zz}_{\rm tree} ( K_1 , K_2 , K_3 , K_4 )  = 
 \frac{
G_0 ( K_1 ) G_0 ( - K_2 )}{ 
 [ 1- \delta ( \omega_3 ) b^{\prime} V^z_{\bd{k}_3 } ]
 [ 1- \delta ( \omega_4 ) b^{\prime} V^z_{\bd{k}_4 } ]}
\Biggl\{   
 \frac{ G_0^{-1} ( - K_2 ) + V^{\bot}_{\bd{k}_2} - V^z_{ \bd{k}_3 + \bd{k}_4 } }{ 
 1 - \beta b^{\prime} V^z_{ \bd{k}_3 +\bd{k}_4 } }   \delta ( \omega_3 ) 
 \delta ( \omega_4 ) M_0 b^{\prime \prime}
 \nonumber
 \\
 &   & \hspace{20mm}  +
 \biggl\{ 
G_0 ( K_1 + K_3 )  \Bigl[ 1 + \delta ( \omega_3 ) b^{\prime} V^{\bot}_{\bd{k}_1 + \bd{k}_3 }
  \Bigr] \Bigl[ 1 - \delta ( \omega_4 ) b^{\prime} \bigl(  G_0^{-1} ( - K_2 )  + V^{\bot}_{\bd{k}_2} \bigr) \Bigr]
 + ( K_3 \leftrightarrow K_4 )  \biggr\}
\Biggr\}.
 \label{eq:G4tree2}
 \end{eqnarray} 
Note that for vanishing exchange couplings 
Eqs.~(\ref{eq:G3tree})--(\ref{eq:G4tree2}) reduce to the corresponding 
generalized blocks given in
Eqs.~(\ref{eq:G3loc})--(\ref{eq:G4loc2}).

Given the connected spin correlation functions, we can construct the
corresponding irreducible vertices generated by
our hybrid functional $\tilde{\Gamma}_{\Lambda} [ \bd{m} , \varphi]$
defined via Eqs.~(\ref{eq:GammaHMdef}) and (\ref{eq:Gammashift})
using the general relations (\ref{eq:G3general1})--(\ref{eq:G4general3}) derived in Appendix~A.
The two-point vertices $\Gamma^{+-}_{\rm tree } ( K )$ and $\Gamma^{zz}_{\rm tree} ( K )$ in
tree approximation are given in 
Eqs.~(\ref{eq:Gammapmtree}) and (\ref{eq:Gammazztree}) of the main text.
Substituting  the tree approximation $G^{+-z}_{\rm tree} ( K_1 , K_2 , K_3 )$
given in Eq.~(\ref{eq:G3tree})  for the mixed three-spin correlation function 
on the right-hand side of Eq.~(\ref{eq:G3general1}),   we obtain
 \begin{eqnarray}
 M_0 \Gamma_{\rm tree}^{+-z} ( K_1  , K_2 , K_3  ) = 
  1  - \left[  G_0^{-1} ( K_2 ) +  V^{\bot}_{ {\bd{k}}_2 }   \right]  \delta (   \omega_3  )  b^{\prime}  
 =  1  - {G}_1^{-1} ( \omega_2 )    \delta (   \omega_3  )  b^{\prime}  ,
 \label{eq:threetree}
 \end{eqnarray}
where $G_1 ( \omega )$ is defined by
 \begin{equation}
 G_0^{-1} ( K ) + V^{\bot}_{\bd{k}}  = 
\frac{ H + M_0  ( V_0^z - V_{\bd{k}}^\bot )  - i \omega}{M_0} + V^{\bot}_{\bd{k}} 
 = 
\frac{ H + M_0  V_0^z - i \omega}{M_0}  \equiv {G}^{-1}_1 ( \omega ).
 \end{equation}
Note that with the substitutions $b \rightarrow M_0$ and  $G_0^{-1} ( \omega_2 ) 
 \rightarrow G_1^{-1} ( \omega_2 )$ the mixed three-legged vertex in tree approximation
can be obtained from the corresponding irreducible vertex 
for vanishing exchange interaction
given in Eq.~(\ref{eq:block3}).
Next, consider the transverse four-point vertex in tree approximation, which according to
Eqs.~(\ref{eq:G4general1})  and (\ref{eq:G4tree1}) is given by
 \begin{eqnarray}
  M_0^2 \Gamma_{\rm tree} ^{++--} ( K_1 , K_2,  K_3 , K_4 ) & = &
  G_0^{-1} ( K_3 ) +  G_0^{-1} ( K_4 )      + V^{\bot}_{\bd{k}_3}  + V^{\bot}_{\bd{k}_4 } 
 - V^z_{ \bd{k}_1 +  \bd{k}_3}   - V^z_{\bd{k}_1 + \bd{k}_4 }   \nonumber
 \\
&   &  \hspace{-29mm}  -  \left\{  \frac{ \delta (  \omega_1 +  \omega_3  )   b^{\prime}  }{ 
 1 - \beta V^z_{\bd{k}_1 + \bd{k}_3 } b^{\prime}} 
 \left(   G_0^{-1} ( K_3 )  + V^{\bot}_{\bd{k}_3} - V^z_{\bd{k}_1 + \bd{k}_3 }
  \right)  \left(  G_0^{-1} (K_4 ) +   V^{\bot}_{\bd{k}_4 }   -   V^z_{\bd{k}_1 + \bd{k}_3 }  \right)     
 +    ( K_3 \leftrightarrow K_4 ) \right\}
 \nonumber
 \\
 & & \hspace{-29mm} 
 + \Bigl\{ M_0 \Gamma^{+-z}_{\rm tree} ( K_1 , K_3 , - K_1 - K_3 ) F_0 ( - K_1 -  K_3 )  
 M_0 \Gamma^{+-z}_{\rm tree} ( K_2  , K_4 , - K_2  - K_4 )
  + ( K_3 \leftrightarrow K_4 ) \Bigr\} .
 \hspace{7mm}
 \label{eq:Gammappmmtree}
 \end{eqnarray}
In the last line we substitute again the tree approximation (\ref{eq:threetree})
for the three-point vertices and  obtain after some re-arrangements
 \begin{eqnarray}
  M_0^2 \Gamma_{\rm tree} ^{++--} ( K_1 , K_2,  K_3 , K_4 ) & = &  
 G_1^{-1} ( \omega_3 )  + G_1^{-1} ( \omega_4 ) 
 - [ \delta ( \omega_1 + \omega_3 ) + \delta ( \omega_1 + \omega_4 ) ]  b^{\prime}
  G_1^{-1} ( \omega_3 )   G_1^{-1} ( \omega_4 ),
 \label{eq:Gammappmmtree}
 \end{eqnarray}
which again can be obtained from the corresponding expression (\ref{eq:Gammappmmlocal})
for vanishing exchange couplings  by replacing $G_0 ( \omega ) \rightarrow G_1 ( \omega )$.
Finally,  for the mixed four-point vertex, we obtain from 
Eqs.~(\ref{eq:G4general2})  and (\ref{eq:G4tree2}),
\begin{eqnarray}
  M_0^2 \Gamma_{\rm tree} ^{+-zz} ( K_1 , K_2,  K_3 , K_4 ) & = & 
 -   \frac{ G_0^{-1} ( - K_2 ) - V^z_{ \bd{k}_3 + \bd{k}_4 } }{ 1 - \beta b^{\prime} V^z_{ \bd{k}_3 +\bd{k}_4 } }   \delta ( \omega_3 ) \delta ( \omega_4 ) M_0 b^{\prime \prime}
 \nonumber
\\
& &  \hspace{-29mm} - G_0 ( K_1 + K_3 )  -  G_0 ( K_1 + K_4 )
  + G_0^{-1} ( - K_2 )  \bigl[     G_0 ( K_1 + K_3 )   \delta ( \omega_4 ) 
+  G_0 ( K_1 + K_4 )   \delta ( \omega_3 ) \bigr] b^{\prime} 
 \nonumber
 \\
 & & \hspace{-29mm}
 + \Bigl\{ M_0 \Gamma^{+-z}_{\rm tree} ( K_1,  - K_1 - K_3, K_3  ) 
 G_0 (  K_2 + K_4 )  M_0 \Gamma^{+-z}_{\rm tree} ( - K_2 -   K_4 , K_2,  K_4 )
 + ( K_3 \leftrightarrow K_4 )  \Bigr\} 
  \nonumber
 \\
 & & \hspace{-29mm}
  + M_0\Gamma_{\rm tree}^{zzz}( K_3 , K_4, -K_3 - K_4 )F_0(-K_3 - K_4)M_0\Gamma_{\rm tree}^{+-z}(K_1 , K_2, -K_1 - K_2 ).
 \end{eqnarray}
Substituting Eqs~(\ref{eq:threetree}) and (\ref{eq:blockzzz})
for the three-point vertices,  we finally obtain 
 \begin{eqnarray}
 M_0^2 \Gamma_{\rm tree} ^{+-zz} ( K_1 , K_2,  K_3 , K_4 ) & = &
 - \left[ \delta ( \omega_3 ) [1 - V^{\bot}_{\bd{k}_2 + \bd{k}_4}G_0(K_2 + K_4)]  + \delta ( \omega_4 )[1 - V^{\bot}_{\bd{k}_2 + \bd{k}_3}G_0(K_2 + K_3)] \right]b^{\prime}
   \nonumber
 \\
 & & \hspace{-29mm}
 + \delta ( \omega_3 ) \delta ( \omega_4 ) G_1^{-1} ( \omega_2 ) \left[
 [2  - V^{\bot}_{\bd{k}_2 + \bd{k}_4}G_0(K_2 + K_4) - V^{\bot}_{\bd{k}_2 + \bd{k}_3}G_0(K_2 + K_3)]( b^{\prime} )^2  - M_0 b^{\prime \prime}
  \right].
 \hspace{7mm}
 \end{eqnarray}
\end{widetext}
In the limit of vanishing exchange interaction, we recover again
the corresponding
single-site irreducible vertex given in Eq.~(\ref{eq:Gammapmzzlocal}).
In a
deformation scheme where initially only 
the transverse exchange interaction is switched off,  
the tree approximation 
defines the initial condition for the SFRG flow provided the range of the 
longitudinal exchange interaction 
is sufficiently large so that the momentum integrations are suppressed by the inverse interaction range.

\end{appendix}


\begin{thebibliography}{99}
\bibitem{footnoteJ} 
For a Heisenberg ferromagnet with nearest neighbor exchange $J$ the bare dispersion of long-wavelength magnons is $E_{\bd{k}} = \rho_0 k^2$ with $\rho_0 = J M_0 a^2$.
This is consistent with our definition $J = \rho_0 / ( M_0 a^2)$ in Eq.~(\ref{eq:Jdef}).
However, as emphasized in the main text after Eq.~(\ref{eq:Vexpansion}),
the tree approximation used to calculate the  initial condition for our flow equations
can only be justified for long-range exchange interaction.
In this case $J = \rho_0 / ( M_0 a^2)$ simply defines  the energy scale associated with the bare spin stiffness.
\end{thebibliography}
\end{document}